Humboldt University of Berlin

School of Library and Information Science

Master Thesis

# The potential of preprints to accelerate scholarly communication

## A bibliometric analysis based on selected journals

by

Valeria Aman

Submitted for the degree of

Master of Arts in Library and Information Science

Referees:   1. Frank Havemann

2. Michael Heinz

Date of submission: 7[th] March 2013

# Table of Contents

**List of Figures**

## List of Abbreviations

| | |
|---|---|
| A&A | Astronomy & Astrophysics |
| AAS | American Astronomical Society |
| AJ | Astronomical Journal |
| API | Application Programming Interface |
| APS | American Physical Society |
| CERN | European Organization for Nuclear Research |
| COROT | COnvection ROtation and planetary Transits |
| DESY | Deutsches Elektronen-Synchrotron |
| DOI | Digital Object Identifier |
| EDP | Édition Diffusion Presse |
| Fermilab | Fermi National Accelerator Laboratory |
| HEP | High Energy Physics |
| IOP | Institute of Physics |
| JASIST | Journal of the American Society for Information Science and Technology |
| JHEP | Journal of High Energy Physics |
| JSON | JavaScript Object Notation |
| LANL | Los Alamos National Laboratory |
| LIS | Library and Information Science |
| NASA-ADS | National Aeronautics and Space Administration - Astrophysics Data System |
| NPB | Nuclear Physics B |
| PHP | PHP: Hypertext Preprocessor |
| PRD | Physical Review D |
| RSS | Rich Site Summary (or nowadays Really Simple Syndication) |
| SISSA | International School for Advanced Studies |
| SCOAP$^3$ | Sponsoring Consortium for Open Access Publishing in Particle Physics |
| SLAC | Stanford Linear Accelerator Center |
| URL | Uniform Resource Locator |
| WoS | Web of Science |

# 1.    Introduction

It is in the nature of science to conduct research and to publish results, mostly as peer-reviewed articles reporting findings, theories, models, or reviews. Research activities often implement techniques and ideas previously established by other scientists in the same research field. Bibliographic references reveal the researcher's dependency on already existing literature. An academic paper thus demands citations - and it is a small effort to cite. Citations can be used to measure the importance and influence of a single article, a journal, an author, or a group of researchers. Besides citations, there are several other metrics reflecting distinct facets of science. Bibliometrics in general, deals with measurable properties of the communication process in science. The communication ground in science is a network of published papers, above all journal articles. However, in the course of time, other means of communication have evolved. Among those are preprints, i.e., manuscripts that have not been peer-reviewed.

In a subject like high-energy physics (HEP) rapid dissemination of information is crucial, so that preprints became a necessity. In the past, scientists used to send printed drafts to colleagues to report the current state of research, and to receive valuable comments. The arXiv emulated this paper-based process that has been in existence for decades but designed it in an automatic way. It was Paul Ginsparg who set up arXiv in 1991, a preprint repository, originally devoted exclusively to high-energy physics. It was expanded to comprise astronomy, computer science, mathematics, quantitative biology, statistics, and quantitative finance. Preprint servers are one of the first choices for physicists and other researchers to find information on current topics, and to keep up with colleagues. Preprints enable not only an unlimited and free access to relevant information; they also allow the convenient dissemination of results. Nevertheless, the final publication in a journal is still inevitable and common practice for most researchers.

This work examines to which extent preprints can accelerate scholarly communication. The longsome process of peer review and journal publication has an adverse effect on science. Hence, any way of speeding up the publication cycle is worth supporting. The investigation of the acceleration of science is quite sensitive to the research field in question. Scholarly communication in social sciences and humanities is different from the communication in sciences because authors prefer to publish books instead of articles, and use resources older than those used by natural scientists.

This work is thus limited to arXiv, which provides subject areas that are known for their speed in the communication of results. Since arXiv includes distinct subject fields, it was only natural to choose relevant journals for the field in question. Nevertheless, the thesis does not cover all of arXiv's subject fields but restricts itself to HEP, Mathematics, Astrophysics, Quantitative Biology, and Library and Information Science (LIS). The bibliometric analysis cannot give an overall picture of data in arXiv. It only operates on the basis of selected journals for which preprints exist.

Little research has been done in bibliometric analysis on the basis of preprints. The overall goal is to study the acceptance of preprints as a scholarly channel, and to investigate their potential effect on the acceleration of scholarly communication. The aim is to analyze the growth of preprint numbers over the years, the publication delay, and the effect of preprints on citation rates. The following questions will be addressed in this work: Has the publication delay decreased over the years? Do publication delays vary by arXiv discipline? Do articles with a preprint deposited in arXiv get more citations than articles that do not have a foregoing preprint? Is the time advantage of preprints used to accumulate citations?

The paper is outlined as follows. After this short introduction, part two deals with a reflection on scholarly communication, and the benefits and drawbacks of accelerated dissemination of research results in science. Part three describes preprints in general. It provides a definition first; explains the meaning of the preprint culture; underscores plenty benefits, and discusses the issue of lacking peer review. Part four is devoted to the materials and methods used. It provides information about the databases used to retrieve the data, and describes the analysis performed. Part five comprises the core of the work. It deals with the bibliometric analysis of journals. For each of the five above-mentioned fields two or three journals were chosen to present findings and to discuss the results. The paper ends with a summarized discussion of the presented results, and a prospect with regard to the development of preprints as a means of scholarly communication.

The work was initiated by Frank Havemann. In 2004, he carried out an investigation to find out in how far preprints are accepted as a scholarly communication means (Havemann, 2004). His findings are overall consistent with the results presented in this paper. I am grateful for his support, and the time and effort he and Michael Heinz invested to improve my approach.

# 2.    Acceleration of scholarly communication

The scholarly communication means still in use within modern times, dates back to 1665, when the first scientific journals were published to report new ideas and discoveries among researchers. In the course of time, several other communication techniques evolved, some of which were improved, substituted, or abandoned. Nevertheless, refereed journal articles are today the primary mode of communication in scientific research.

Scholarly communication serves to boost the progress of research by disseminating knowledge and discoveries. It has both formal and informal manifestations. Formal communication can be based on printed or online journals, monographs, reports, or conference papers. Informal communication means are correspondence through mails or e-mails, blogs, face-to-face debates, or discussions among colleagues.

The rise of the Internet and the digitization of information led to a quantum leap in our communication culture. Although science has always been disseminated through distinct means, the methods of dissemination changed fundamentally in the past 20 years. Oral and written communication got intrinsically tied to each other. Mobile devices such as computers and phones allow communication at every time and place. The scholarly communication in a networked era enhanced not only its openness but also its speed.

What we associate with accelerated communication today are e-mails, blogs, Twitter, discussion groups, repositories, and electronic journals. Electronic journals made it possible "to become much more rapid, global and interactive" (Harnad, 1992, p. 90). Internet technologies enable to speed up the publication in journals because they are capable of reducing the time delays in the communication between authors, editors, reviewers, publishers, and readers (Kling & Swygart-Hobaugh, 2002). Undoubtedly, natural sciences are far ahead of social sciences when it comes to the implementation of technologies that boost the speed of communication.

It is certain that the Internet increased the velocity of the publication process and is likely to increase it in future. To which extent, depends not only on the discipline, but also on other factors, such as the publication output, the need for rapid communication, the peer review process, the mode of submission, etc.

For authors who are frustrated with long publication delays, the fact that the Internet accelerates scholarly communication may sound appeasing. It is obvious that acceleration in communication is essential when it comes to studies in medicine, or chemistry. Moreover, in almost every research discipline, accelerated communication can prevent from doing double research, and consequently saves time and money.

If communication is accelerated it is apparent that the output increases proportionally. This leads to a plethora of drawbacks. Firstly, an accelerated communication gives rise to information overload and is prone to errors. Since every field of scientific research requires being the first, overhasty publications of results can flood science with fallacious information. Lacking the capability to tell valuable information from misinformation, the normal reader or even a scholar might be misled. This problem could be faced with an open peer-review process, where every single researcher can act as a referee. However, even with a better peer-review system, errors are not unpreventable.

Another point of criticism is that not every participant in the research cycle can keep up with the speed of communication. A scholar might have a good idea for a paper but is scooped by another researcher, because he was not quick enough to publish his results. Furthermore, peer reviewers need more time to scrutinize a paper and publishers cannot cope with the flood of submissions. These few examples may illustrate why the wish for slower communication grows and why scholarly communication should abstain from speed. To get ahead, we might need to slow down.

Nevertheless, it is obvious that the solution to the above mentioned problems cannot be simply to slow down scholarly communication. We rather need new reliable methods of scholarly communication that speed up science, but thoroughly. The goal remains to accelerate progress in science by a transformation in scholarly communication. One way of improving the efficiency of scholarly communication are preprints, which developed into an acceptable form of academic publishing. They are capable of speeding up the communication process by making results accessible prior to journal publication. Preprints are also freely available to the entire community and the knowledge can be immediately used and cited. Nonetheless, the circulation of preprints should be restricted to certain subject fields or to scientists who can judge them correctly (Delamothe, 1998). Peer-reviewed publications should remain a standard in fields such as medicine, where wrong papers could have disastrous effects.

# 3. Preprints in scholarly communication

## 3.1 Defining a preprint

With the advent of arXiv and the World Wide Web, scientific literature used the terms e-print and preprint almost interchangeably. For the purpose of this work it is useful to distinguish these two terms. An e-print describes the general category of an electronic manuscript. This term can be used for any work, which an author makes electronically available. It may thus refer to a peer-reviewed paper, an unpublished paper or a preprint (Luce, 2001). According to Tomaiuolo & Packer (2000) preprints are:

> Papers that authors have submitted for journal publication, but for which no publication decision has been reached, or even papers electronically posted for peer consideration and comment before submission for publication. In fact, preprints can also be documents that have not been submitted to any journal.

Harnad (2003) distinguishes preprints from postprints simply by emphasizing that "the former are published before peer review, whereas the latter are research papers after peer review." The present work uses the term preprint to refer to a digital document that has been submitted to a repository without peer review. In the form of individual papers, a preprint server such as arXiv helps scientists to share their results immediately with the community. Preprint servers are mainly hosted at universities and professional institutions. Physics, astronomy, computer science, mathematics, chemistry, and medicine are leading research fields in preprint publication. This stems from the long-existing preprint culture in those fields.

## 3.2 The preprint culture

Paul Ginsparg coined the term "preprint culture" in 1994 to describe the way communication worked in high-energy physics for the decades before arXiv was set up (Ginsparg, 1994, p.157). More than 50 years ago the high-energy physics community developed into a preprint culture to boost scholarly communication beyond of the cumbersome journal publication process. According to Ginsparg, physicists realized that refereed journals were irrelevant for continuous research (ibid.).

As O'Dell et al. (2003) state, "Research into the tiniest of particles requires some of the biggest machines in the world […]. When research is this expensive, there is simply no room to do the same science twice." To avoid double research publication of the same outcomes, institutes printed their results as preprints, and distributed those copies of papers among researchers in this field. At the same time, they were sent to journals for publication. With preprints, researchers were able to share their findings before they had been refereed. Since the publication process of scientific journals was characterized by delays and inefficiency (Kling & Swygart-Hobaugh, 2002), physicists did not hesitate to cite findings before the journal article was published. The advent of faxes quickened the distribution of preprints but was not capable of reducing the workload.

The Internet was a true option to bypass the delay between submitting a manuscript and its peer-reviewed publication. As soon as e-mail became available, authors rather preferred to use this medium for sharing preprints than the slow distribution via fax or mail (Harnad, 2003). With the advent of the Web in 1989, the procedure was simply to deposit the preprint and to advise interested readers to its URL via e-mail or alerting lists.

Ginsparg's preprint server improved the speed of preprint distribution significantly. After some years of existence, a number of institutes announced that they would stop the expensive distribution of paper preprints (O'Dell et al., 2003).[1] At present, preprints are common practice, but their role varies among subject fields. Whereas the HEP community makes extensive use of preprints, other communities still rely on refereed articles.

To sum up, the preprint culture had existed for ages in high-energy physics and was simply transferred into the online world. The transition to a preprint server happened firstly in physics because physicists felt the strong urge for fast dissemination of results and they had high-performance servers at their disposal.

## 3.3    Benefits of preprints

In the first place, preprints bridge the time gap between submission and publication. They can be circulated immediately among scholars to make research quickly available and to establish priority. Preprints are used as an early warning system to keep colleagues away from research that may take several months to be published in a journal (Delamothe, 1998). In an era of accelerated communication, it is important to make the work publicly available as soon as the research results exist. Preprints are also a way to reduce the likelihood of avoidable parallel research because they help to identify quickly any correlations (Pinfield, 2001). Nevertheless, some authors are concerned that the early distribution of results enables other researchers to publish similar results in a journal. It may sound contradictory, but intellectual property is established by open publications, as is the case with preprints. The easier the publication is accessible to the public, the more it is protected from plagiarism.

---

[1] Ginsparg mentions in his article from 1994 that "larger high-energy physics groups typically spent between $15,000 and $20,000 per year on copying, postage, and labor costs for their preprint distribution" (p. 157).

In Pinfield's opinion (2001) preprint archives have democratized the scholarly communication process, because anyone with access to the Internet can enter preprint literature. An important benefit is that search engines can lead Internet users easily to preprints. Furthermore, any researcher can submit papers and participate actively in the progress of science. With preprint servers, comments can be received from a much wider community, and those comments can be included in the refined formal journal article. Undoubtedly, it is of high value for authors to receive critical comments that stimulate their drive to research as well as their outcomes. Finally, preprint archives enable scholars to increase their visibility and impact by self-archiving their papers. If their results are open to the community, they can be accessed, used, and cited.

## 3.4    Preprints vs. peer-reviewed articles

Peer-reviewed journals are a well-established medium in scholarly communication. Interestingly enough, the peer-review process has not changed significantly over the last years and remains the principal procedure of quality judgment. Without peer review, "research would not be reliable, or controlled" (Harnad, 2004, p. 82). However, the traditional peer review process is not designed to detect mistakes, fraud or plagiarism. According to Ginsparg (2003) this task is left to future readers. Publishers can only guarantee that the author is who he claims to be, and that the article is not fundamentally wrong, and of relevance to the reading community.

A negative aspect of peer review is that it delays the presentation of research results, and may be full of bias towards authors and their views (Tomaiuolo & Packer, 2000). Even back in 1994, Ginsparg criticized the peer-review process, arguing that HEP community members do not want to "rely on the alleged verification of overworked or otherwise careless referees" (Ginsparg, 1994, p. 157). Furthermore, he is of the opinion that the small filtering of peer-reviewed journals plays no significant role in HEP because researchers are able to draw their conclusion from the author's name, title, and abstract, judging if a paper is worth reading and approvable (ibid.).

Since there is no reliable certification for preprints, scientists are aware that they have to assess the credibility of the papers they are reading and have to decide if they want to cite those works in their own papers (Rodriguez et al., 2006). Undoubtedly, opponents of preprints claim that without a strict peer review, preprints would contain erroneous information and cause confusion. The threat of circulating poor research is especially worth considering in medicine. Prior versions of manuscripts may include errors that could endanger both physicians and patients (Tomaiuolo & Packer, 2000).

Observing the successful growth of the preprint culture and the arXiv one could assume that peer review has become obsolete in physics. However, there is no evidence that peer review is likely to disappear in future. As Harnad (2003) emphasized, it is "peer review that keeps preprints in physics to their high standards." This explains why a high amount of preprints is still simultaneously sent to journals for peer review and publication. There is nothing wrong about submitting a preprint to a journal for peer review, given the fact that arXiv is not able to implement conventional peer review because of cost and labor needed (Ginsparg, 2011).

It may be time, though, to alter the peer-review process, especially because peers review for free and their expenditure of time is not compensated. The following question arises. What is more useful for the scholarly community at large? Is it of higher value to have a single reviewer assuring the quality of a paper for the whole community or is it rather the community who has to ensure that a paper satisfies the requirements in a field? Rodriguez et al. (2006) are optimistic about the assumption that a community would be able to turn a preprint into a formal publication, and additionally to give a diversified review of the preprint.[2] This post-publication peer review involves a fair dialogue between the author, the editors, and the community. The more transparent the presentation of peer review is, the more it will help to comprehend critical review.

---

[2] Rodriguez et al. (2006, p. 150) propose to apply the interactive journal concept to preprints to use the Web for public discussions. This concept envisages a two-stage-review process in which papers are first reviewed in an open forum. After beneficial comments and the author's revision, the paper can undergo the standard peer-review process. This procedure reduces the work of referees and fosters the participation of the community. However, it can also lead to needless comments.

# 4. Materials and methods

## 4.1 Primary sources

### 4.1.1 Scopus

Numbers of publications as well as citations are easily available through databases such as Thomson Reuters' Web of Science (WoS) or Elsevier's Scopus. For the bibliometric analysis in this paper Scopus was used. It is considered as the largest abstract and citation database of peer-reviewed scientific literature.[3] The publisher Elsevier introduced Scopus in 2004 and nowadays it can be considered as a good alternative to the competitor Web of Science. Scopus covers more than 20,500 source titles from more than 5,000 publishers all over the world. The database not only supports research in scientific, technical, and medical fields, but also in social sciences, and arts and humanities.

Scopus extended its coverage over the years and nowadays also offers the search in "Articles-in-Press" and Open Access journals. It contains more than 49 million records, and about 2 million new records are added each year on a daily basis.[4] Scopus indexes journals, book series, and conference materials that have an ISSN assigned to them. It captures also conference papers which are not published in a serial publication with an ISSN. According to a study from 2007, Scopus "offers about 20% more coverage than Web of Science […] Scopus covers a wider journal range, of help both in keyword searching and citation analysis, but it is currently limited to recent articles (published after 1995) compared with Web of Science" (Falagas et al., 2007). Furthermore, Falagas et al. (2007) state that Scopus' "citation analysis is faster and includes more articles than the citation analysis of Web of Science."

Just as WoS, Scopus is a commercial service that requires an access fee. The registration for SciVerse Applications makes it possible to benefit from several applications such as the personalization of the website or the key for the API (Application Programming Interface).

---

[3] SciVerse (2012). What does Scopus cover? http://www.info.sciverse.com/scopus/scopus-in-detail/facts
[4] Ibid.

### 4.1.2   arXiv

In August 1991, the xxx.lanl.org preprint server for the high-energy physics community was announced by the Los Alamos National Laboratory (LANL) in New Mexico (Ginsparg, 1994). Three years later, Paul Ginsparg wrote about his invention: "Having concluded that an electronic preprint archive was possible, I spent a few afternoons during the summer of 1991 writing the original software" (Ginsparg, 1994, p. 159). The preprint server would not have been conceivable without the concourse of certain circumstances. On the one hand, the World Wide Web was introduced by Tim Berners-Lee in 1989, and the availability of computers was growing since, along with greater bandwidth (Lucas-Stannard, 2003). On the other hand, the computer program LaTeX, created by Donald Knuth in 1977, and improved in the 1980s by Leslie Lamport, became indispensable for typesetting documents, especially mathematical formulas and equations (ibid.).

Ginsparg designed the software as an automated system, which researchers could maintain without any intervention. With this digital archive, Ginsparg wanted to facilitate not only the search for information but also the submission and replacing of papers. What has begun as an experiment to outsmart the cumbersome publication process in journals became within a short time, the preferred communication means in high-energy physics. In 1999, it changed its address into arXiv.org. The spelling has its origin in the Greek letter "X" (chi), which also imitates Donald's Knuth usage in the typesetting language LaTeX.[5] The arXiv is now hosted at Cornell University in New York, comprising nine mirror sites all over the world.[6]

Initially, arXiv covered only high-energy physics, but up to the present day it has grown outside HEP, including astronomy, computer science, mathematics, physics, quantitative biology, quantitative finance, and statistics. The number of preprints in arXiv for the period August 1991 to December 2012 is 810,705.[7] It receives up to 8,000 new papers each month.[8]

Scientists can conveniently disseminate their manuscripts and share their results with a wide community of researchers. Every researcher is welcome to upload their paper if it is of any value to the community. There is no refereeing system in arXiv; instead there are moderators who determine what is of potential use for the community. They can move preprints from one section to a more appropriate or withdraw junk papers. Ginsparg also set up a plagiarism check to scan papers and to find out whether they resemble already existing papers (Bernstein 2008, p. 151). The copyright is maintained by the author publishing in arXiv.

---

[5] arXiv (2012). What's been New on the arXiv.org e-print archives. http://de.arxiv.org/new
[6] arXiv (2012). arXiv mirror sites. http://arxiv.org/help/mirrors
[7] arXiv (2013). arXiv submission rate statistics. http://arxiv.org/help/stats/2012_by_area/index
[8] arXiv (2013). arXiv submission statistics in graph form. http://arxiv.org/show_monthly_submissions

The arXiv is well-adopted and serves a wide community. Although arXiv functions successfully without peer review, scientists still regard it as worthwhile to be finally published in a peer-reviewed journal. Consequently, a high amount of papers is submitted for journal publication. Nevertheless, arXiv has the potential to act as a platform for open peer review.

### 4.1.3   INSPIRE HEP

INSPIRE HEP is a high-energy physics information system that serves scientists as a research tool.[9] It is run by the Deutsches Elektronen-Synchrotron (DESY) in Hamburg, the Fermi National Accelerator Laboratory (Fermilab) in Illinois, the Stanford Linear Accelerator Center (SLAC) in California, and the European Organization for Nuclear Research (CERN) in Geneva. INSPIRE interacts with HEP publishers, arXiv, NASA-ADS[10] and other information providers. The database offers detailed record pages and searchable full text for arXiv documents.

The predecessor of INSPIRE was the Stanford Physics Information Retrieval System (SPIRES) which was set up by SLAC in 1969. It was designed as a database management system to deal with preprints in high-energy physics and was regarded as the first grey literature database.[11] Shortly after, the DESY library joined the SLAC library to cover the complete HEP literature consisting of preprints, reports, journal articles, conference papers, and books (Heuer et al., 2008). Since 1974, SPIRES covered all existing HEP literature in both preprint and published form (Gentil-Beccot et al., 2009). In 1985, the database included more than 185,000 metadata records. Today it is close to one million records, boasting more than 994,000 papers.

SPIRES became in 1991 the first database accessible through the World Wide Web.[12] In 1992, SPIRES was linked to arXiv for full text. In return, SPIRES provided detailed indexing and citation data for preprints in arXiv. Besides arXiv, it is also interlinked with other databases offering information on authors, institutions, experiments, and conferences (Heuer et al., 2008). In 2012, SPIRES, which was curated at DESY, Fermilab and SLAC, was combined with CERN's Invenio digital library technology to enable a community-based information system under the name INSPIRE HEP.[13] It added functionalities such as search speed, full-text search, and capture of user-generated content.

Since the communication in HEP is based on preprints, INSPIRE collects citations to and from preprints. As soon as a preprint is published in a journal, the citations to the two versions are treated as a single entity. It is worth mentioning that INSPIRE only tracks content relevant to HEP. It is used mainly by universities, colleges, and research institutions.

## 4.2    Data collection

Basically, a MySQL database has been set up with the help of a computer scientist[14] in order to enable me to query the data retrieved. The data collection was performed by using the three databases presented above. First of all, a key for the Scopus API was requested, to allow data retrieval in the response format JSON (JavaScript Object Notation). The text-based format JSON enables human-readable data exchange. The starting point was to search for the journals in question and to download their articles. Scopus provides several fields on article-level. For the bibliometric analysis only the following fields were of interest: title, document type, cited by count, source title, ISSN, volume, issue, page, publication date, DOI (Digital Object Identifier), first author, and affiliation. Thus, articles of a given journal were downloaded from Scopus via the API with the above listed metadata fields.[15] The Scopus API was also used to collect the metadata of citing articles, i.e., the future articles citing the journal article in question.

---

[12] SLAC (2006). The Early World Wide Web at SLAC: Early Chronology and Documents. http://www.slac.stanford.edu/history/earlyweb/history.shtml
[13] INSPIRE HEP (2011). INSPIRE Project Information. http://www.projecthepinspire.net/
[14] I owe special thanks to Daniel Lunow, who wrote the full programming code and made it possible to retrieve automatically data to this extent. The complete data collection was performed by him according to my instructions.
[15] To avoid server utilization, and to restrict the performance, a break of one or two seconds between each request was embedded.

This top-down-procedure started with the journals, retrieving the articles, and then the related citing articles. It worked without any complications but was time-consuming for large journals. The next step was to match the articles downloaded from Scopus with corresponding preprints in arXiv. The arXiv API was used to grant programmatic access and to extract metadata. As it is stated on the website, "the goal of the interface is to facilitate new and creative use of the vast body of material on the arXiv by providing a low barrier to entry for application developers."[16] The arXiv returned a long list of results with potential preprints in the Atom 1.0 format.[17]

After the potential preprints were collected, a matching was applied under consideration of some characteristics. The primary criterion was the publication date. Only those preprints were selected whose publication date is prior to the date of article publication. It is worth mentioning here that a large amount of documents in arXiv are in a strict sense postprints. A further important criterion was the first author of an arXiv preprint, who had to coincide with the first author retrieved from Scopus. Since the authors using arXiv come from different countries, they have various names and spellings. Several Unicode characters had to be substituted by ASCII characters. Apparently, authors who submit papers to arXiv spell their names in their native language, whereas publishers prefer to spell the author's name in Latin alphabet. As an example, the German "ß" had to be substituted by "ss".

In case the titles matched but the author's name did not appear in a preprint because an institution was listed instead, the match was counted as valid. To match titles, the PHP[18] similar text function was applied to the title. The similarity value was set to 85%. If there was more than one preprint matching the article, the one with the greater similarity was chosen. However, before the matching process could work accurately on titles, all letters, especially Greek letters, had to be substituted by small letters. Additionally, all Greek letters had to be substituted by their equivalent written words. Some mathematical symbols were substituted by other symbols, e.g., the Unicode symbol for a function arrow was substituted by a hyphen and a greater than symbol. The rendering of mathematical symbols and formulas may remain incomplete, but these additional regulations, made it possible to retrieve from arXiv the preprints that fitted the articles best. In case there were multiple versions of the same manuscript, the most up-to-date preprint version was chosen. Though, in certain cases, an older version was preferred, so that the criterion of a preprint was still fulfilled (because the most up-to-date version would be a postprint in these cases).

---

[16] arXiv (2012). arXiv API. http://arxiv.org/help/api/index
[17] Atom is a lightweight xml-based format that is used in website syndication feeds. arXiv (2012). arXiv API. http://arxiv.org/help/api/index
[18] PHP stands for PHP: Hypertext Preprocessor and is a scripting language to create dynamic Web pages.

The final step was to use the INSPIRE HEP database to retrieve citation metadata for papers related to HEP. The search for the arXiv ID was feasible with the RSS feed.[19] There were two different search strings to search for items in INSPIRE HEP that relate to a preprint in arXiv. One way of searching was to pick only the arXiv ID without the version number. The other way was to choose the string "arXiv:" followed by the ID, again without the version number. In INSPIRE HEP's metadata scheme the field "description" consists only of plain text and includes the arXiv ID in the first paragraph.[20] With these retrieval strategies potential data for the preprints in arXiv was gained. On the basis of the ID search, the probability was very high to find arXiv preprints in INSPIRE HEP. Nevertheless, it is possible that either a wrong arXiv preprint was matched, or has not been found at all because the search was only based on the "description" field in INSPIRE HEP. Furthermore, the INSPIRE HEP record ID was used to search for citing preprints. For this purpose the same RSS feed was used. Since the search for citing preprints was performed on the basis of the record ID, mistakes at this stage are almost excluded.

A short note on data reliance: The following example shall demonstrate the completeness of the data retrieved. In the Annals of Mathematics, volume 175 (2012), issue 1, 13 articles were published. For 11 of these articles a preprint was matched according to the above-described search strategy. Of the two remaining articles, one has according to arXiv no preprint, whereas for the other article a preprint has been found manually in arXiv. The preprint's title is: "Every ergodic transformation is disjoint from almost every IET".[21] This title has been changed in the Annals of Mathematics to the following: "Every ergodic transformation is disjoint from almost every interval exchange transformation".[22] Since in the final paper the acronym has been dissolved, the title is unfortunately below the similarity threshold of 85%,

---

and consequently not counted as an appropriate preprint. Evidently, the whole search procedure described here is prone to errors, but the utmost has been done to avoid any mistakes.[23]

## 4.3    Data analysis

Bibliometric analysis may be considered as primarily valid for larger sets. Nevertheless, the following bibliometric analysis is limited to a small number of journals. The journals were primarily chosen according to the criterion that preprints exist in arXiv. To decide on the journals, the arXiv was browsed beforehand in order to find striking journals in arXiv's Journal-ref. field. Since arXiv offers several disciplines, it was self-evident to compare different subject areas. To make a thorough analysis, some areas had to be omitted, mainly because their short existence in arXiv forbids a long-term analysis.[24]

The study is based on 13 journals, examined on article-level. The dataset does not only contain articles, but also errata and reviews because preprints were found for these items, as well as citations. For the following analysis a preprint is defined as a manuscript in arXiv that has been published at a later date in a journal. Two groups of data are mainly compared to each other. On the one hand, articles with a previous preprint in arXiv, and on the other hand, articles without a preprint in arXiv. It was not checked whether preprints deposited in arXiv have been also made available in any other institutional or disciplinary repository. Although all existing articles for a journal were downloaded from Scopus, the MySQL queries were limited to the following period of time: $1^{st}$ January 1996 to $31^{th}$ December 2012. This limitation is imposed by Scopus which does not provide cited references for publications prior to 1996.

The bibliometric analysis deals first and foremost with the growth of the publication output of a journal, and the number of articles published with a previous preprint in arXiv. After that, the publication delay is examined, which is originally defined as the time between the submission of a manuscript and its appearance in print or electronic form (Kling & Swygart-Hobaugh, 2002). However, for the purpose of the following analysis the publication delay is defined as the chronological distance between the deposit of a preprint in arXiv and its formal publication in a print or online journal. It is thus the date difference between the date of deposit in arXiv and the date of article publication according to Scopus. With the aim to examine whether the publication speed has decreased or increased over the years, the median publication delay as a function of time will be analyzed for journals that provide sufficient data. In addition, the distribution of preprints over months prior to publication will be considered.

---

[23] One example is that articles were found that were cited from the past, citations dating back more than 30 years. This is due to the fact that review articles can be updated years later, citing recent literature but keeping their original year of publication.
[24] Statistics is included in arXiv since April 2007, Quantitative Finance since December 2008.

With regard to the impact of preprints and the speed of communication, total citation numbers, and the time it takes until the first citation will be analyzed. For HEP and Astrophysics, citation data is based on INSPIRE because these fields are related to HEP. For journals not related to HEP the analysis is based on Scopus. This allows comparing articles having a previous preprint in arXiv with articles that do not feature a previous preprint. Welch's t-test will be applied to examine if there is a significant citation advantage for articles with a previous preprint. This test is also used to examine the citation delay of articles with a foregoing preprint, and of those without a preprint. In this context, the citation delay is defined as the chronological distance between the date of publication and the date of the first citation. In addition, for journals that provide sufficient data to operate with, the average citation rates will be compared within a fix citation window. Finally, it has to be mentioned that the amount of data retrieved allows for far more analyses than the ones presented in the following part.

# 5.    Bibliometric analysis

## 5.1    High-energy physics

The total number of submissions for arXiv's HEP, since the establishment in 1991 until the year 2012, is 20.2% (164,075).[25] Thus, every fifth paper deals with the field, the arXiv was originally aimed for. Since 1991, arXiv's growth in submissions has been linear. As Harnad (2003) illustrates, "This was fast enough so that within a few years virtually the entire annual high-energy physics output was archived there and accessible to all." The linkage of INSPIRE HEP and arXiv guarantees full coverage of HEP literature. These two databases are used as the primary points of entry to the literature by high-energy physicists. The majority of research articles in HEP are published by six leading journals: Physical Review D, Journal of High Energy Physics, Nuclear Physics B, Physics Letters B, European Physical Journal C, and Physical Review Letters. It is often claimed that their content is entirely in arXiv. The following bibliometric analysis is restricted to just three of the six main journals in the field: Physical Review D, Journal of High Energy Physics, and Nuclear Physics B. Gentil-Beccot et al. (2009) write that Physical Review D and the Journal of High Energy Physics cover about 50% of all HEP literature. Thus, the three chosen journals are representative of the entire scholarly communication in HEP.

### 5.1.1    Physical Review D

Physical Review D - Particles, fields, gravitation, and cosmology (PRD) is a leading journal in elementary particle physics, field theory, gravitation, and cosmology. It is published by the American Physical Society (APS) and appears monthly since 1970. In a stricter sense, Physical Review D is split into D1 and D15. Whereas D1 reports on experimental HEP, cosmic-ray physics, and electroweak interactions, D15 covers general relativity, cosmology, particle astrophysics, and string theory.[26] Physical Review D's Impact Factor for the year 2011 is 4.558. The journal appears in two volumes a year, with 12 issues each. From 1998 onwards (volume 57), Physical Review D changed its publication process and started to publish articles first electronically, with the date of posting as the publication date.[27] At monthly intervals these articles are collected to make up a printed issue. Since 2011, authors have the option to publish Open Access in PRD at a fee of $1,700.[28] During the analysis of the data it became clear that Scopus' provision of the publication date was insufficient for the years 1996 to 2006.[29] Hence, for the years prior to 2006 the date of publication was deduced from the volume and number of the issue. The publication date was always set to the 15th of a month. As a result, the date of publication may differ ±15 days from the actual date of publication.

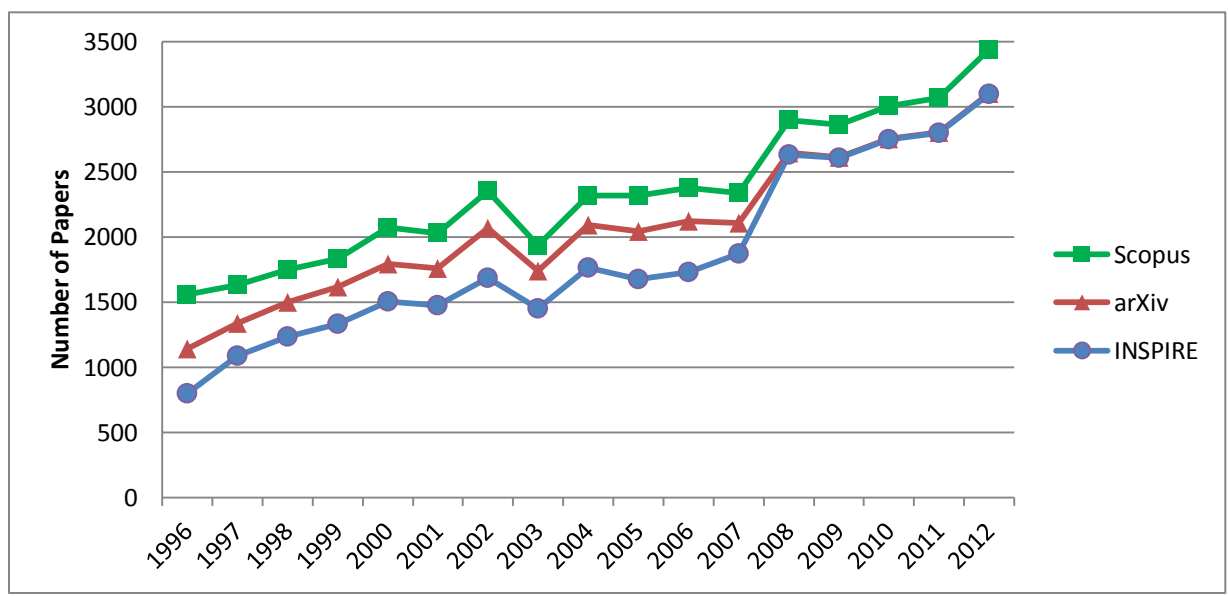

**Figure 1.** Growth of publication numbers from 1996 to 2012 for Physical Review D.

In Figure 1 the publication output for Physical Review D can be seen. The green line graph shows the number of articles retrieved from Scopus.[30] The numbers may slightly sheer from the actual numbers of articles published in PRD, but in regard to the total number it is a negligible quantity. It is visible that the number of journal articles has doubled from 1996 to 2010, and is still on the rise. The red line graph represents the number of articles with a preceding preprint in arXiv. It is evident that the Scopus graph and the arXiv graph are similar for the last 17 years. The percentage of articles with a previous preprint in arXiv ranged from 73.3% in 1996 to 91.6% in 2010. In contrast, the blue line graph, representing the number of articles with a previous preprint in INSPIRE HEP, keeps his quotient until the year 2007. From 2008 onwards, we can see that the arXiv graph and the INSPIRE graph conflate into a single one. This can be explained by the fact that since 2008, INSPIRE HEP is no longer selective and includes every single preprint that is published in arXiv and related to HEP. The percentages for the last five years show that PRD's content is almost completely covered by INSPIRE HEP. To quantify Figure 1, from 1996 to 2012, overall 39,777 articles were published in Physical Review D. 35,236 of them have a preceding preprint in arXiv, which results in a share of 88.6%. In INSPIRE HEP, 89.4% (31,486) of all preprints in arXiv that were submitted to PRD at a later date, can be found. It is common practice not only to upload preprints in arXiv but also postprints. For the time period 1996-2012, 683 postprints (1.7%) can be found for PRD in arXiv, of which the majority (56.7%) is posted within the first 30 days after journal publication.

---

[27] APS (2012). Early Electronic Publication of Physical Review D Articles.
http://www.aps.org/publications/apsnews/199802/prd.cfm
[28] Open Access describes published papers that are available to the reader at no cost. The "green" Open Access functions without any payment because the authors simply make their articles freely available, such as depositing them on arXiv. For the "gold" Open Access way either the author or the research funder pays, so that the reader has free access to the articles.
[29] For the year 2004, Scopus provides only 7 distinct publication dates, instead of 24 (all of them in the second half of the year). Without any amendment, these wrong publication dates would distort the analysis of the publication delay as well the complete citation analysis.
[30] Articles refer in this context to all single papers published in PRD, such as brief reports, rapid communications, comments, errata, and genuine articles. The same applies to all other journals.

In 2009, Ingoldsby found an exact match of 97.1% for articles published in Physical Review D and the equivalent in arXiv.[31] The number is much higher than the one presented here. This is due to the fact that the number presented above only takes preprints into account and is calculated on the basis of a longer time period. Moreover, Ingoldsby restricted his analysis to articles, whereas above, all kind of papers are included because arXiv proves that preprints exist for comments, errata, and reviews as well.

Havemann (2004, pp. 6-7) analyzed 211 theoretical articles published in PRD, in volume 58 (1998), and 59 and 60 (1999). With 203 articles, he found a share of 96% of articles, which were also available in arXiv. This shows that either the inclusion of comments and errata reduces the percentage of preprints found, or that the threshold for the similarity function was set too high. As a consequence, one and the same paper may be instantly visible for the trained human eye, but not for the machine that operates too strictly. Nevertheless, the number of preprints matched for PRD is relatively high and can be explained as follows. The HEP community is a very small one compared to other fields and above all, the arXiv was aimed for this community. It is common practice among physicists to submit their preprint to arXiv and simultaneously to a journal. In addition, Physical Review D prefers web-based submissions, over other methods. Submissions in electronic formats such as LaTeX, which is also appreciated in arXiv, even qualify for waiver of publication charges.[32] In the mid-nineties, PRD introduced accelerated publications, called Rapid Communications, which are intended for important new results and which are given priority in peer review and editorial processing. Since Physical Review D operates under accelerated circumstances, it is of interest to see the publication delay of all articles published in PRD between 1996 and 2012.

---

[31] Ingoldsby (2009, p. 8) studied 2,100 articles published by the APS of which he found 97.1% from Physical Review D, 40.0% from Physical Review B (Condensed Matter and Materials Physics), and 55.0% from Physical Review Letters in arXiv.

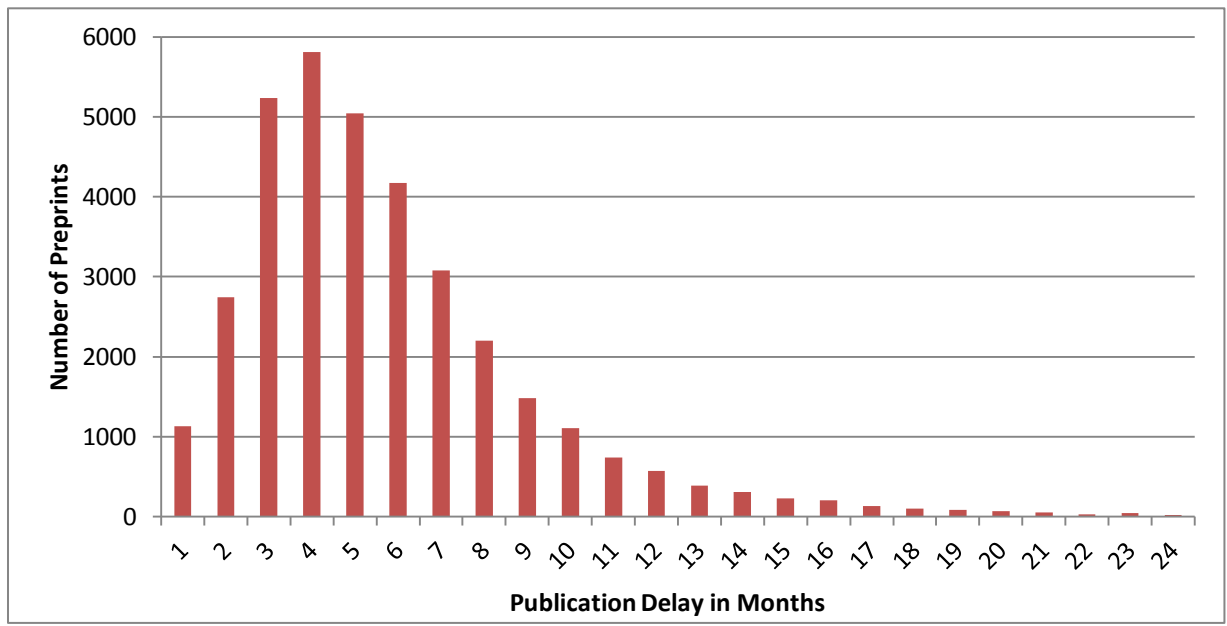

**Figure 2.** Distribution of preprints over months prior to journal publication. Those preprints are considered whose articles were published between 1996 and 2012 in Physical Review D.

Figure 2 illustrates the publication delay of all preprints that were published in PRD as articles between 1996 and 2012. The publication delay is split into blocks of 30 days, thus approximately one month. 217 additional preprints are not displayed in Figure 2 because they are scattered over the months 25 to 94. It is visible that the mode of the distribution is four months. It means that a majority of 16.5% of all preprints are published in arXiv 91 to 120 days before the article appears in PRD. More than half of all preprints are uploaded within five months prior to publication (56.7%). A high amount of 94.6% of all preprints in the reference period is placed in arXiv within one year prior to formal publication.

---

[32] APS (2013). Web Submission Guidelines. http://publish.aps.org/authors/web-submission-guidelines-physical-review

Due to improved Internet technologies, one could assume that the publication delay has decreased in recent years. Therefore, the next figure investigates if this is really the case. On the one hand, it shows bars representing the median number of days that elapses between the submission in arXiv and the journal publication in the respective year. On the other hand, it shows again the growing number of articles with a previous preprint in arXiv, because the calculation of the median publication delay is based on these numbers.

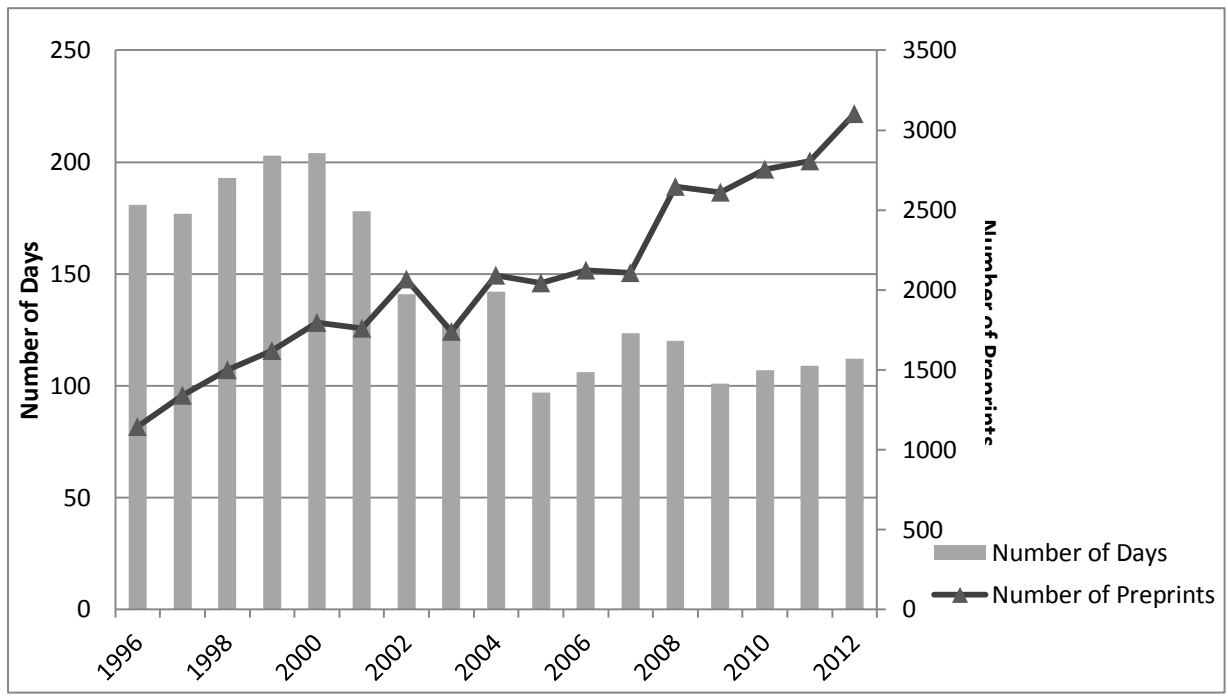

**Figure 3.** Time series of the median publication delay in days for articles that were published between 1996 and 2012 in Physical Review D and have a previous preprint in arXiv.

We can infer from Figure 3 that the median publication delay has used to be around six months from 1996 to 2001 and started to decrease since 2002.[33] It ranged between three and four months in recent years and was surprisingly consistent. It is noteworthy that with a growing number of articles published, the median publication delay has gone down over the years, either because the publisher cut down any delays in the publication process, or because authors upload their preprint in arXiv closer to the anticipated date of journal publication.

Another important question is if the publication delay influences the impact of preprints. The first citation is an indicator of the visibility of a preprint. It shows that another author makes use of the content and disseminates it to the interested reader. The shorter the time gap between the completion of a preprint and its first citation is, the more it is likely to be of value. This time gap is primarily dependent on the publication delay. The next figure depicts a histogram of all preprints that are included in INSPIRE HEP and were published in article form between 1996 and 2010. The citation window is set to two years after journal publication.

---

[33] Havemann (2004) found a publication delay of seven months for 203 articles published in PRD between 1998 and 1999.

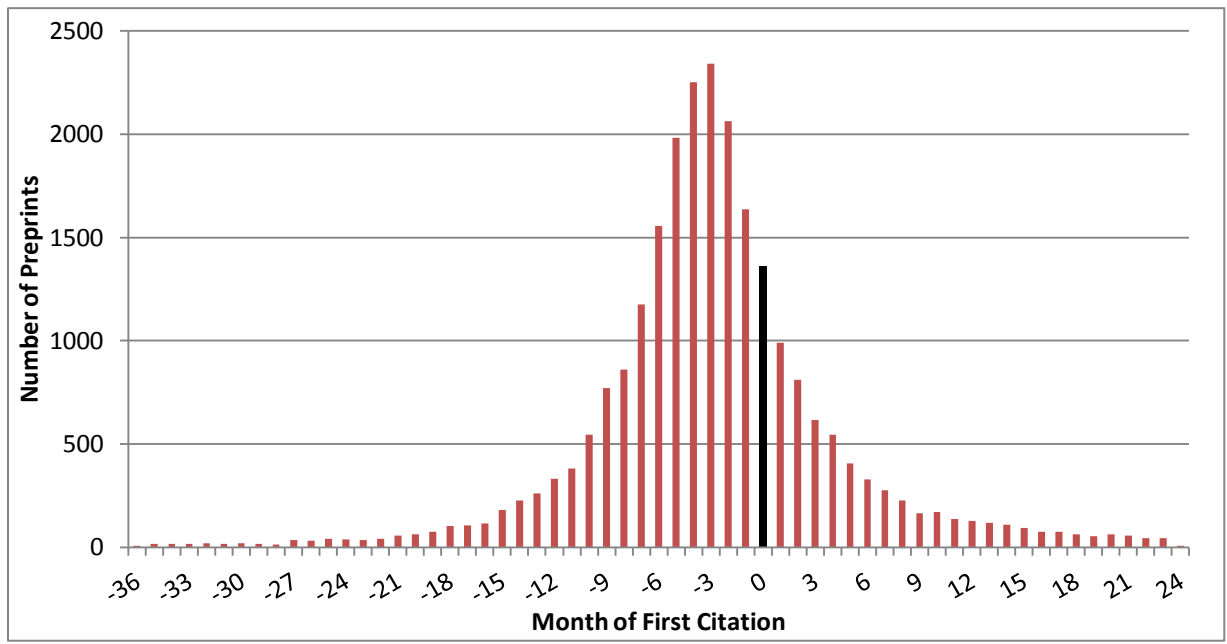

**Figure 4.** Distribution of first citations over months after journal publication for papers in INSPIRE HEP that were published between 1996 and 2010 as articles in Physical Review D. The citation window is two years after article publication and the citation data is based on INSPIRE HEP.

The results in Figure 4 demonstrate a citation advantage for preprints submitted to arXiv and published in PRD at a later date. INSPIRE HEP tracks a total of 25,591 preprints, which appeared between 1996 and 2010 as articles in PRD. 960 of them remained uncited within the first two years after publication, and are consequently not displayed in Figure 4. The black bar signifies the number of citations within the first 30 days after journal publication. A majority of 9.1% preprints (2,341) received their first citation 61 to 90 days prior to journal publication. This is visible as the mode of the distribution. On the whole, 69.1% of all preprints received their first citation before they were published in the journal. This number is consistent with the one Havemann (2004) presented in his paper. He found that for articles published between 1998 and 1999 in PRD, 75% of preprints were cited before the online journal article was available. Does the advantage derive from a wider accessibility or an earlier distribution of preprints? The publication delay has shown that there is a time advantage for preprints of several months to be read earlier. The first citation implies that the preprint is not only read but also reused in another paper. It suggests that researchers in HEP do not particularly distinguish between preprints and articles, and do not hesitate to cite preprints in their papers.

To receive the first citation prior to publication does not necessarily mean that the majority of citations will be awarded to preprints. Therefore, the next figure provides a time series of all citations awarded to papers published in PRD between 1996 and 2010.

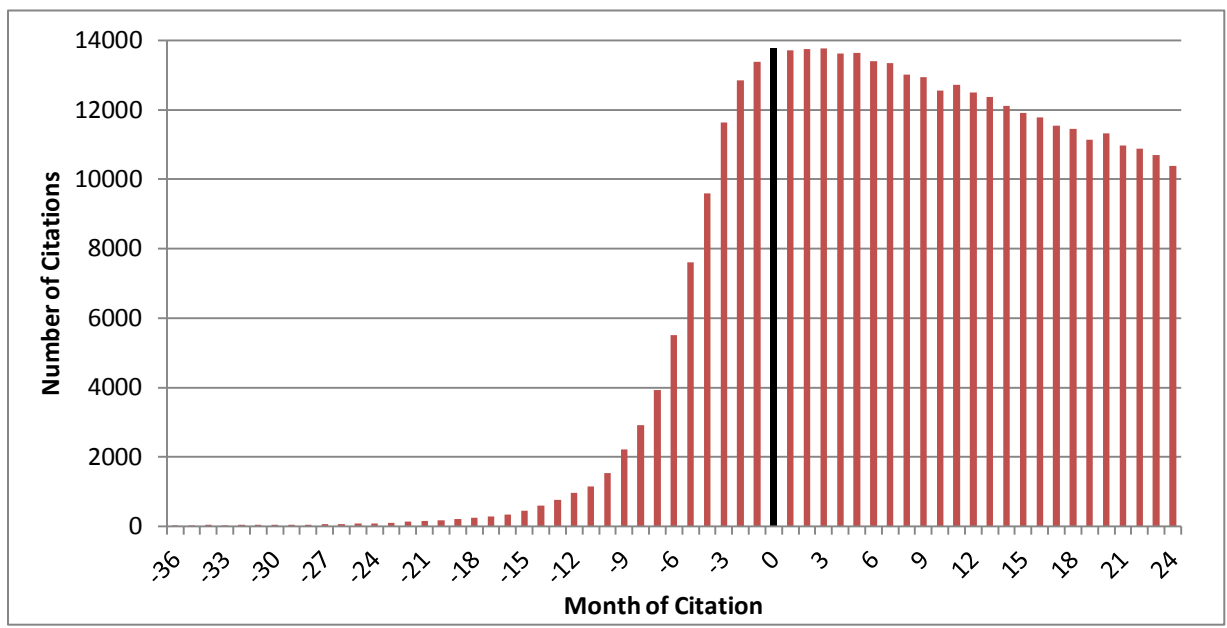

**Figure 5.** Distribution of all citations over months after journal publication for papers in INSPIRE HEP that were published between 1996 and 2010 as articles in Physical Review D. The citation window is two years after article publication and the citation data is based on INSPIRE HEP.

Figure 5 shows all citations for papers as a function of time, in relation to journal publication. The black bar represents the period from 1 to 30 days after journal publication. It is evident that preprints start to receive many citations prior to publication. This is illustrated by the bars left to the black bar. The area right to the black bar, thus after the journal publication, is of course much larger. The mode of the distribution is three months after publication. Since the number of papers covered by INSPIRE is high, the bars have a smooth distribution and little variation. It is furthermore visible that the accumulation of citations has an exponential behavior. As a result, 20.1% of all citations within two years after publication fall on the time prior to publication. According to INSPIRE HEP the citations received for preprints and their published article version are treated as one entity. From this it follows, that citation data tracked by Scopus is to some degree overlapping with the data tracked by INSPIRE HEP.

### 5.1.2 Journal of High Energy Physics

The Journal of High Energy Physics (JHEP) is owned by the International School for Advanced Studies (SISSA - Trieste, Italy) and is published by Springer.[34] This online journal was established in 1997 and works with an electronic peer-review system. It is complemented by the preprint distribution via arXiv, "that has so successfully replaced the conventional system."[35] Its Impact Factors for the year 2011 is 5.831. The journal is currently published monthly (1 volume a year, with 12 issues). According to Mele et al. (2006) JHEP contributes 19.2% to the total HEP literature. Since 2007, JHEP offers Open Access and has an embargo policy of two years. Authors that are affiliated to a subscribing institution can publish Open Access without paying any charge.

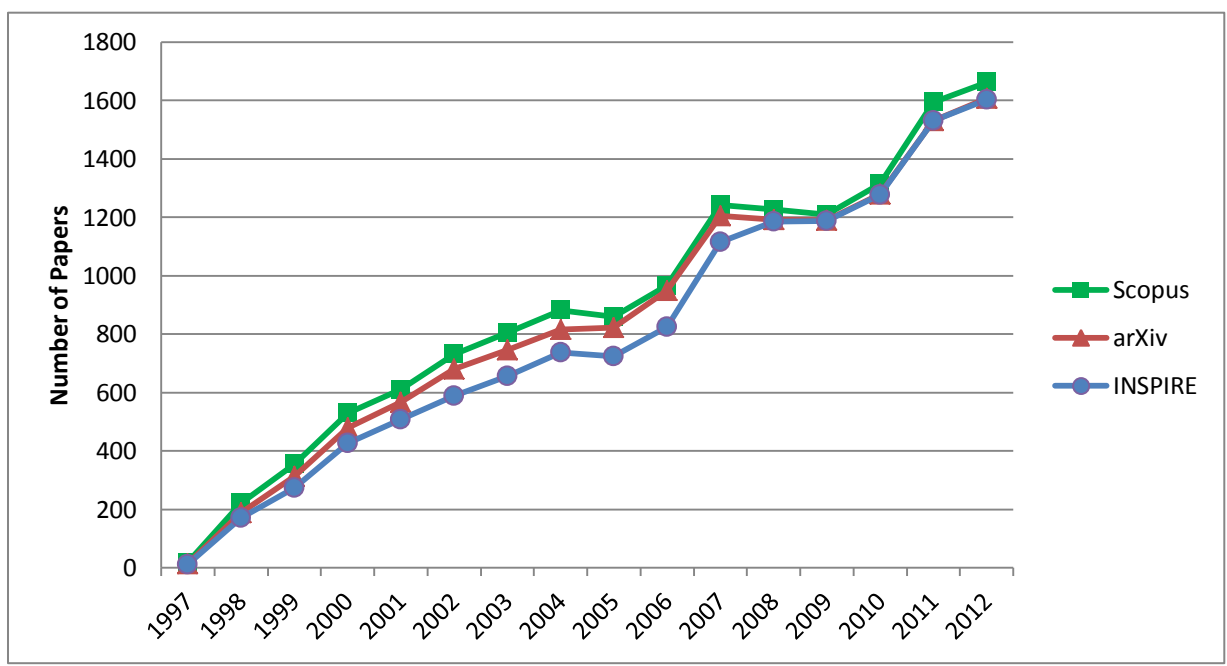

**Figure 6.** Growth of publication numbers from 1997 to 2012 for the Journal of High Energy Physics.

As can be seen in Figure 6, the number of papers has been growing almost linearly since the journal's formation. It is obvious that all three graphs are very close to each other, which means that a lot of preprints exist in arXiv that were later published in JHEP, and that many of them are included in INSPIRE HEP. Just as for PRD, the line graphs representing arXiv and INSPIRE HEP conflate into a single one in 2008. This is not surprising, since the title of the journal reflects exactly INSPIRE's mission to include literature related to HEP. The overlap of preprints is close to 100% from 2008 to 2012, with an exact match of 100% in 2011. Overall, between 1997 and 2012, 14,226 articles were published in JHEP of which 13,575 have a previous preprint in arXiv. This makes up 95.4% of all articles published until 2012. In addition, arXiv features 345 postprints (2.4%), which were deposited in arXiv shortly after their publication in JHEP. The number of preprints that can be found in INSPIRE HEP is comparatively high, with a share of 94.4% of all arXiv preprints, published between 1997 and 2012 in JHEP. The high amount of preprints is explainable by the fact that the journal was founded in 1997, after the arXiv has been long established, and has become standard practice in HEP. Moreover, the publisher permits to publish prior versions of articles "on non-commercial pre-print servers like arXiv.org […] and they can be updated with the author's accepted version. Springer requires a link to be inserted to the published article on Springer's website."[36] Surprisingly, only 132 e-prints in the reference period have a note on the journal publication, within the Comments or the Journal-ref. field in arXiv. Since the publisher invites authors to publish prior versions in arXiv, it is of interest to see how much time lies between the upload in arXiv and the final publication in JHEP.

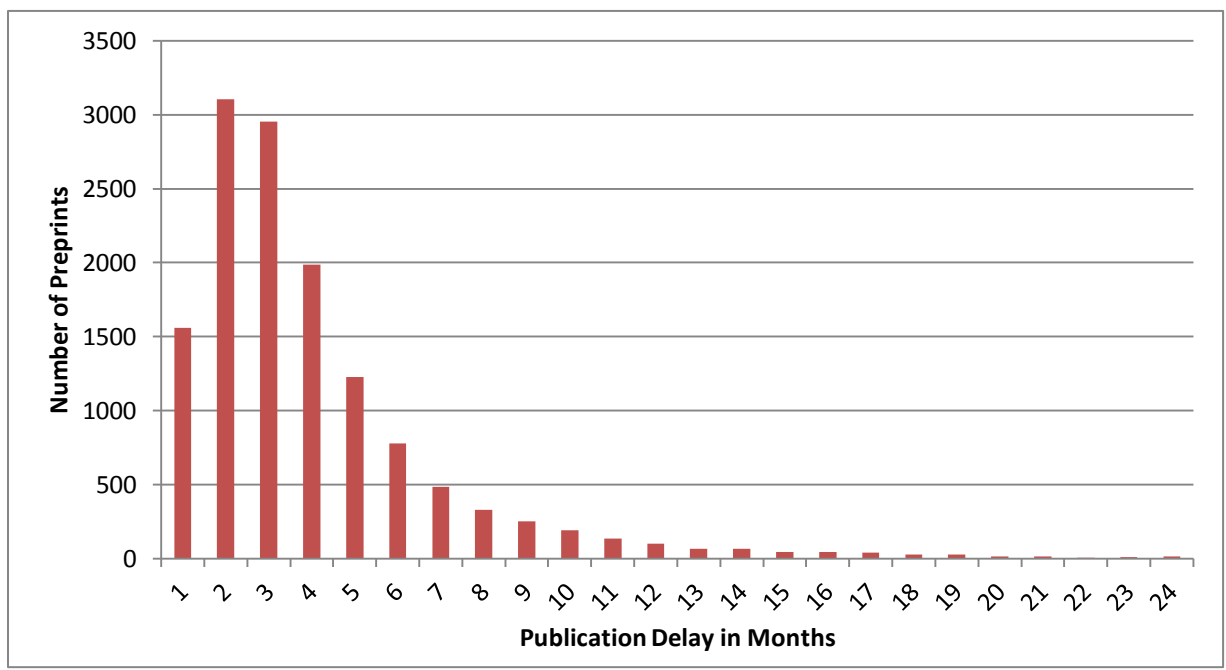

**Figure 7.** Distribution of preprints over months prior to journal publication. Those preprints are considered whose articles were published between 1997 and 2012 in the Journal of High Energy Physics.

---

[36] Springer (2010). Copyright Transfer Statement. www.springer.com

We can gather from Figure 7 that the mode of the distribution is two months and thus shorter than for Physical Review D. The majority of preprints are deposited in arXiv 31 to 60 days prior to formal publication (22.9%). Within five months before journal publication 79.7% of all articles are available as preprints in arXiv, within one year even 96.5%. Figure 7 does not include the long tail of 89 preprints, which are scattered over the months 25 to 105 prior to publication. The visualization of the publication delay of all preprints in the respective period suggests that it is lower than for PRD; but has the median publication delay been decreasing over the years, just as for Physical Review D?

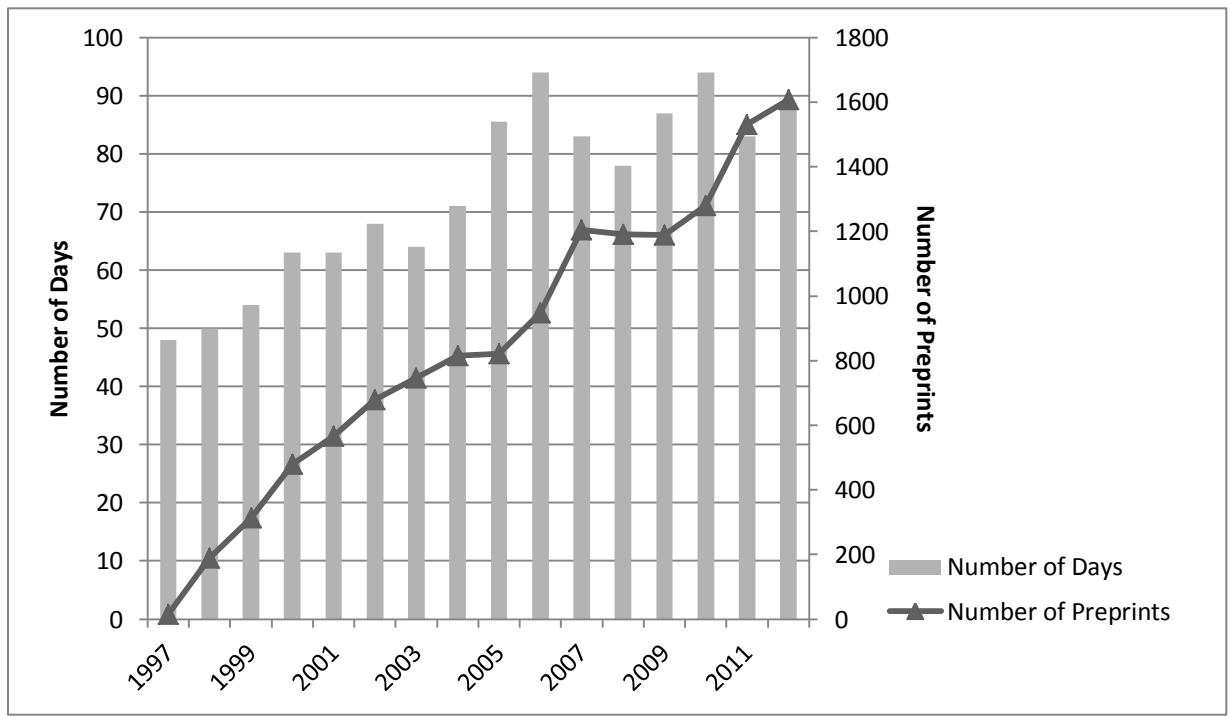

**Figure 8.** Time series of the median publication delay in days for articles that were published between 1997 and 2012 in the Journal of High Energy Physics and have a previous preprint in arXiv.

Figure 8 displays the median publication delay as a function of time. At first glance, we can see that for each year the median publication delay for articles has been below 100 days. Different from PRD, the median publication delay did not decrease over the years, but on the opposite increased in line with the growing number of articles published having a preprint. The median publication delay nearly doubled from 48 days in 1997 to 94 days in 2006. This increase in time could be explained by the fact that with a higher number of articles submitted, it takes more time to edit and review them. Another reason might be that authors realized over the years that it is more advantageous to submit a preprint as early as possible, and not to wait until acceptance. Authors might have experienced that their preprints are noticed and cited as soon as they are available in arXiv. However, the publication delay is short and makes us curious to see whether this time advance is used for the citation of preprints.

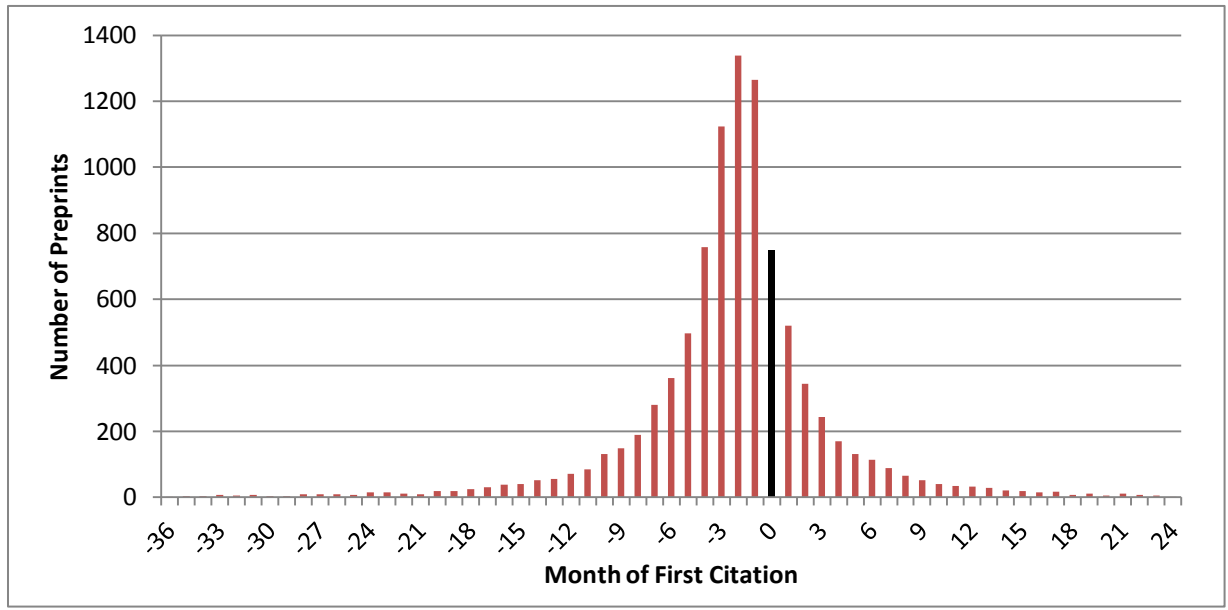

**Figure 9.** Distribution of first citations over months after journal publication for papers in INSPIRE HEP that were published between 1997 and 2010 as articles in the Journal of High Energy Physics. The citation window is two years after article publication and the citation data is based on INSPIRE HEP.

Figure 9 illustrates the months, in which papers covered by INSPIRE HEP received their first citation. In total, 9,684 preprints are analyzed, which were published in JHEP between 1997 and 2010. In their first two years of article existence and prior to that, 219 papers remained uncited. Not surprisingly, 69.5% of all preprints, which were published as articles in JHEP between 1997 and 2010, received their first citation before the respective article appeared. The mode of the distribution is two months prior to publication. A majority of 13.8% of all preprints received their first citation within 31 to 60 days prior to publication.

The next figure illustrates the distribution of all citations over months that were awarded to papers covered by INSPIRE HEP, and published in JHEP between 1997 and 2010. At first glance, the histogram seems similar to the one for PRD. Different from PRD, the mode of the distribution is zero, which means that most citations are awarded shortly after the journal publication. Furthermore, Figure 10 shows that the citation numbers after journal publication decline more rapidly than for PRD. As a result, 16.9% of all citations are accumulated by preprints prior to their publication in JHEP.

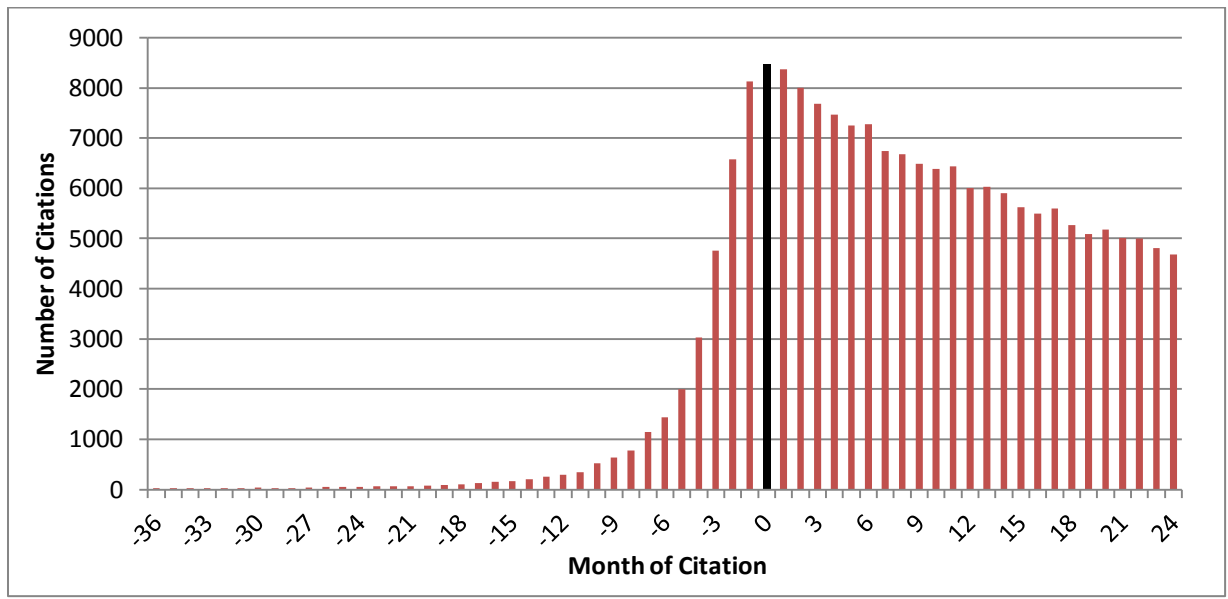

**Figure 10.** Distribution of all citations over months after journal publication for papers in INSPIRE HEP that were published between 1997 and 2010 as articles in the Journal of High Energy Physics. The citation window is two years after article publication and the citation data is based on INSPIRE HEP.

### 5.1.3 Nuclear Physics B

Nuclear Physics B (NPB) exists since 1956 and focuses on the domain of high-energy physics and quantum field theory. The journal puts its emphasis on original research papers and includes particle physics, field theory and statistical systems, and physical mathematics.[37] Nuclear Physics B is published by Elsevier and has an Impact Factor of 4.661 (2011).[38] Currently, 36 issues are published a year (12 volumes, with 3 issues each). Elsevier is known for its Open Access option, where the author pays a fee of $3,000 for an article.[39] This fee covers all costs for peer review and publication, and in the end every reader has free access to the article.

In Figure 11 we can see that different from PRD and JHEP, the number of papers in Nuclear Physics B has been significantly decreasing over the last 13 years. The number of articles published per year has nearly halved from 815 in 2000 to 417 in 2006. From 2007 on, fewer than 400 articles were published per year.

---

[37] Elsevier (2013). Nuclear Physics B. http://www.journals.elsevier.com/nuclear-physics-b/
[38] Ibid.
[39] Elsevier (2013). Sponsored articles. http://www.elsevier.com/about/open-access/sponsored-articles#publication-costs

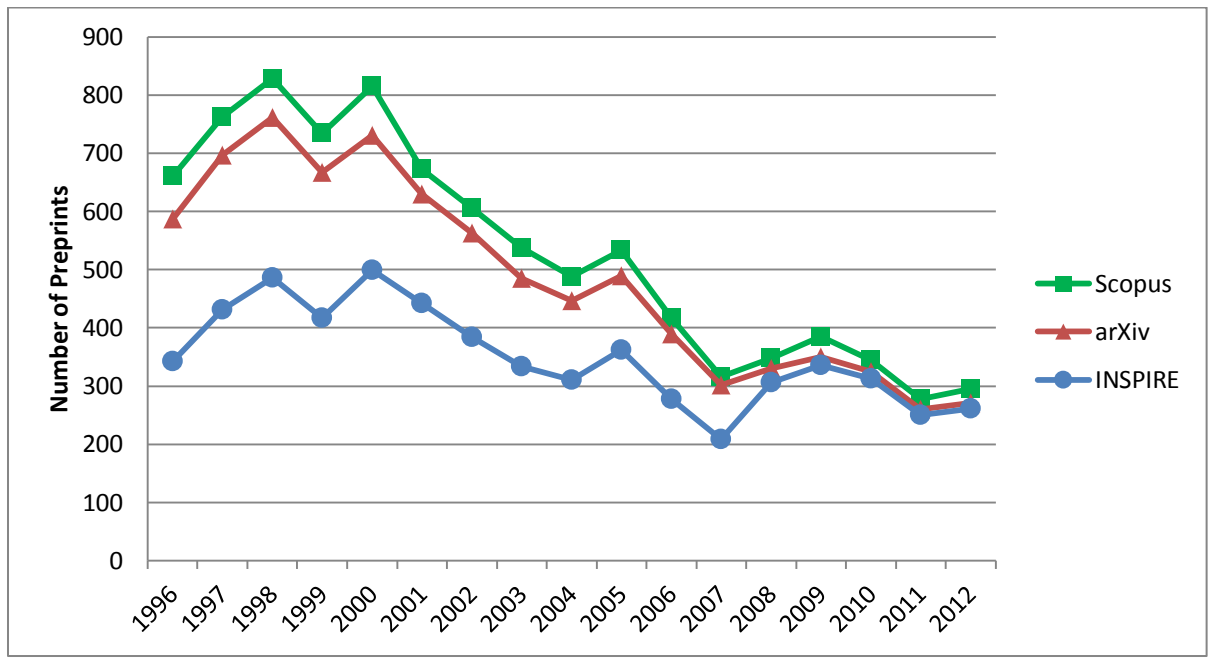

**Figure 11.** Growth of publication numbers from 1996 to 2012 for Nuclear Physics B.

The arXiv line graph reveals that the number of preprints converged more and more to the whole output of journal articles in recent years. The share of preprints included in INSPIRE HEP has grown as well, and coincides with the number of preprints in arXiv. The overlap has been constantly at 96% for the last four years. This shows that, unlike articles published in PRD and JHEP, not all articles in NPB are relevant to INSPIRE HEP. Presumably, preprints dealing with physical mathematics are not completely covered by INSPIRE HEP. In total, 9,023 articles have been published between 1996 and 2012 in NPB. 8,284 of these articles have a previous preprint in arXiv, which makes up 91.8% of all articles published in the last 17 years. The inclusion of preprints in INSPIRE HEP is lower than for the two other HEP journals. Overall, 71.9% of preprints in arXiv can be also found in INSPIRE HEP. The high number of preprints in arXiv can be explained by the fact that the publisher accords to place a preprint on a public server prior to its submission to a journal. Elsevier does "not require authors to remove electronic preprints of an article from public servers should the article be accepted for publication in an Elsevier journal."[40] Furthermore, Elsevier allows authors to distribute their manuscripts, e.g., to post them on their websites or their repository. The next figure shows a histogram of the publication delay of preprints that were published as articles between 1996 and 2012 in NPB. The shape reminds of a normal distribution with a long tail of preprints that were published more than one year prior to publication.

---

[40] Elsevier (2013). Article posting policy. http://www.elsevier.com/about/open-access/open-access-policies/article-posting-policy

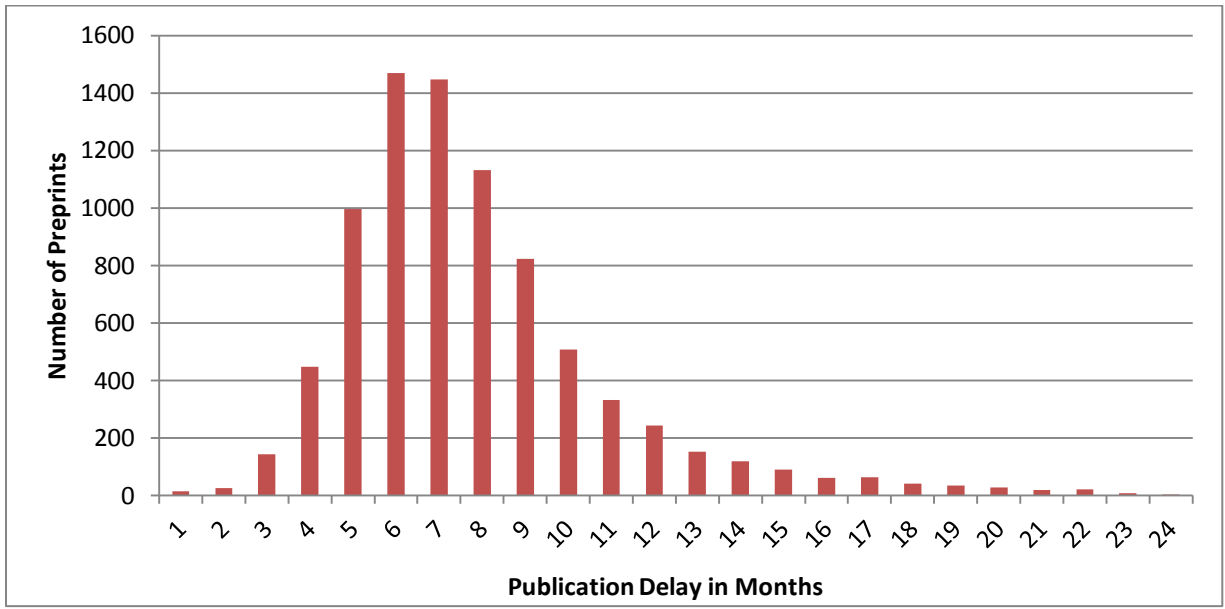

**Figure 12.** Distribution of preprints over months prior to journal publication. Those preprints are considered whose articles were published between 1996 and 2012 in Nuclear Physics B.

Not all preprints are included in Figure 12, since 67 additional preprints are scattered over the months 25 to 56 prior to publication in Nuclear Physics B. As can be seen, the mode of the distribution is six months. A majority of 17.7% of preprints appear in the journal 151 to 180 days after their upload in arXiv. On the whole, 91.5% of all preprints are submitted to arXiv at most one year prior to final publication.

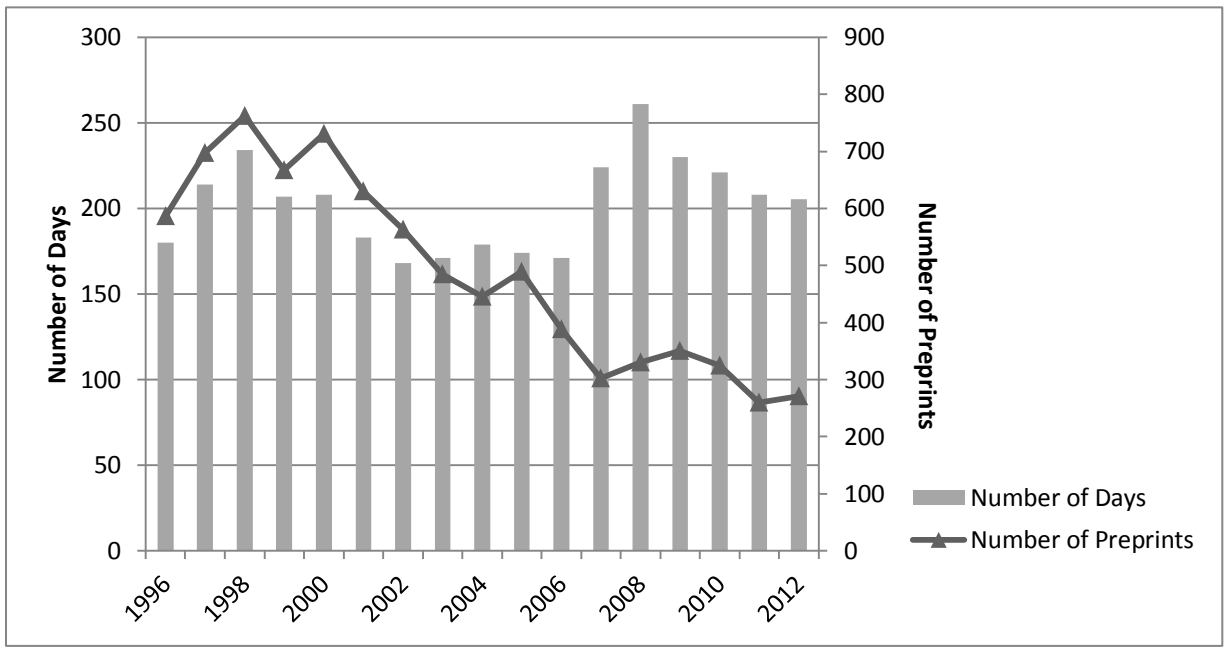

**Figure 13.** Time series of the median publication delay in days for articles that were published between 1996 and 2012 in Nuclear Physics B and have a previous preprint in arXiv.

Figure 13 indicates that the median publication delay for Nuclear Physics B has been much longer than for the two other HEP journals. It used to be eight months in 1998 and went down until 2006, in line with the number of articles published having a previous preprint. Interestingly enough, with a sinking number of articles published, the publication delay increased again and had a maximum of 261 days in 2008. Since then, the median publication delay decreased gradually. The long publication delay is either caused by the publisher, or authors let deliberately time elapse between the upload in arXiv and the submission to NPB, in order to receive useful comments that may improve their paper. Due to this longer publication delay it is obvious that there must be a citation advantage for NPB. Therefore, the next figure depicts the distribution of the first citation for articles that were published between 1996 and 2010. The citation window is set to two years after journal publication.

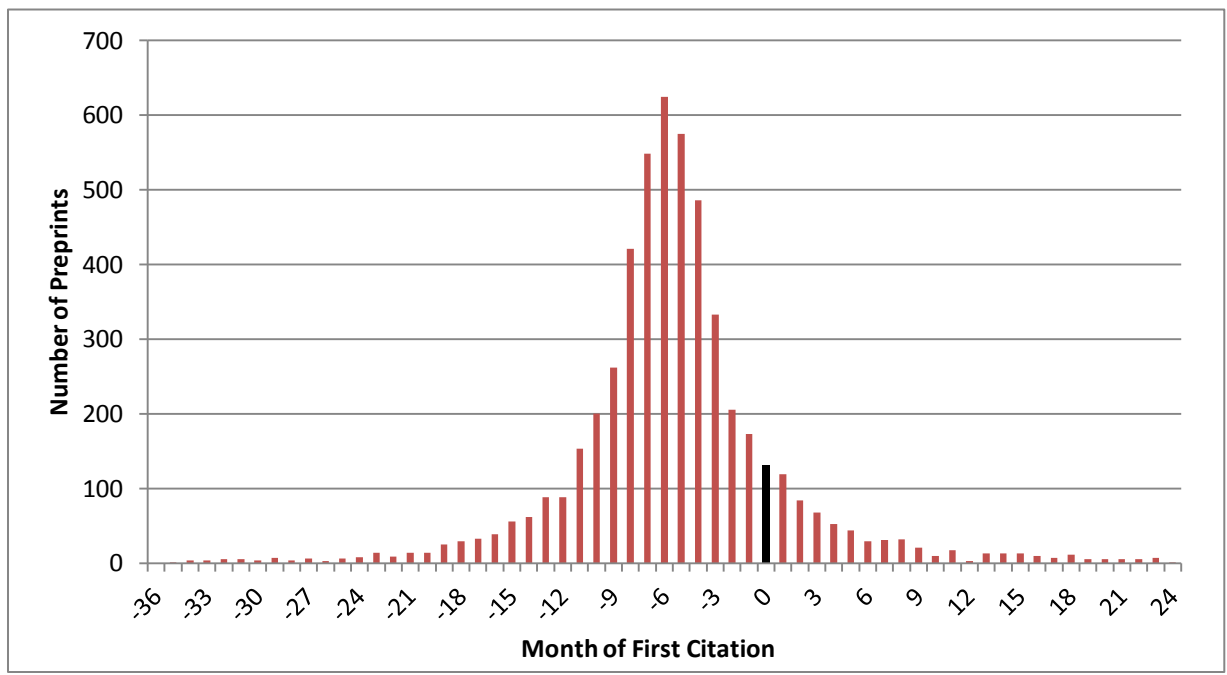

**Figure 14.** Distribution of first citations over months after journal publication for papers in INSPIRE HEP that were published between 1996 and 2010 as articles in Nuclear Physics B. The citation window is two years after article publication and the citation data is based on INSPIRE HEP.

It is striking that the black bar, representing papers that are cited for the first time shortly after their publication in NPB, is lower than for the other two HEP journals. On the whole, 5,446 preprints are considered, of which 151 remain uncited in their first two years after journal publication. The mode of the distribution of the first citation is six months and thus far ahead of the article publication in NPB. Most preprints (11.5%) receive their first citation 151 to 180 days prior to formal publication. A high share of 83.7% papers receives their first citation before the respective article is published in NPB. Hence, in percentage terms, more preprints are cited prior to publication in Nuclear Physics B than in PRD or JHEP. This should not surprise, since we have seen that the publication delay is much longer, and predicts a higher citation advantage. This citation advantage also becomes clear if we have a look on the distribution of all citations awarded to papers that were published between 1996 and 2010 in Nuclear Physics B. Figure 15 shows that the area before the journal publication is larger than for PRD or JHEP. At the time of publication in NPB, preprints covered by INSPIRE HEP have attained 27.5% of all citations they will receive by the end of the following two years. Because of the smaller amount of papers for NPB the behavior of the bars is not as smooth as for PRD or JHEP.

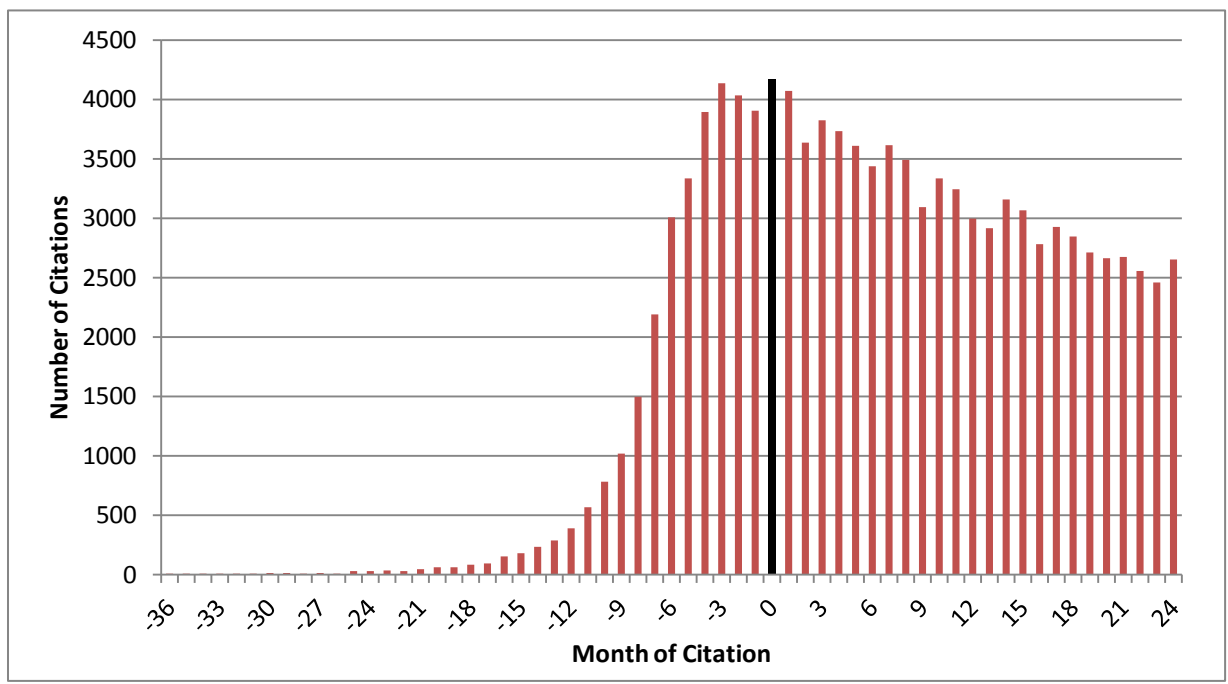

**Figure 15.** Distribution of all citations over months after journal publication for papers in INSPIRE HEP that were published between 1996 and 2010 as articles in Nuclear Physics B. The citation window is two years after article publication and the citation data is based on INSPIRE HEP.

### 5.1.4   Discussion

In 2009, Ingoldsby presented one of the first systematic studies of articles deposited in arXiv. He investigated 2,800 articles published by the American Institute of Physics and 2,100 by the American Physical Society (Ingoldsby, 2009, p. 6). The percentage of overlapping articles found in arXiv, varied among the disciplines. In a field such as elementary particle physics he found an overlap of nearly 100% (ibid., p. 10). In General Physics he found 82.3%, in Gravitation and Astrophysics 90.9%, and in Nuclear Physics 66.7% of preprints that were later published in journals. Davis (2009) criticized Ingoldsby for the claim that "everything published in physics can be found in arXiv." Davis satirized Ingoldsby's stance and is of the opinion that this myth serves the purpose to appease publishers that arXiv does not render publishers useless.

When it comes to the publication delay, the results prove that arXiv has accelerated scholarly communication considerably in the past two decades. Ginsparg wrote in 1994 that it takes six months to a year for a paper to get published in a journal after submission (Ginsparg 1994, p. 157). ArXiv helped to bypass the long publication delay, by inviting researchers to instantly deposit their results and submit them simultaneously for journal publication. As has been shown, the publication delay varies among journals. Whereas JHEP's median publication delay has never surpassed 100 days; the publication delay for NPB is twice as long nowadays. The median publication delay for PRD decreased over the years and is today around 100 days. The importance of publication delay arises with HEP as a fast developing field which requires large investments. A delay in the process of several months can lead to a waste of time and effort. As Goldschmidt-Clermont (1965) illustrated, "a certain amount of duplication is needed; however, because of the costs involved, it has to be controlled - not blind - duplication."

The citation analysis showed that the immediate access to preprints results in a citation advantage, as far as the speed of visibility is concerned. HEP researchers do not wait for the published article to appear, in order to cite it. The communication begins as soon as the preprint is available, which accelerates the scientific communication immensely. It has been shown that 69% to 84% of papers receive their first citation prior to publication. Additionally, around 16% to 28% of all citations within a time frame of two years after journal publication occur before the articles are published. How is this advantage explainable? The submission of preprints directly to arXiv and to a journal is common practice in HEP because of the long existing preprint culture. Since there is no commercial incentive for research results in HEP, in comparison to medicine or biology, researchers can share their results immediately. Hence, the fast discourse in HEP is boosted by an early and free dissemination.

If authors search for literature in INSPIRE HEP it only takes them one click to get to the full text in arXiv. According to Gentil-Beccot et al. (2009) HEP scientist prefer to access arXiv (82%) instead of the publisher server (18%) when they have the choice from the INSPIRE HEP database.

According to Harnad (2003), the rejection rate in HEP journals is lower than in other fields because authors disburden peer reviewers by submitting almost finished articles. Hence, the arXiv version is often similar to the published article version. The question arises why someone should pay for journal subscriptions if the literature in HEP is accessible for free in arXiv. Nevertheless, journals are important to guarantee quality through their peer review, and as a reward for authors. Because the process of peer review is of immeasurable value, libraries continue to pay for subscriptions to HEP journals. SCOAP$^3$ (Sponsoring Consortium for Open Access Publishing in Particle Physics) grew out of the idea that most HEP literature is available for free in arXiv. Consequently, SCOAP$^3$ wants to convert the leading journals in HEP to an Open Access model, by redirecting the subscription fees to pay for peer review.[41] Authors do not have to pay any charges and readers have free access in return. PRD, JHEP, and NPB are regarded as core journals to participate in the SCOAP$^3$ initiative because they publish the majority of HEP articles. In 2012, Elsevier announced that it will participate in SCOAP$^3$ with Nuclear Physics B. Elsevier's Physics Letters B will also change from a subscription based model into an Open Access model in 2014.[42]

A last word on the citation impact: It does not make sense to compare the set of articles having a previous preprint with the set of articles having no preprint in regard to citation counts, because the size of the two sets differs greatly. It should be rather assumed that almost each article has a preceding preprint in arXiv so that a comparison of the two sets gets meaningless, and was not drawn.

---

[41] SCOAP$^3$ (2012). Redirecting the current expenditures on HEP journals. http://scoap3.org/whichjournals.html
[42] Elsevier (2012). Elsevier to Participate in Scoap3 with the journals Physics Letters B and Nuclear Physics B. http://www.elsevier.com/about/press-releases/science-and-technology/elsevier-to-participate-in-scoap3-with-the-journals-physics-letters-b-and-nuclear-physics-b

## 5.2 Mathematics

Another fifth of all papers in arXiv, 19.4% (157,284), deals with mathematics and mathematical physics.[43] Mathematics was included in arXiv in February 1992. For the following bibliometric analysis three mathematical journals have been chosen: the Annals of Mathematics, Advances in Mathematics, and the Journal of Mathematical Analysis and Applications.

### 5.2.1 Annals of Mathematics

The Annals of Mathematics was founded in 1884 at the University of Virginia. It was transferred in 1899 to Harvard and in 1911 to Princeton University.[44] Since 1933, the journal is published bimonthly by the Department of Mathematics at Princeton University and the Institute for Advanced Study. It is a prestigious journal with an Impact Factor of 2.928 (2011). Since 1998, it is disseminated in an electronic edition next to its print edition. The Annals of Mathematics was also available as an Open Access journal from 2001 (volume 153) until 2007 (volume 166).[45] Articles are freely available five years after publication. Currently, two volumes a year are published, with three issues each. During the analysis of the articles on the publisher's website it became clear that Scopus lacked some articles.[46] Since these errors cannot be bypassed, the following data should be regarded as an almost complete set.

In total, 959 articles were published between 1996 and 2012. 472 of them have a preprint in arXiv, which results in a percentage of 49.2%. For the time period in question, Annals of Mathematics reveals a high number of postprints, which is 174 (18.1%). It implies that 67.3% of all articles published in the Annals of Mathematics are available for free in arXiv. Figure 16 presents the publication output for the Annals of Mathematics. We can see at first glance that the number of articles published per year is low in comparison to HEP journals. On average, 56 articles a year were published in the analyzed time period. Over the years, the number of articles has increased markedly, from 38 in 1996 to 91 in 2010. At the same time, the number of articles published having a preceding preprint has increased steadily.

---

[43] arXiv (2013). arXiv submission rate statistics. http://arxiv.org/help/stats/2012_by_area/index
[44] Princeton University & the Institute for Advanced Study. About the Journal | Annals of Mathematics. http://annals.math.princeton.edu/about
[45] JSTOR. Annals of Mathematics. http://www.jstor.org/journals/0003486x.html
[46] For instance, for the volume 172 (second half of 2010) Scopus lists only 42 articles, whereas on the Princeton University's website 53 articles are listed. As a result the number of articles for 2010 is 103.

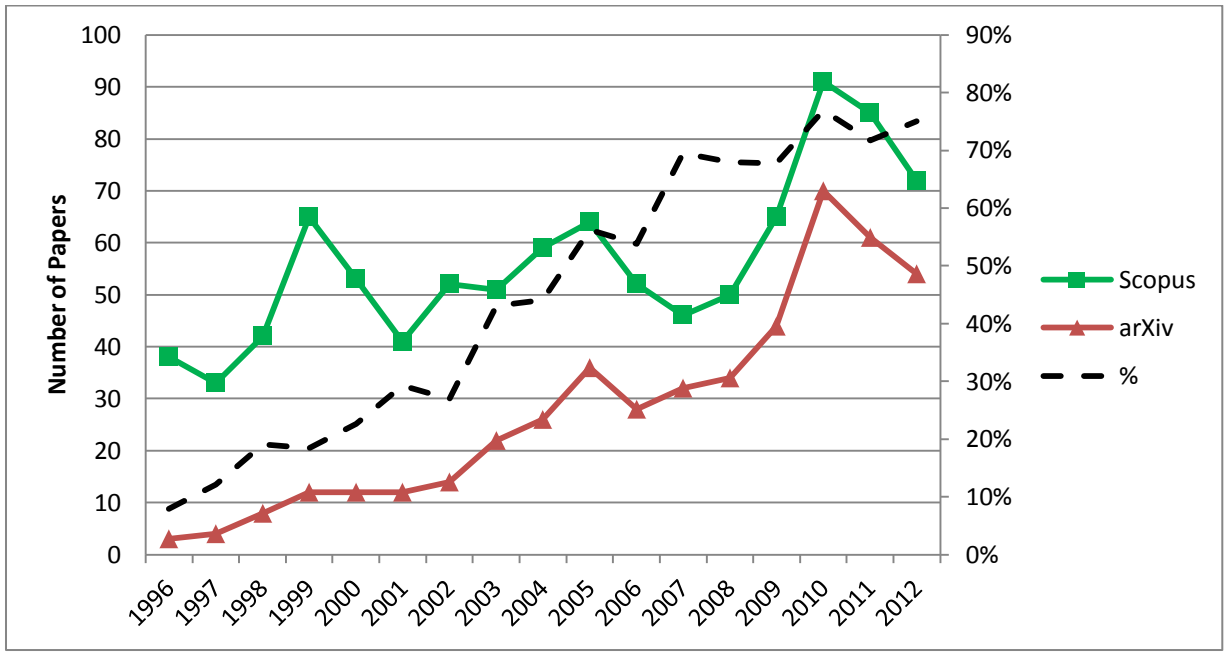

**Figure 16.** Growth of publication numbers between 1996 and 2012 for Annals of Mathematics.

This is well expressed by the percentage line graph in Figure 16. Whereas in 1996, fewer than 10% of all articles published in the Annals had a preceding preprint in arXiv, today's share is around 70%. The high number of preprints is not surprising if we have a look on how much time elapses between the upload of a preprint and its publication in the Annals of Mathematics. In Figure 17 the publication delay of all preprints, which appeared as articles between 1996 and 2012 in the Annals of Mathematics, is displayed; scattered over a period of 5 years (60 months) prior to publication.

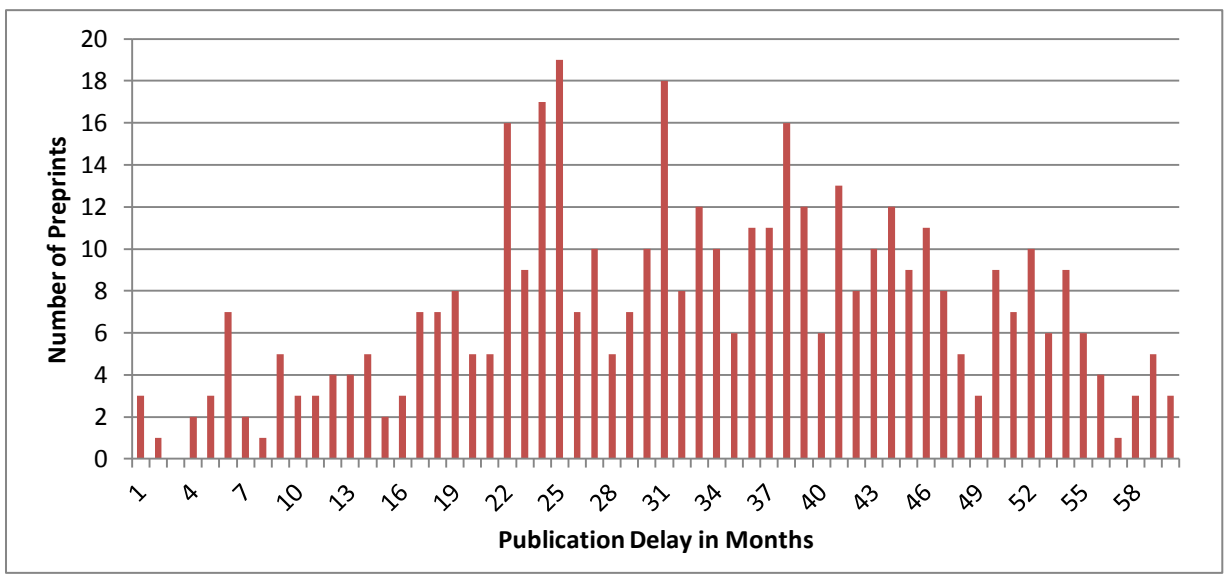

**Figure 17.** Distribution of preprints over months prior to journal publication. Those preprints are considered whose articles were published between 1996 and 2012 in Annals of Mathematics.

Only the first 432 preprints are depicted. A long tail of 40 additional preprints is scattered over the months 61 to 94 prior to journal publication, and thus is not included in Figure 17. We can see that the mode is 25 months, which means that the majority of preprints are deposited in arXiv two years prior to formal publication. Within one year prior to journal publication, only 6.4% of all preprints are available in arXiv, within two years 25.0%, and 51.1% within three years prior to publication. The median value for the publication delay of all preprints is 1065.5 days, thus approximately three years. The question arises whether the publication delay has always been so long or whether it may have decreased in recent years. Therefore, the next figure displays the median publication delay as a series of time.

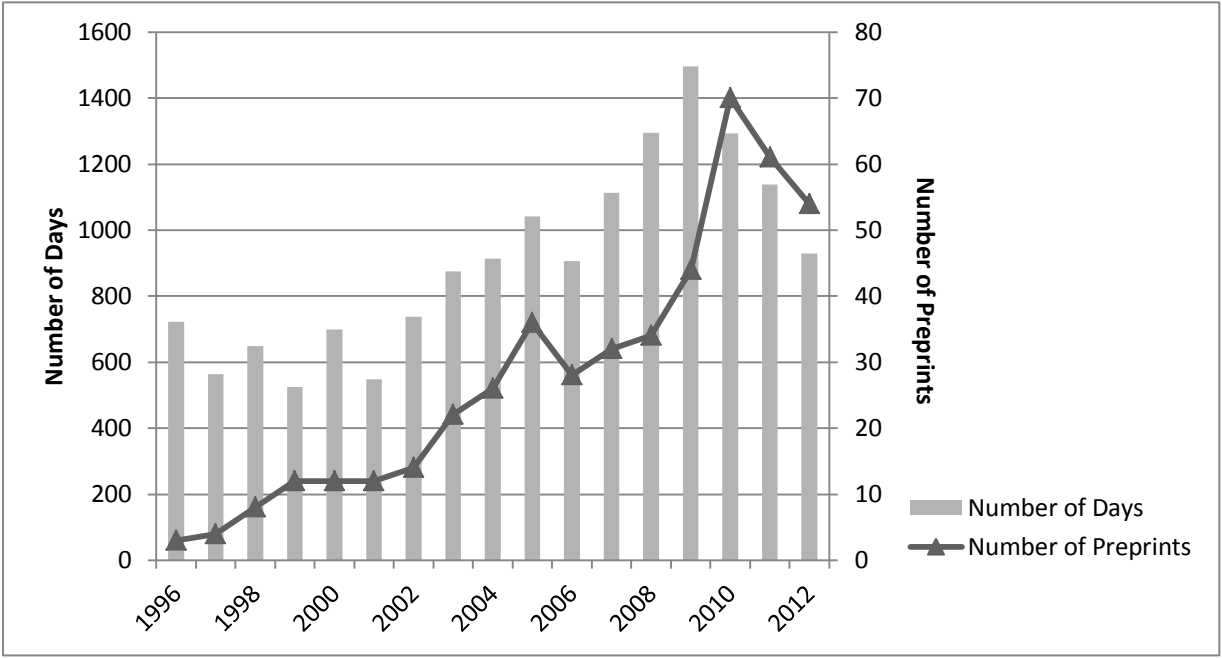

**Figure 18.** Time series of the median publication delay in days for articles that were published between 1996 and 2012 in Annals of Mathematics and have a previous preprint in arXiv.

We can gather from Figure 18 that the median publication delay has increased strongly, having its peak in 2009 with 1496.5 days. The shortest publication delay of 526 days occurs in 1999. It is noticeable that the line graph bears resemblance to the shape of the bars. It seems like the publication delay has grown in line with the number of articles published having a preprint. This might be explained by the fact that the more articles are published, the higher the probability is that very old preprints are among them. Moreover, it is evident that with advanced time, preprints can be older and are likely to influence the publication delay due to their high age. The longest publication delay of a preprint in this dataset is eight years and eight months. It shows that time does not really matter, as long as the subject of the preprint is relevant to the Annals of Mathematics. It might be also the case that authors intentionally let time lapse away between the deposit in arXiv and the submission to the Annals of Mathematics because the preprint may ripen and they can incorporate valuable comments in their manuscript before submitting it. On the other hand, it might be that a long time elapses on the publisher's site because of profound peer review and revision. If the submission rate is high, peer reviewers have to filter carefully good papers from substandard papers, which can prolong the publication process. Figure 18 indicates that since 2010, the publication delay declines gradually.

In the following analysis, citation numbers for the Annals of Mathematics will be discussed. Different from the analysis in HEP, the data is based on citations tracked by Scopus. First of all, it is of interest to examine whether articles with a foregoing preprint are more cited than articles that do not have a preprint in arXiv. It is necessary to find out to which extent the long publication delay affects citation rates. Therefore, the citation numbers of articles having a preprint in arXiv and those that do not have a preprint have been compared. To verify the numbers, Welch's t-test has been administered, which compares the means of unpaired groups. The assumption is that both groups of data are samples of Gaussian distributions but do not have the same standard deviation. Welch's t-test is useful to determine a confidence interval for the distance of two expected values. The test reports a P-value, which indicates significant difference if $P < 0.05$.

To examine any citation advantage for articles with a previous preprint, the citation window was set to one year after the date of journal publication. The publication period is thus from 1996 to 2011. With a confidence value of 96.64% it can be said that articles with a previous preprint in arXiv get 0.3 times more cited than articles without a preprint, in their first year after journal publication (P=0.034). Hence, there is a citation advantage for articles with a preceding preprint.

Since the citation half-life in mathematics is long compared to other fields, the citation window for the following analysis was set to three years. Table 1 presents a comparison of citation numbers for articles having a preprint and those that do not have a preprint in arXiv. The table consists of a column for the publication year, a column showing the number of articles published in the respective year without a preprint, and the average citation number of those articles. The two further columns show the number of articles with a preceding preprint in arXiv, and their average citation number. It is obvious that articles with a foregoing preprint get more cited within the first three years after publication.

**Table 1.** Average citation rates for articles published between 1996 and 2009 in Annals of Mathematics. The citation window is three years after journal publication and the citation data is based on Scopus.

| Publication year | Number of articles without preprint | Average citation number | Number of articles with preprints | Average citation number |
|:---:|:---:|:---:|:---:|:---:|
| 1996 | 35 | 4.49 | 3 | 4.00 |
| 1997 | 29 | 3.93 | 4 | 3.00 |
| 1998 | 34 | 3.82 | 8 | 6.88 |
| 1999 | 53 | 3.89 | 12 | 5.33 |
| 2000 | 41 | 4.68 | 12 | 4.75 |
| 2001 | 29 | 3.55 | 12 | 5.00 |
| 2002 | 38 | 4.53 | 14 | 4.00 |
| 2003 | 29 | 4.62 | 22 | 7.77 |
| 2004 | 33 | 4.73 | 26 | 6.12 |
| 2005 | 28 | 6.50 | 36 | 6.97 |
| 2006 | 24 | 8.71 | 28 | 8.68 |
| 2007 | 14 | 4.50 | 32 | 9.09 |
| 2008 | 16 | 7.38 | 34 | 9.47 |
| 2009 | 21 | 7.76 | 44 | 7.05 |
| Total | 424 | 4.95 | 287 | 7.19 |

Indeed, for articles published between 1996 and 2009, and a citation window of three years, a statistically significant difference arises. On average, articles with a previous preprint in arXiv receive one more citation than articles without a preprint (P=0.011). Table 1 also suggests that the average citation number has increased over the years.

It is of further interest to see if articles with a previous preprint are earlier cited than articles without a preprint in arXiv. For that reason only the first citation is considered, on the basis of Scopus. Again, the citation window was set to three years, because of the long cited half-life in mathematics and in order to gain a larger dataset to operate with. As a consequence, articles published between 1996 and 2009 were considered. The number of articles published without a previous preprint in this time period is 424, of which 53 remained uncited in their first three years (12.5%). The number of articles with a foregoing preprint is 287, of which 21 remained uncited in their first three years (7.3%). To include the uncited articles into the examination, their citation delay was set to 1,095 days. The median value of the citation delay for articles having a preprint is 304 days. Articles without a preprint receive their first citation on average three months later; the median is 395 days. Welch's t-test, on the other hand, shows that articles with a former preprint in arXiv have a citation advantage of six weeks (P=0.026). Figure 19 displays a cumulative distribution function of the first citation for articles with a previous preprint and those without. It becomes evident that articles with a foregoing preprint are earlier cited, especially in the first two years after publication.

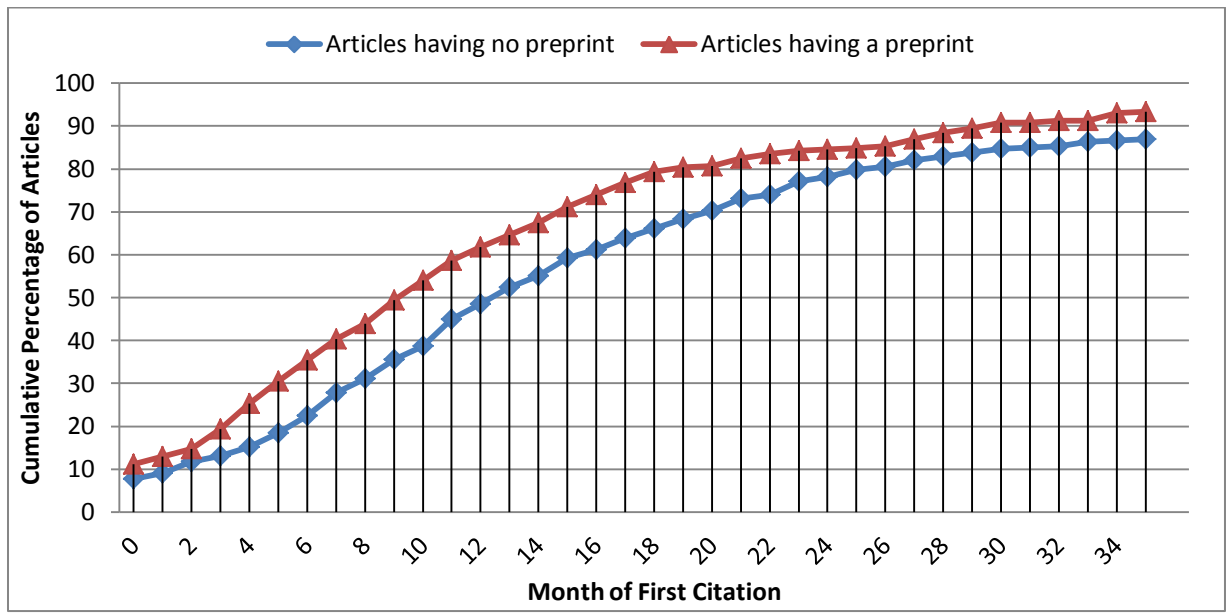

**Figure 19.** Cumulative distribution function of the first citation for articles published in Annals of Mathematics between 1996 and 2009. The citation data is based on Scopus. Articles with a previous preprint are on average six weeks earlier cited than articles without a preprint.

Finally, the search in arXiv's Comments or Journal-ref. field lists 130 records for Annals of Mathematics. This means that only one in four papers provides a reference to the peer-reviewed article. Probably, because a long period of time elapses between the submission and the actual publication, so that authors simply forget about updating the data on the publisher's version.

### 5.2.2 Advances in Mathematics

Advances in Mathematics was founded in 1961 and presents research on pure mathematics. The aim of the journal is to give a "complete survey of results in special topics in mathematics, published in the last 10-30 years that have not yet appeared in handbooks."[47] Advances in Mathematics is published by Elsevier in three volumes a year, one volume consisting of six issues. The Impact Factor for the year 2011 is 1.177. According to Elsevier, authors publishing in Advances in Mathematics "retain the right to post their own version of their accepted manuscript in the arXiv subject repository."[48]

---

[47] Elsevier (2013). Advances in Mathematics . http://www.journals.elsevier.com/advances-in-mathematics/
[48] Elsevier. Manuscript posting in arXiv - Advances in Mathematics. http://www.journals.elsevier.com/advances-in-mathematics/news/voluntary-posting-arxiv-subject-repository-is-permitted/

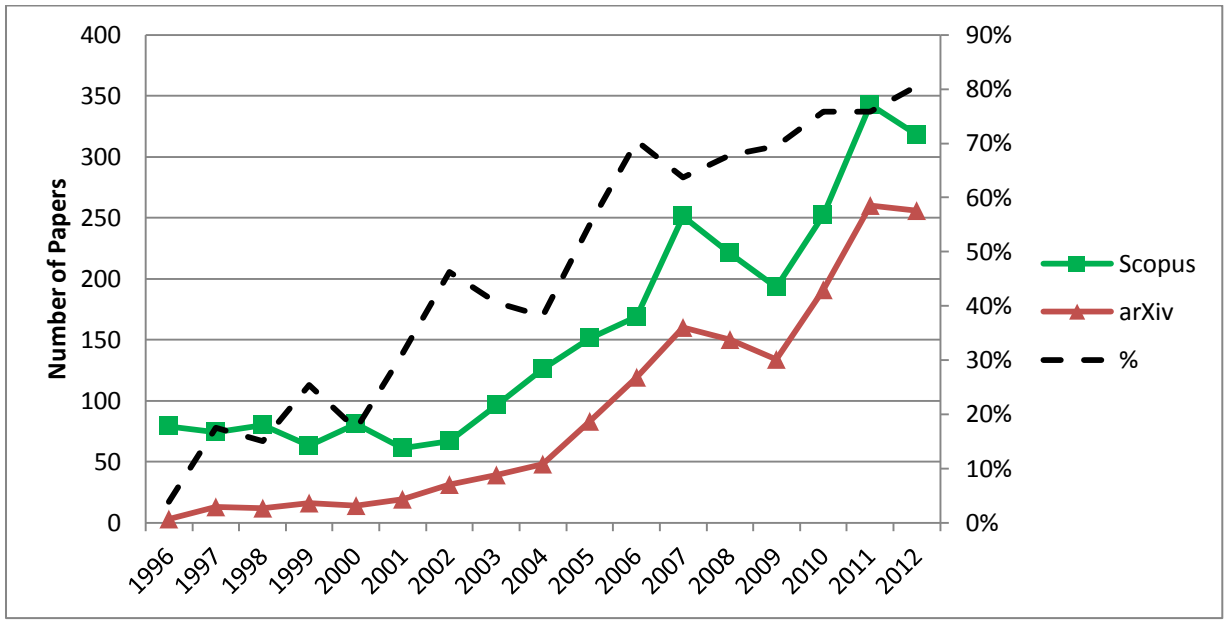

**Figure 20.** Growth of publication numbers between 1996 and 2012 for Advances in Mathematics.

Figure 20 displays the growth of the number of articles and preprints in arXiv. It is obvious that the number of articles published per year increased greatly. Until 2003, the number of articles was below 100 per year, whereas in the late two years it surpassed 300. Just as the number of articles grew over the years, so did the number of preprints. The percentage of articles having a preprint used to be less than 10% in 1996. It grew enormously over the years and reached a share of nearly 80% in 2012. From 1996 to 2012, 2,625 articles were published; of which 1,548 have a preprint. This results in a share of 59.0%. Different from the Annals of Mathematics, only 25 articles were deposited as postprints in arXiv, less than 1%.

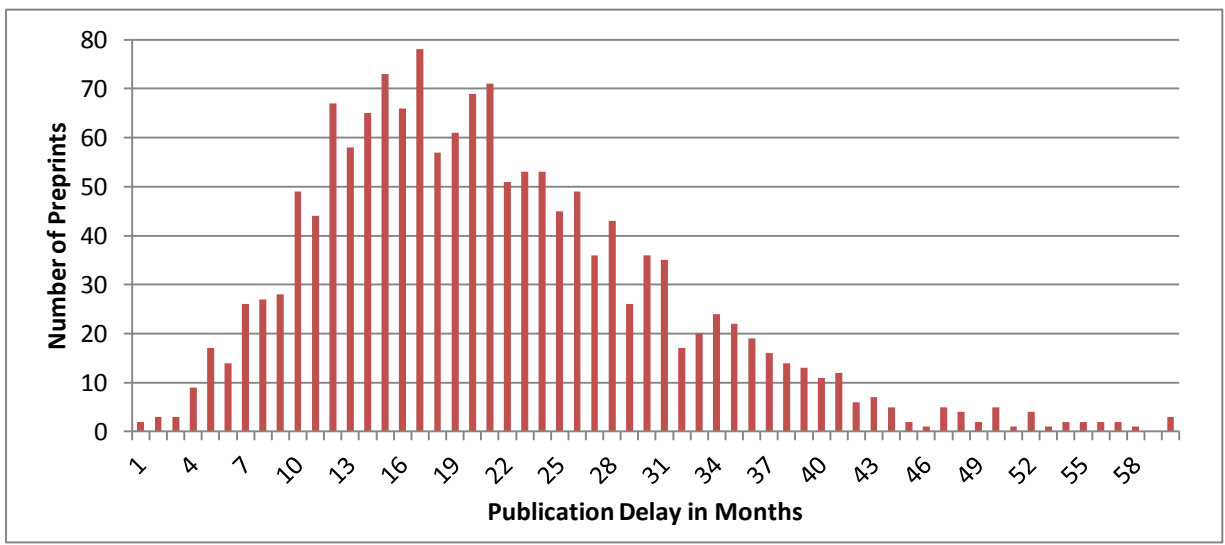

**Figure 21.** Distribution of preprints over months prior to journal publication. Those preprints are considered whose articles were published between 1996 and 2012 in Advances in Mathematics.

The histogram of the publication delay of all preprints published in Advances in Mathematics between 1996 and 2012 is depicted in Figure 21. Eleven preprints are not included in the histogram because they are scattered over the months 61 to 77. The displayed values seem normally distributed, with a long tail of preprints having a longer publication delay. The mode of the distribution is 17 months. A majority of 78 preprints are deposited in arXiv one year and five months prior to journal publication. Within one year prior to journal publication, 18.7% of all preprints are deposited in arXiv, within two years 67.4%, and 91.5% within three years prior to publication. The median publication delay of all 1,525 preprints is 581 days. The publication delay of all preprints ranges between 18 and 2,297 days (more than six years).

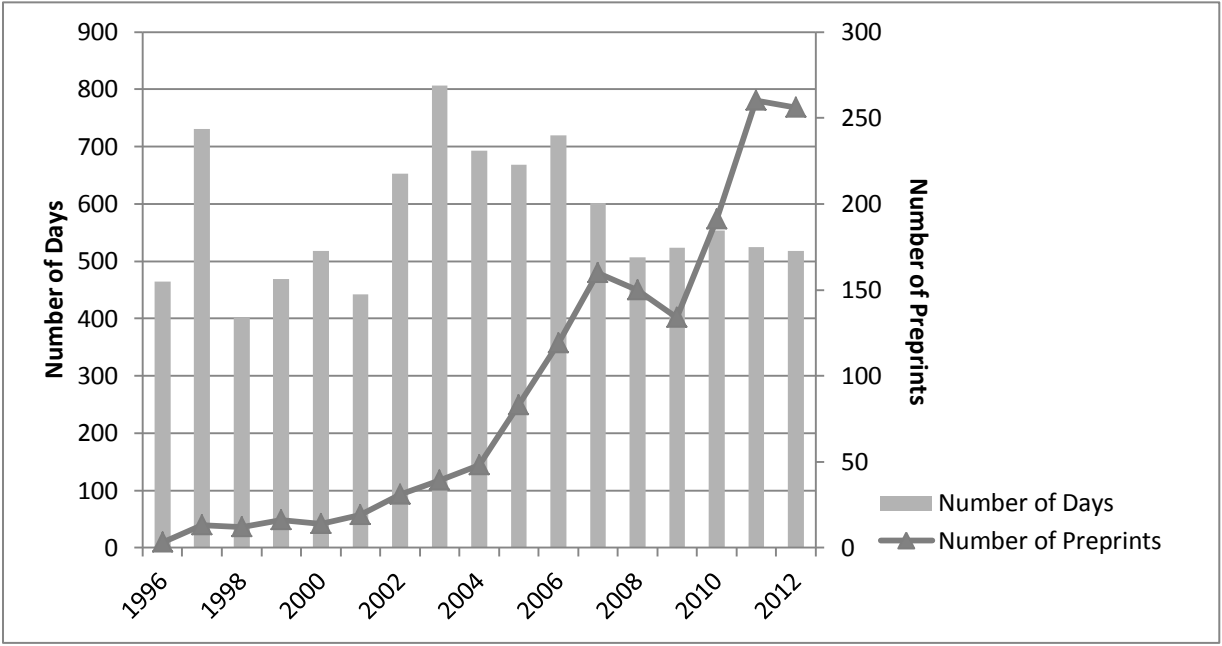

**Figure 22.** Time series of the median publication delay in days for articles that were published between 1996 and 2012 in Advances in Mathematics and have a previous preprint in arXiv.

Figure 22 reveals that the median publication delay is not as widespread as for the Annals of Mathematics, and overall much shorter. It has its peak in 2003 with 807 days, whereas the shortest median publication delay occurs in 1998 with 401 days. Different from the Annals of Mathematics the publication delay did not grow with the number of preprints published. On the contrary, whereas the number of articles increased in the past five years, the median publication delay stayed unchanged at around 500 days. With the aim to examine if preprints accelerate the communication in mathematics, citation numbers have been analyzed. The publication period is 1996 to 2009 and the citation window is set to three years after journal publication. Welch's t-test shows that articles with a previous preprint get on average 0.5 times more cited than articles without a preprint deposited in arXiv (P=0.005).

In regard to the citation delay, the results show that 24.5% of articles without a preprint in arXiv remain uncited within three years after journal publication. From the set of articles that have a previous preprint in arXiv, 14.4% remain uncited in their three-year existence after journal publication. Therefore, the date difference for the uncited articles was set to 1,095 days. As a result, the median value of the day of the first citation for articles without a preprint is 574 days after publication, whereas for articles with a preceding preprint it is only 426 days, thus five months less. Welch's t-test has proven that articles with a previous preprint are on average ten weeks earlier cited than articles without a preprint (P=0.018). This becomes evident in Figure 23, where the cumulative distribution function is depicted. The citation advantage becomes especially evident one year after article publication.

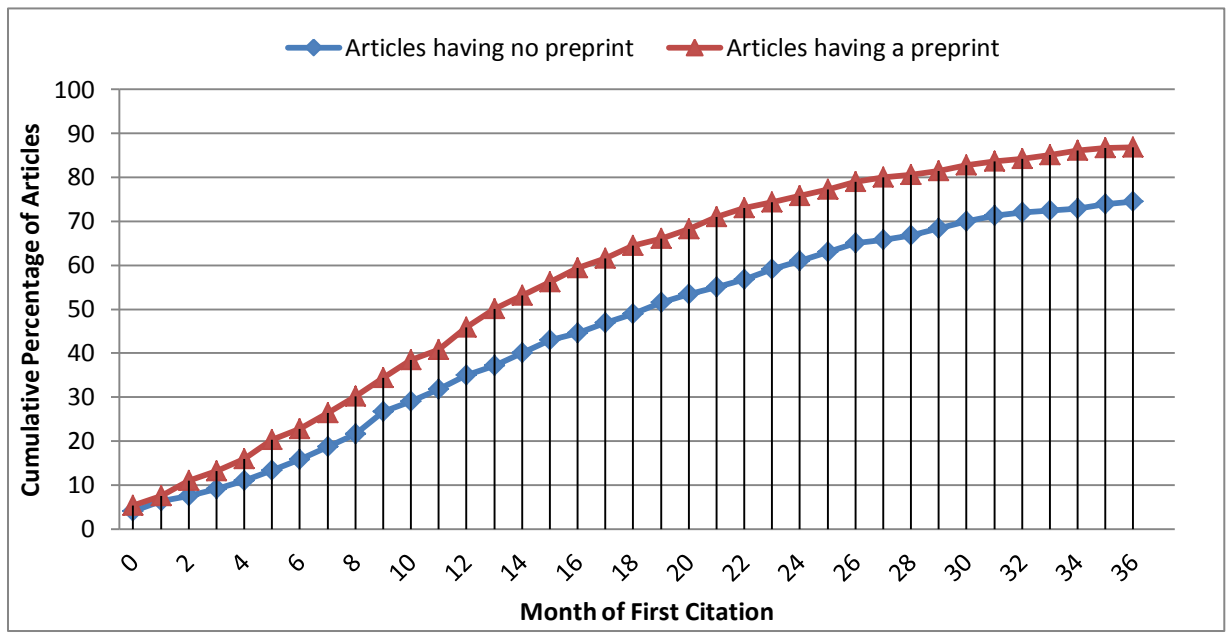

**Figure 23.** Cumulative distribution function of first citation for articles published in Advances of Mathematics between 1996 and 2009. The citation data is based on Scopus. Articles with a previous preprint are on average ten weeks earlier cited than articles without a preprint.

Finally, arXiv shows that approximately 399 e-prints provide a note on the journal publication in Advances in Mathematics. This makes up around one in four papers, just like for the Annals of Mathematics. It is worth mentioning that a remarkably high number of articles published in Advances in Mathematics cite arXiv. According to Scopus, 344 articles published between 1996 and 2012 cite both preprints and postprints in arXiv, either because preprints are earlier available or because authors do not have access to the licensed publisher's version.

### 5.2.3 Journal of Mathematical Analysis and Applications

The Journal of Mathematical Analysis and Applications puts emphasis on articles devoted to the mathematical treatment of interdisciplinary questions. It covers following areas of classical analysis: functional, complex, and numerical analysis, dynamic systems, probability, mathematical biology, and combinatorics.[49] The journal's Impact Factor for the year 2011 is 1.001. It is published by Elsevier in 12 volumes a year, with 2 issues each.

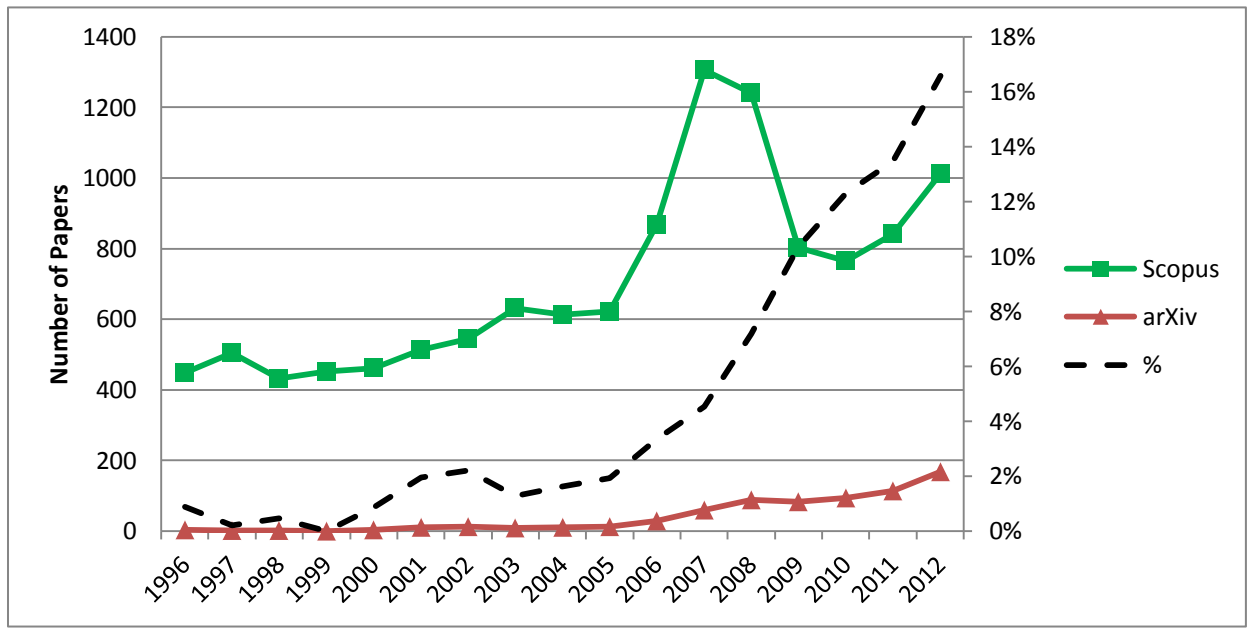

**Figure 24.** Growth of publication numbers between 1996 and 2012 for the Journal of Mathematical Analysis and Applications.

Figure 24 demonstrates that the number of articles published used to be steady from 1996 to 2005, and doubled unexpectedly from 621 in 2005 to 1,305 in 2007. The number of articles published went down in 2009, and is slightly growing since then. We can gather from Figure 24 that the number of articles with a previous preprint in arXiv is very low, compared to the other two mathematical journals. Nonetheless, as the percentage graph proves, the relative number of preprints has grown steadily from 3.3% in 2006 to 17.0% in 2011. The total number of articles published in the Journal of Mathematical Analysis and Applications in the time period 1996-2012 is 12,014, of which 696 articles have a foregoing preprint. This makes up a share of 5.8% of all articles published in the reference period. The number of postprints

---

published between 1996 and 2012 is 65, which is less than one percent of all available articles from this time period. Although Elsevier demands to provide a note on the publisher's version, only 95 e-prints were found in arXiv with a note on the formal publication. This is fewer than one in seven. Since the journal is not that dominated by preprints and the publication of preprints seems to be a recent trend, it is of interest to see whether the publication delay is as long as for the other two mathematical journals or not.

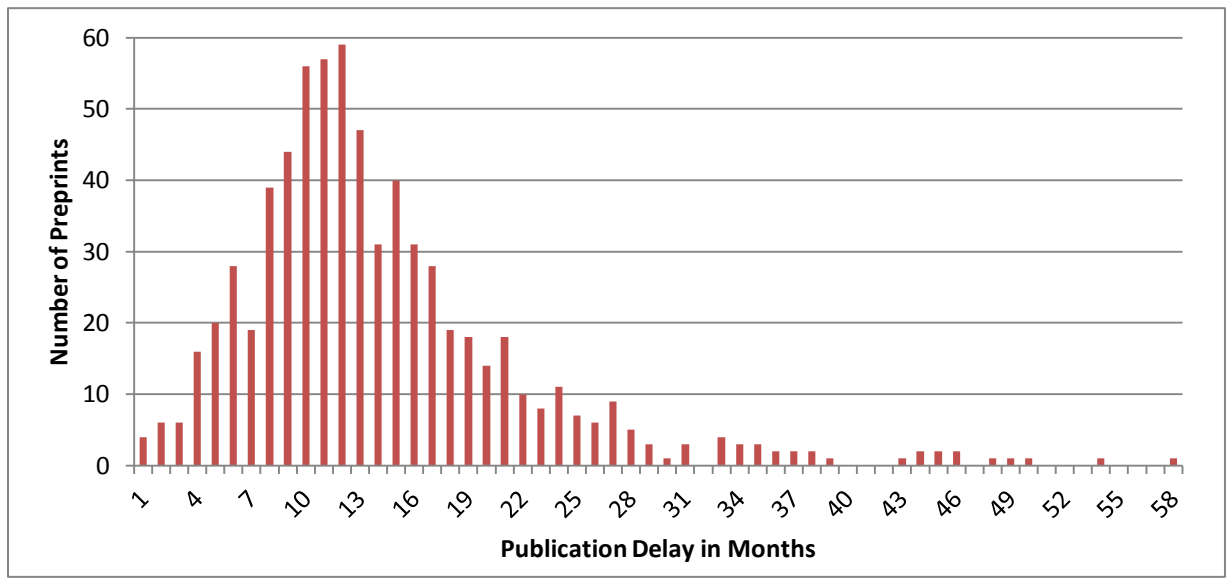

**Figure 25.** Distribution of preprints over months prior the journal publication. Those preprints are considered whose articles were published between 1996 and 2012 in the Journal of Mathematical Analysis and Applications.

In Figure 25 we can see the publication delay of all preprints that were published between 1996 and 2012 as articles. The mode value is 12 months, which means that a majority of 8.5% of all preprints is deposited in arXiv one year prior to their peer-reviewed publication in the journal. Four additional preprints are scattered over the months 61 to 70, and are consequently not depicted. The median publication delay of all 696 preprints is 358.5 days. The shortest publication delay is 2 days, whereas the longest is 2,092 days (more than five years). The median value of the publication delay is consistent with the mode in the histogram.

It is of further interest to see whether the publication delay has decreased in the course of time or not. Figure 26 presents a time series of the median publication delay. It is visible that the median publication delay rose until the year 2005, where it had its peak with 541 days. After 2005, the median publication delay went gradually down. In the past five years it has been 300-odd days. Even if the number of articles with a preceding preprint nearly doubled from 2008 to 2012, the median publication did not particularly change.

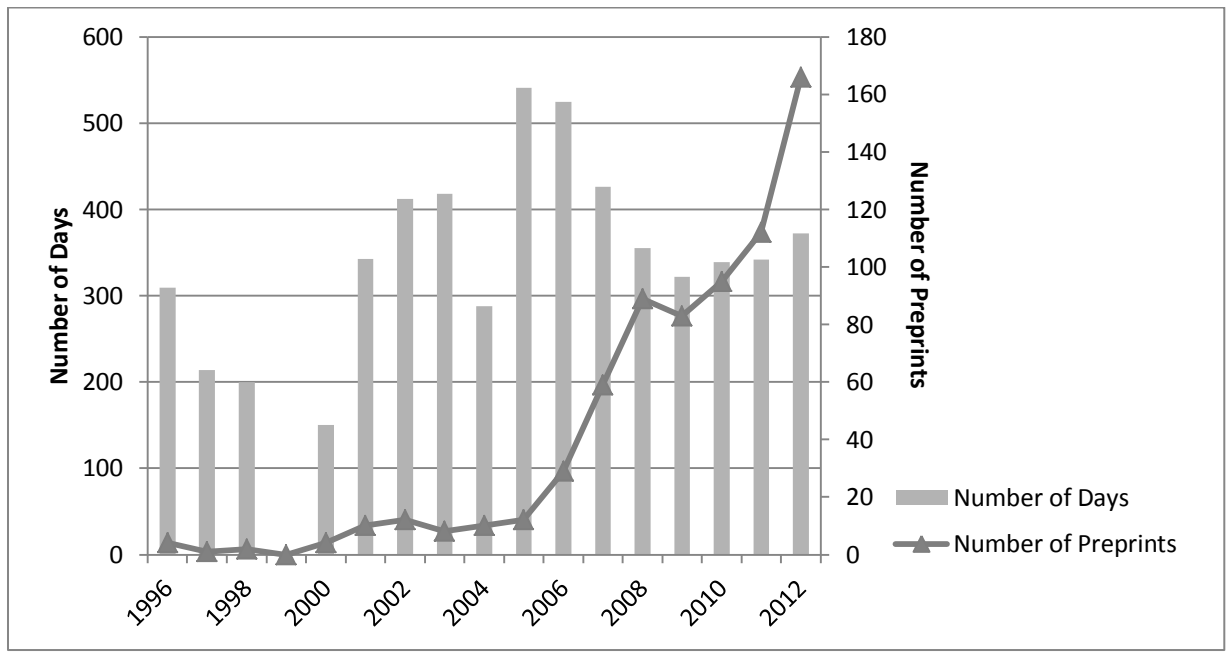

**Figure 26.** Time series of the median publication delay in days for articles that were published between 1996 and 2012 in the Journal of Mathematical Analysis and Applications and have a previous preprint in arXiv.

Since the median publication delay of all preprints published in the Journal of Mathematical Analysis and Applications is much shorter than for the two other journals, it is of interest to have a look on the impact of articles with a foregoing preprint. Therefore, articles published between 1996 and 2009 have been examined, again with a citation window of three years after journal publication. In this reference period, 323 articles were published that have a former preprint in arXiv. Welch's t-test shows that articles with a preceding preprint in arXiv are on average 0.3 times more cited than articles without a preprint (P=0.049).

To determine the citation delay of the first citation, articles published in the same reference period were examined. As a result, 87 out of 323 articles with a preceding preprint remained uncited in their first three years after journal publication (26.9%). From the set of 9,114 articles published in the same time frame and that do not have a preprint in arXiv, 3,237 remained uncited (35.5%). According to Welch's t-test, articles with a foregoing preprint receive their first citation twelve weeks earlier than articles without a preprint in arXiv (P=0.044). The advance of three months is observable in Figure 27, where the graphs for the two sets of articles are depicted. Different from the Annals of Mathematics and the Advances in Mathematics, the advance to receive the first citation at an earlier stage is visible for all three years after journal publication.

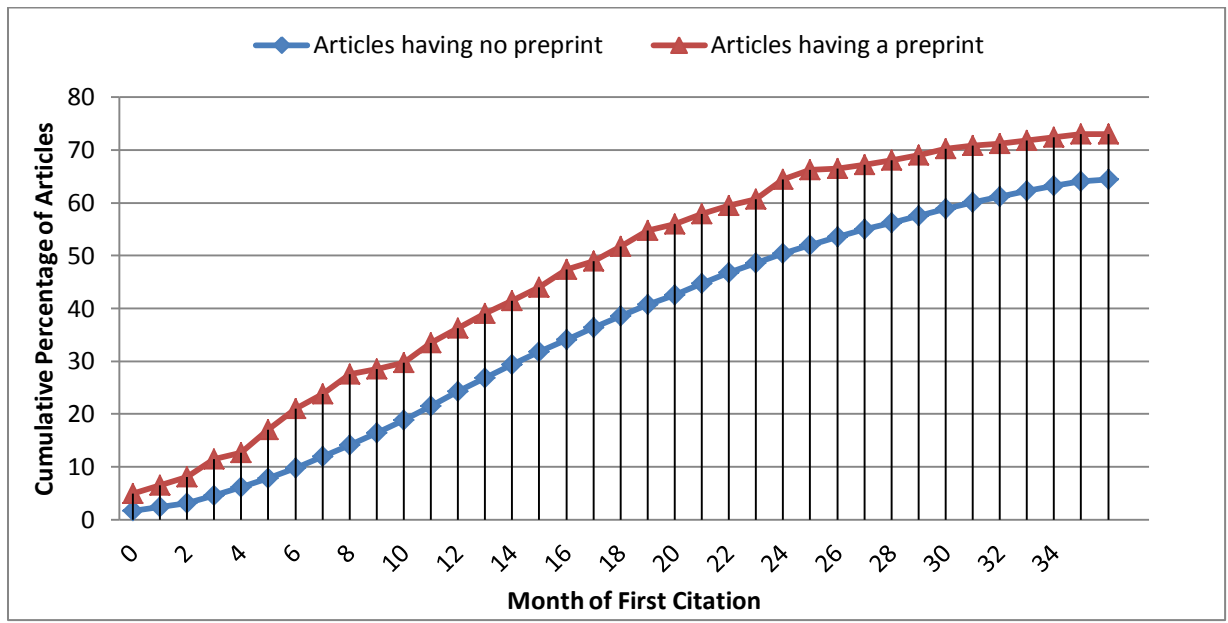

**Figure 27.** Cumulative distribution function of first citation for articles published in the Journal of Mathematical Analysis and Applications between 1996 and 2009. The citation data is based on Scopus. Articles with a previous preprint are on average 12 weeks earlier cited than articles without a preprint.

### 5.2.4 Discussion

The results presented show that there is no real speed pressure in mathematics. The manuscripts can spend one year or even more between the journal submission, which is mostly consistent with the deposit in arXiv, and the reviewed publication. The mathematics community has its "own standards of rigorous peer review, which they call refereeing" (Acheampong, 2011, p. 4). Refereeing has to be detailed because mathematical papers once proven can be cited for all time. Davis & Fromerth (2007, p. 214) write that mathematical journals behave more like those in humanities than those in sciences. Their citation half-life is more than ten years, and thus requires a longer period to investigate in order to draw correct conclusions. Even if this bibliometric analysis in mathematics covered a relatively long period of 17 years, the majority of preprints were submitted to arXiv in recent years.

The results show that the publication delay can range from one to three years. None of the above journals showed a significant decrease of the median publication delay over time. However, long publication delays are not a real problem because arXiv exists as a venue for almost final publications. At the same time, the long publication delay may be due to the fact that preprints uploaded in arXiv are far away from the formal journal publication. The time gap between the upload in arXiv and the publication in a journal is possibly used for various revisions. Nevertheless, since literature in mathematics is valid for a long period of time, a publication delay becomes meaningless.

Davis & Fromerth (2007) have described in their paper that articles deposited in arXiv receive more citations than non-deposited articles. The mean difference they found for articles published in four mathematical journals between 1997 and 2005 was 1.1 citations per article.[50] A similar advantage has been presented for the Annals of Mathematics within a citation window of three years. Articles with a preceding preprint receive on average one more citation than articles without a preprint. On equal terms, the Advances in Mathematics showed that articles with a foregoing preprint are 0.5 times more cited, whereas articles published in the Journal of Mathematical Analysis and Applications are 0.3 times more cited. The citation advantage can be supported by the early availability of preprints, since they appear on average one to three years prior to publication.

All in all, the results suggest that mathematicians will continue publishing in arXiv because they are more likely to publish their results openly than researchers in other fields. This is due to the fact that the field is not ruled by patents, and no economic interests for mathematical findings exist. In few cases, papers posted in arXiv can establish priority, especially in regard to the long referee time in mathematics.

In addition, the preprint submissions in mathematics vary among areas. The front for the arXiv homepage shows that areas such as Algebraic Geometry, Combinatorics, Differential Geometry, Mathematical Physics, and Probability Theory are more popular than General Topology, Metric Geometry, or Spectral Theory.[51]

Mathematical journals do not have to fear arXiv because they insist on peer review and printed versions. They would put themselves at risk if they dropped the print publication of journal articles. Each of the three presented journals maintains an online and a print version. The publisher's task remains to promote the final refereed version and to make it distinguishable from the arXiv preprint. Since many authors fail to update their record in arXiv with the note on the journal article, publishers keep their important role in leading the reader to the final published version.

---

[50] The citation window was not mentioned in the paper.
[51] UC Davis (2013). Front: math Mathematics. http://front.math.ucdavis.edu/math

## 5.3 Astrophysics

Astrophysics is included in arXiv since April 1992. It is represented by 17.4% (141,205) of all preprints for the period 1992-2012.[52] According to Google Scholar[53], arXiv Astrophysics (astro-ph) has an h5-index of 176, and is thus on first position.[54] It is followed by The Astrophysical Journal. On position four, with an h5-Index of 111, is the journal Astronomy & Astrophysics, which will be presented in the following bibliometric analysis. Also, The Astronomical Journal, on position eight with an h5-Index of 76, is central to the investigation.

### 5.3.1 Astronomy & Astrophysics

Astronomy & Astrophysics (A&A) is an international journal that is published by EDP Sciences (Édition Diffusion Presse).[55] It was founded in 1969 after the merging of six former national journals on astronomy. The journal includes theoretical, observational, and instrumental aspects of astronomy and astrophysics. The current Impact Factor is 4.587 (2011). A&A offers Open Access, where authors pay a fee of 400 € to make their article freely available. Interestingly enough, it is stated "that the fee covers only a fraction of the editorial and production costs."[56] Since 2011, 12 volumes a year are published. Before that, 16 volumes a year appeared. During the analysis it became clear that issue 3 from volume 485 (2008) is missing in Scopus. In addition, issue 1 from volume 506 (2009) is not included in Scopus, which is a special feature on COROT.[57]

Figure 28 illustrates the publication output for A&A. What we can infer from Figure 28 is that the number of articles published per year was constant between 1996 and 2000. In 2001, the number of articles increased because the Astronomy & Astrophysics Supplement Series has merged with A&A. Since then, the number of articles rose slightly from 1,869 articles in 2001 to 2008 articles in 2007. For the years 2008 and 2009 a decrease is visible. The decrease in publication numbers results from the above mentioned missing issue in 2008 and the special issue on COROT in 2009.

---

[52] arXiv (2013). arXiv submission rate statistics. http://arxiv.org/help/stats/2012_by_area/index
[53] Google Scholar Metrics. Astronomy & Astrophysics.
http://scholar.google.com/citations?view_op=top_venues&hl=en&vq=phy_astronomyastrophysics
[54] The presented number h signifies that h articles published between 2007 and 2011 were at least h times cited.
[55] EDP Sciences. Astronomy & Astrophysics (A&A). http://www.aanda.org/
[56] EDP Sciences - Administrator. Astronomy & Astrophysics (A&A).
http://www.aanda.org/index.php?option=com_content&view=article&id=863&Itemid=295
[57] COROT stands for COnvection ROtation and planetary Transits.

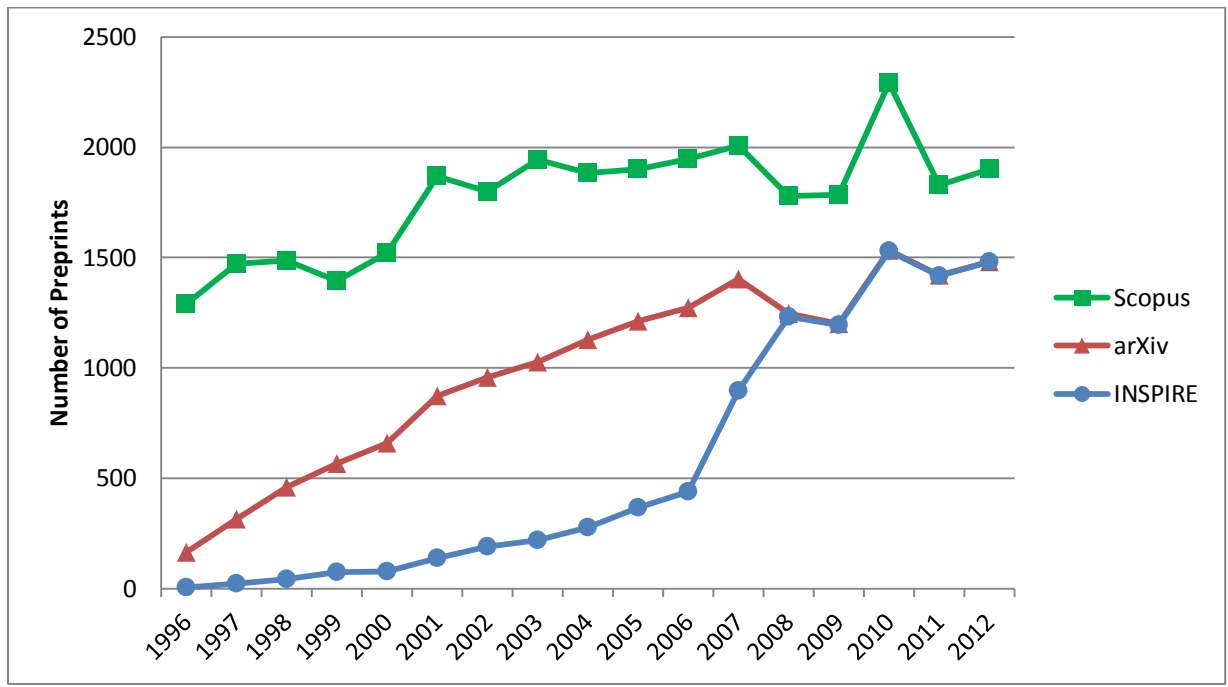

**Figure 28.** Growth of publication numbers between 1996 and 2012 for Astronomy & Astrophysics.

Consequently, the numbers are supposed to be higher, so that there is a continuous growth until 2010. Nevertheless, from 2010 on, the number of articles published declined. Overall, 29,697 articles have been published in A&A between 1996 and 2012. 16,776 of these articles have a previous preprint in arXiv, which is 56.5%. It is visible that the number of articles with a foregoing preprint grew linearly, having a sudden peak in 2001, which suggests that preprints were also meant for the Astronomy & Astrophysics Supplement Series. Just as for HEP journals, it can be seen that from 2008 onwards, almost every preprint in arXiv is covered by INSPIRE HEP. For the past 17 years, 56.6% of all preprints in arXiv can be found in INSPIRE HEP. In addition, 649 postprints were posted in arXiv, thus 2.2% of all articles published in A&A in the reference period.

The next figure displays the histogram of the publication delay for A&A. 60 preprints are not depicted in the histogram because they are scattered over the months 25 to 104 prior to publication. We can gather from Figure 29 that the mode is three months. A majority of 26.9% preprints is published 61 to 90 days prior to publication. Within four months prior to publication 77.6% of all preprints are deposited in arXiv, within one year even 97.6% of all preprints. The median publication delay of all preprints, published as articles between 1996 and 2012, is 76 days. Now, is this short publication delay due to the journal's fast peer-review process or are astrophysicists submitting their papers in arXiv after they have been accepted for journal publication?

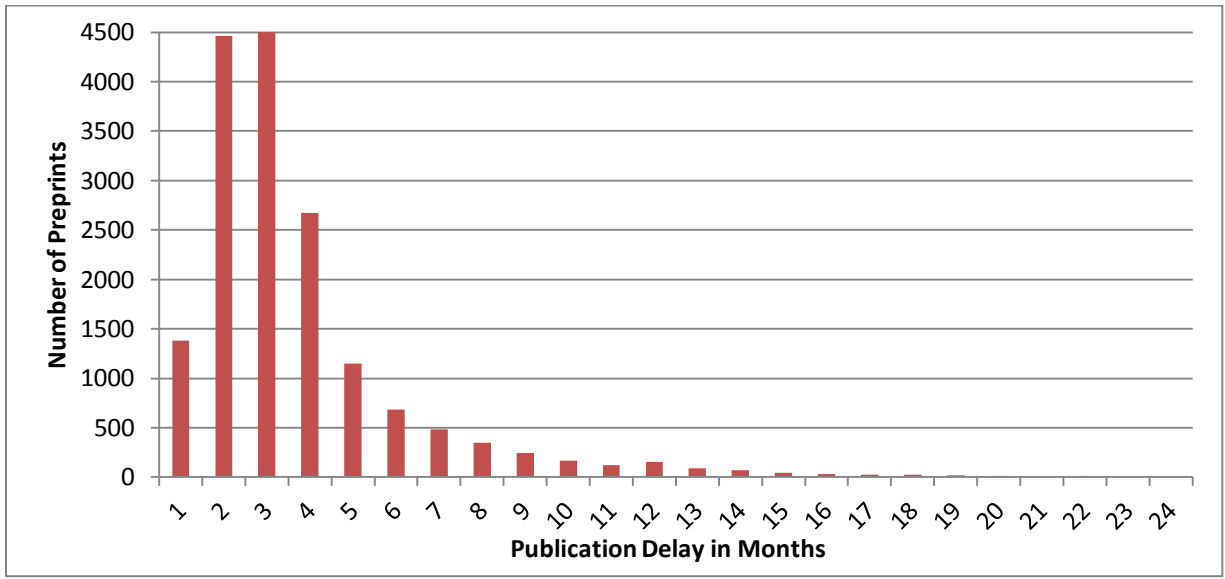

**Figure 29.** Distribution of preprints over months prior to journal publication delay. Those preprints are considered whose articles were published between 1996 and 2012 in Astronomy & Astrophysics.

According to Bertout et al. (2012, p. 29) the peer-review process for Astronomy & Astrophysics takes three months on average. The median acceptance time in 2010 was 81 days, and three-fourths of papers are accepted within four months prior to publication (ibid.). These numbers are consistent with the results presented above, although the latter are only based on preprints in arXiv. It is worth mentioning that Bertout et al. (2012) write that the duration of peer review depends on the author revising his paper and not on the referee. The next figure reveals how the duration of the median publication delay developed over time.

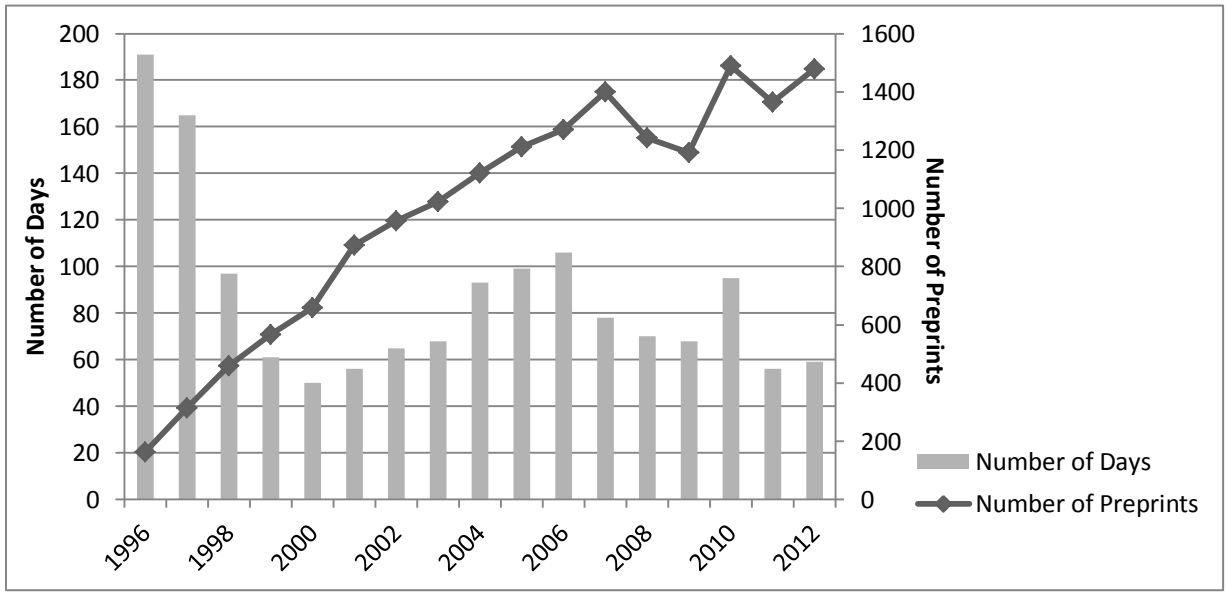

**Figure 30.** Time series of the median publication delay in days for articles that were published between 1996 and 2012 in Astronomy & Astrophysics and have a previous preprint in arXiv.

In Figure 30 it can be seen that in 1996 the median publication delay was more than six months. With advanced Internet technologies, the delay decreased and reached a minimum of 50 days in 2000. From 2000 to 2006 the median publication delay rose, in line with a growing number of preprints published, and decreased since then. All in all, the publication delay is short and ranged in the past 15 years between two and four months. It is striking, though, that with a higher number of preprints published, the delay increased from 2000 to 2006. For the following years there is no clear coherence visible. One attempt to explain the growth of the publication delay, in line with an increasing number of articles published having a foregoing preprint in arXiv is the following. Usually, a certain number of articles is published on current research. If the publication output rises, then probably because articles are included that were postponed to later issues, in which they fit better. This might increase the preprints' publication delay. Nevertheless, the publication delay is shorter than for many other journals. It is thus relevant to see if it is of any use for the preprints' early visibility in INSPIRE HEP.

The next figure provides a histogram of first citations of papers, in relation to publication in A&A. Only those preprints are considered whose articles were published between 1996 and 2010. The citation window is set to two years after journal publication. This applies to a total of 6,654 papers, of which 45% remained uncited prior to journal publication plus two years after. From Figure 31 we can infer that the mode is one month prior to journal publication.

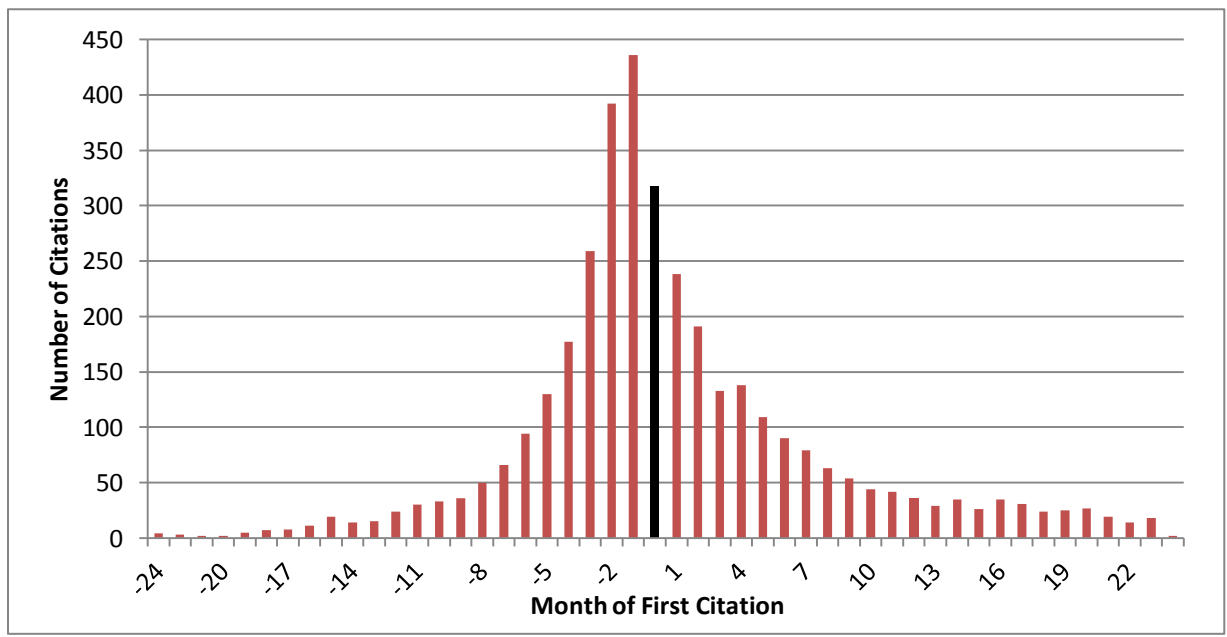

**Figure 31.** Distribution of first citations over months after journal publication for papers in INSPIRE HEP that were published between 1996 and 2010 as articles in Astronomy & Astrophysics. The citation window is two years after article publication and the citation data is based on INSPIRE HEP.

According to INSPIRE HEP, 6.6% of all preprints receive their first citation 1 to 30 days before the formal publication. On the whole, 27.7% of all preprints in the reference period receive their first citation before the peer-reviewed article appears. It suggests that the early visibility of preprints allows instant communication because researchers do not have to wait for the peer-reviewed article.

Furthermore, all citations were considered for articles published between 1996 and 2010. The citation window was again set to two years. As a result, 19.9% of all citations are awarded to preprints, thus to papers that are cited before they are published in A&A. Because of the early reuse of preprints it is central to see whether preprints can predict higher citation rates after their publication in A&A. Therefore, on the basis of citations in Scopus, Welch's t-test has been applied to examine if articles that have a previous preprint in arXiv get more cited during their first two years after journal publication. The t-test shows that articles with a previous preprint receive on average 2.7 more citations in their first two years after publication in A&A (P=0.017). It suggests that the early availability of preprints in arXiv can boost the impact of their formally published analogues. The following figure demonstrates the advantage in citation rates for articles with a previous preprint. The data is based on the year 2002, because for this year the number of articles with a previous preprint, and the number of articles without a preprint is almost equal. Figure 32 shows the month of the citation after journal publication and the cumulative citation number.

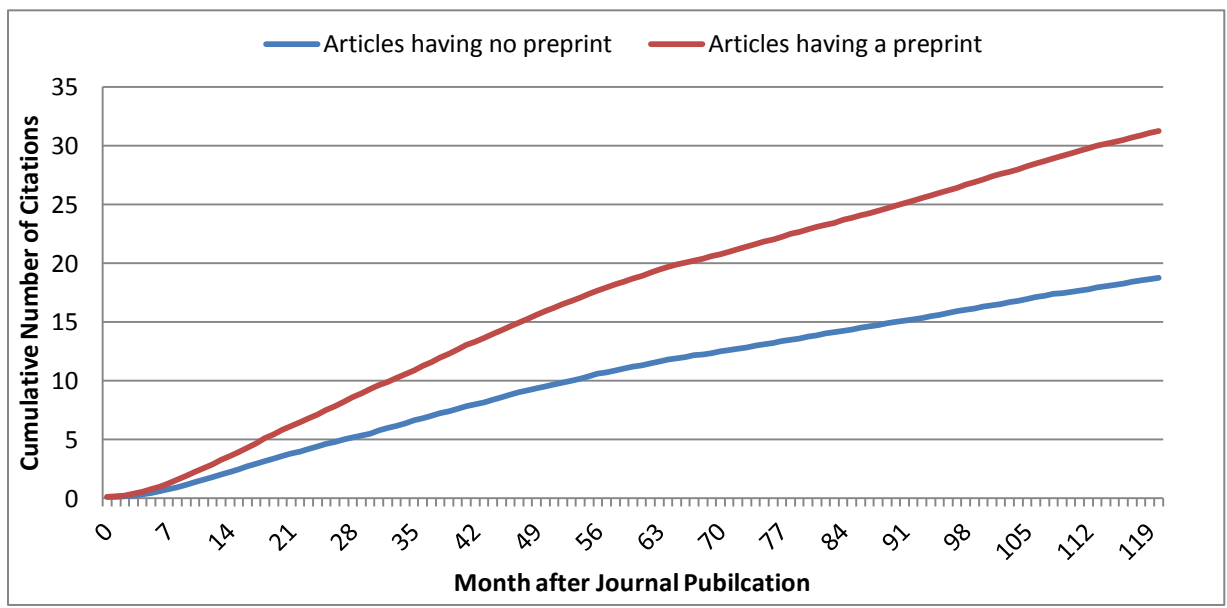

**Figure 32.** Cumulative number of citations after publication in Astronomy & Astronomy for articles with and without a previous preprint in arXiv. The publication year is 2002 and the citation window is ten years after publication. The citation data is based on Scopus.

For the year 2002, 956 out of 1,800 published articles have a foregoing preprint in arXiv (53.1%). It is visible that the red line graph, representing articles with a previous preprint, is far ahead of the blue line graph, which represents articles without a preprint in arXiv. If we have a look on month 24, we can see that articles without a preprint have on average four citations, whereas articles with a previous preprint accumulated on average seven citations. The distance of around three citations per article signifies the result of Welch's t-test, which showed that articles with a previous preprint receive on average 2.7 more citations, two years after publication. The result of Welchs's t-test, which was calculated for all articles published between 1996 and 2009, is thus consistent with the publication data for the year 2002.

### 5.3.2   The Astronomical Journal

The Astronomical Journal (AJ) is a monthly journal owned by the American Astronomical Society (AAS) and published by IOPscience (Institute of Physics). The Astronomical Journal was founded in 1849 and was acquired by the AAS in 1941.[58] AJ has a broad view of astronomy, from the solar system to observational cosmology. Its Impact Factor for the year 2011 is 4.035. The first electronic edition of AJ appeared in 1998. In July 2006 AJ started e-first publications which are available independently of the printed journal version. The journal is published in two volumes a year, with six issues each. For the year 2009 Scopus provides only 260 articles, although 402 exist. Apparently, three issues (volume 137, issue 4, 5, 6) are not tracked by Scopus. This explains the unexpected decrease for the year 2009 in Figure 33.

As can be seen in Figure 33, the number of journal articles was uniform from 1996 to 2007, fluctuating between 446 and 528 articles a year. Even if Figure 33 shows an incorrect decrease in 2009, the number of articles published did go down in the past four years. In 2012, 345 articles were published. On the other hand, the number of preprints in arXiv rose considerably from 1996 to 2001, where it reached its saturation for the following years. About 60% of articles published per year from 2001 to 2008, have a previous preprint in arXiv. The share of articles with foregoing preprints rose again in 2009 and reached 70% in recent years. The inclusion in INSPIRE HEP grew slowly from 1 preprint in 1996 to 158 in 2007. Again, it is visible that from 2008 on, almost every preprint in arXiv is also included in INSPIRE.

---

[58] IOPscience. The Astronomical Journal. http://iopscience.iop.org/1538-3881

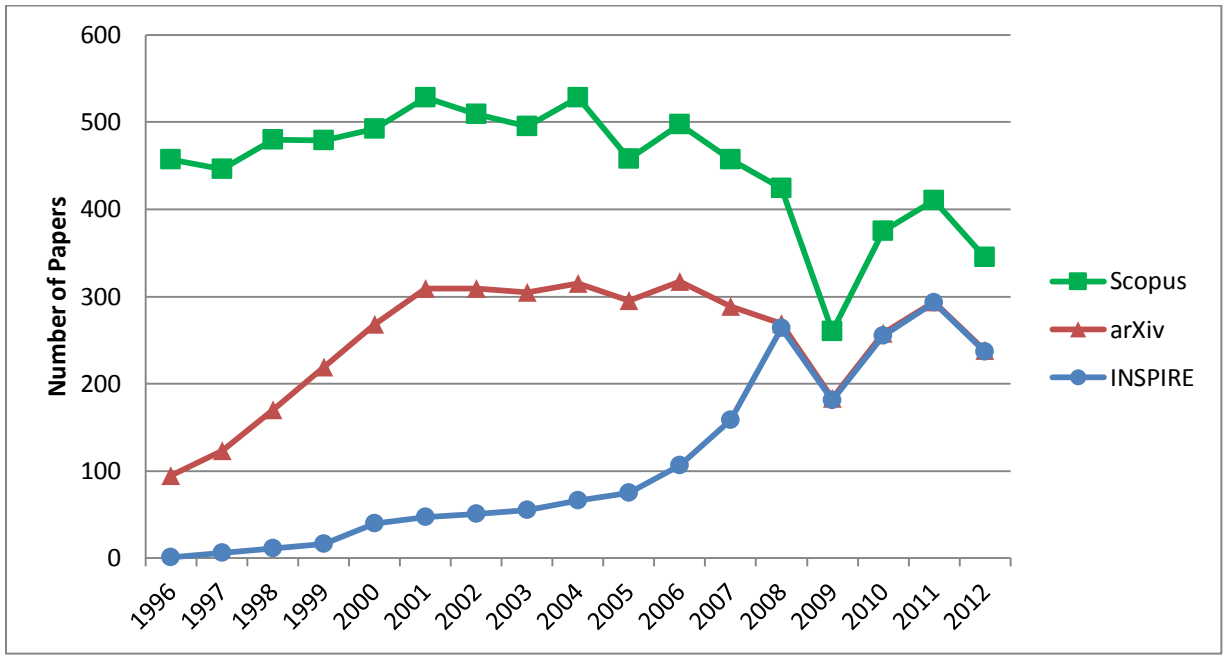

**Figure 33.** Growth of publication numbers between 1996 and 2012 for the Astronomical Journal.

Of all 7,638 articles published between 1996 and 2012 in AJ, 4,245 have a previous preprint in arXiv, which makes up 55.6%. INSPIRE HEP tracks 43.8% of all preprints in the same reference period. The next figure illustrates the publication delay of all preprints published in AJ. Just as the mode of the publication delay for A&A was three months, the mode of the publication delay for AJ is the same. 14 additional preprints are not included in Figure 34 because they are scattered over the months 25 to 48 prior to publication.

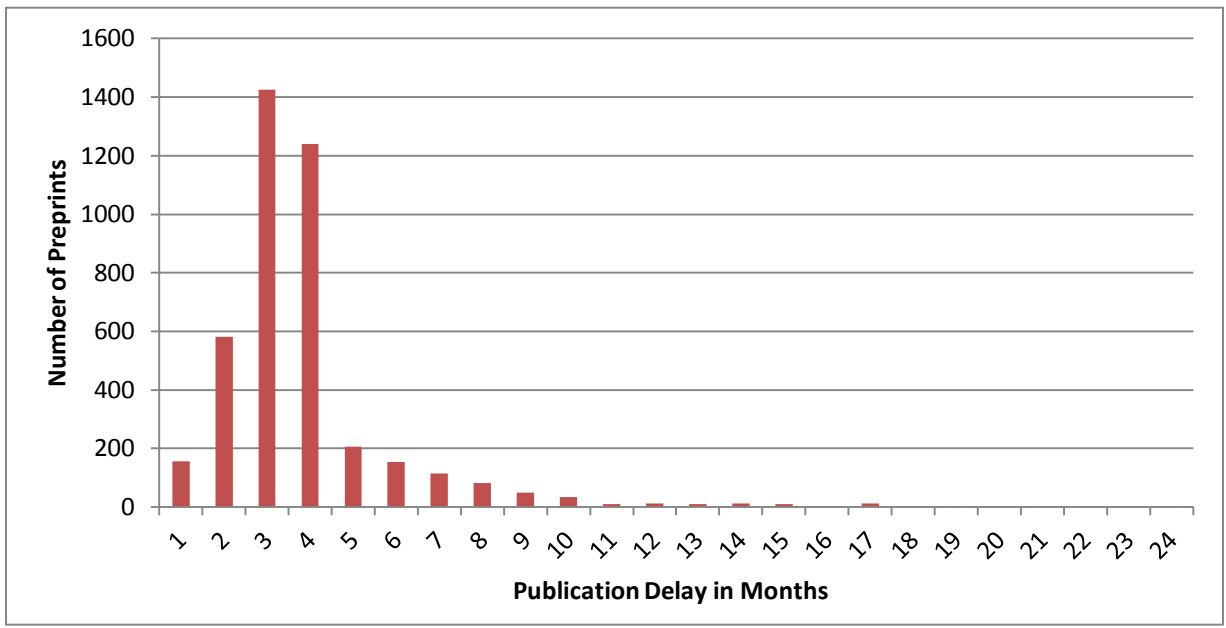

**Figure 34.** Distribution of preprints over months prior to publication. Those preprints are considered whose articles were published between 1996 and 2012 in the Astronomical Journal.

A majority of 36.3% preprints are placed in arXiv 61 to 90 days before the journal publication. Within four months prior to publication 84.6% of all preprints are deposited in arXiv. Within one year prior to publication, even 98.2% of all preprints. The following figure demonstrates how the median publication delay evolved over the years.

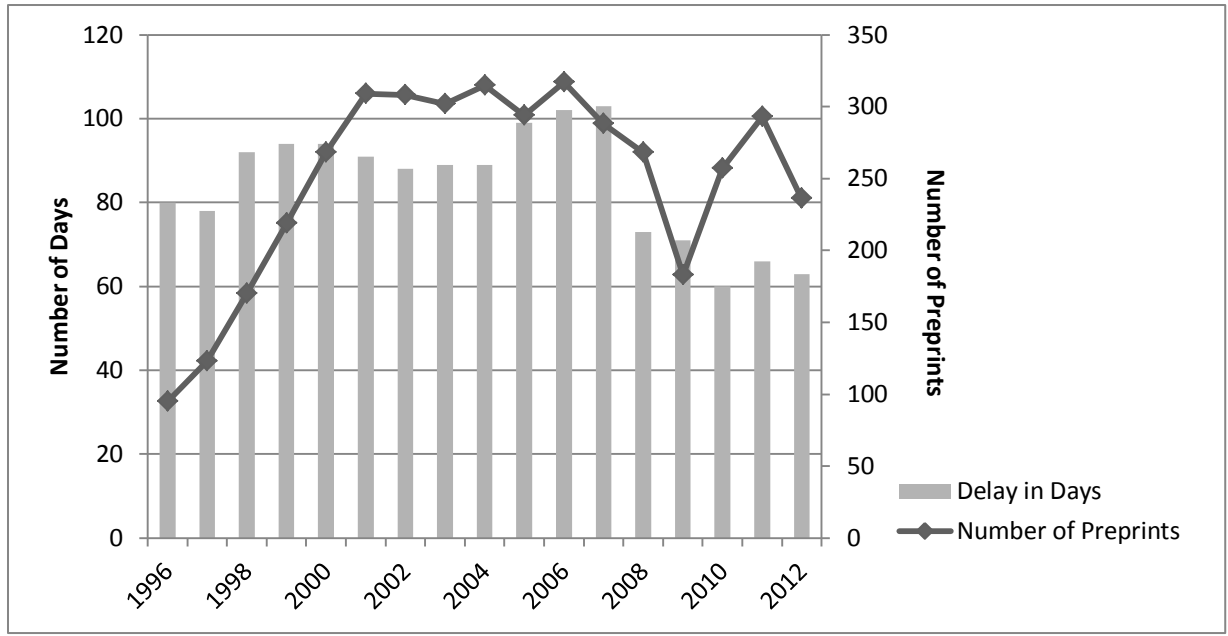

**Figure 35.** Time series of the median publication delay in days for articles that were published between 1996 and 2012 in the Astronomical Journal and have a previous preprint in arXiv.

It can be seen in Figure 35 that the median publication delay has been always shorter than four months. It had its peak in 2007 with 103 days and a minimum of 60 days in 2010. The median publication delay decreased by one month from 2007 to 2008 and has been about two months in recent years. The reason for the shorter publication delay since 2008 might be due to the new publisher IOPscience. Until 2008, AJ was published by the University of Chicago Press and AAS. With the new publisher the median publication delay went remarkably down to two months. Is this short publication delay of any use in regard to citations?

The next figure displays the distribution of first citations over months in relation to publication in the Astronomical Journal. It depicts preprints that were published between 1996 and 2010 as articles in AJ, and are also covered by INSPRIE HEP. This applies to a total of 1,332 preprints. 38.7% of those remained uncited before journal publication plus two years after. Thus, only 61.3% of all papers are included in Figure 36. The histogram indicates that the mode of the first citation is the month after publication. 5.9% of all papers are cited, shortly after their publication in AJ.

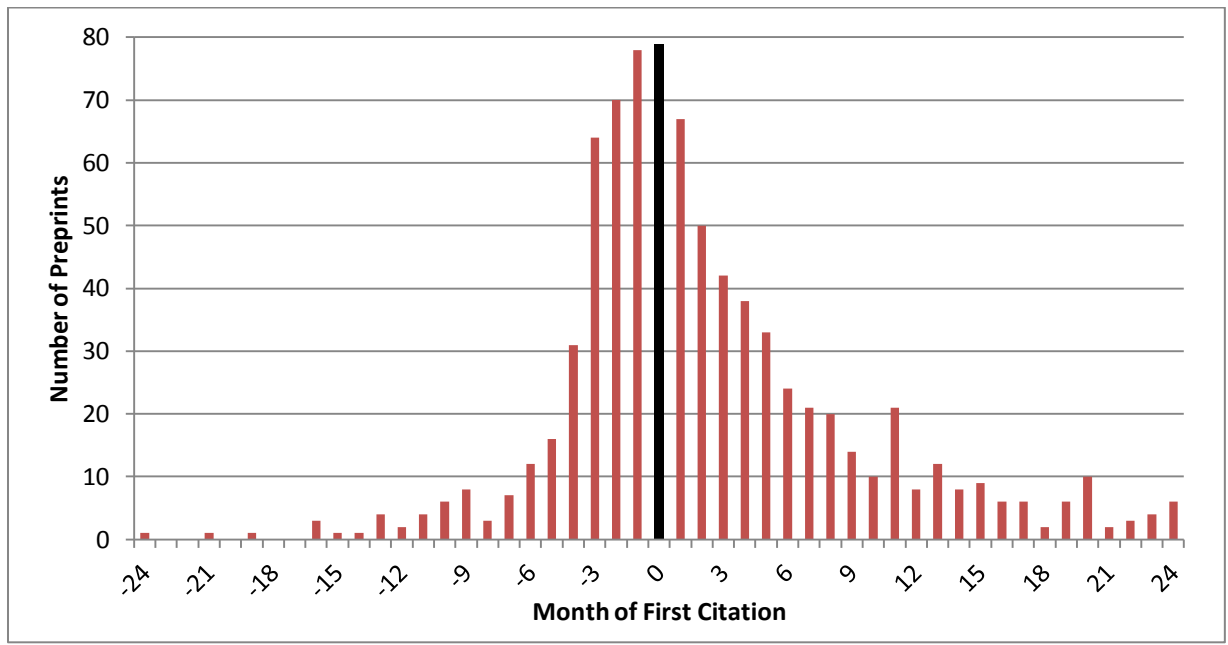

**Figure 36.** Distribution of the first citations over months after journal publication for papers in INSPIRE HEP that were published between 1996 and 2010 as articles in the Astronomical Journal. The citation window is two years after article publication and the citation data is based on INSPIRE HEP.

Overall, 23.7% of all papers receive their first citation before they are published in AJ. This share is consistent with the one for A&A. In regard to all citations received for articles published between 1996 and 2010, and a citation window of two years, 7.5% of all citations are awarded to papers that are cited before their publication in AJ. Preprints published in AJ are not only earlier cited, their article version reaches also a higher average citation rate. Welch's t-test has shown that articles published in the Astronomical Journal, having a preprint in arXiv, are significantly more cited. On average, they receive three more citations in their first two years after publication in AJ than articles without a preprint in arXiv (P=0.013).

The following figure displays the advantage in average citation rates for articles published in 2000 in AJ. The year has been chosen because the number of articles with a previous preprint is almost equal to the number of articles without a preprint. Thus, 268 out of 492 articles have a preceding preprint (54.5%) for this publication year. In Figure 37, it is visible that 24 months after journal publication, articles with a preceding preprint are on average three times more cited. After four years of publication in AJ, articles with a previous preprint are on average six times more cited than those without a preprint. The citation advantage grows gradually over time.

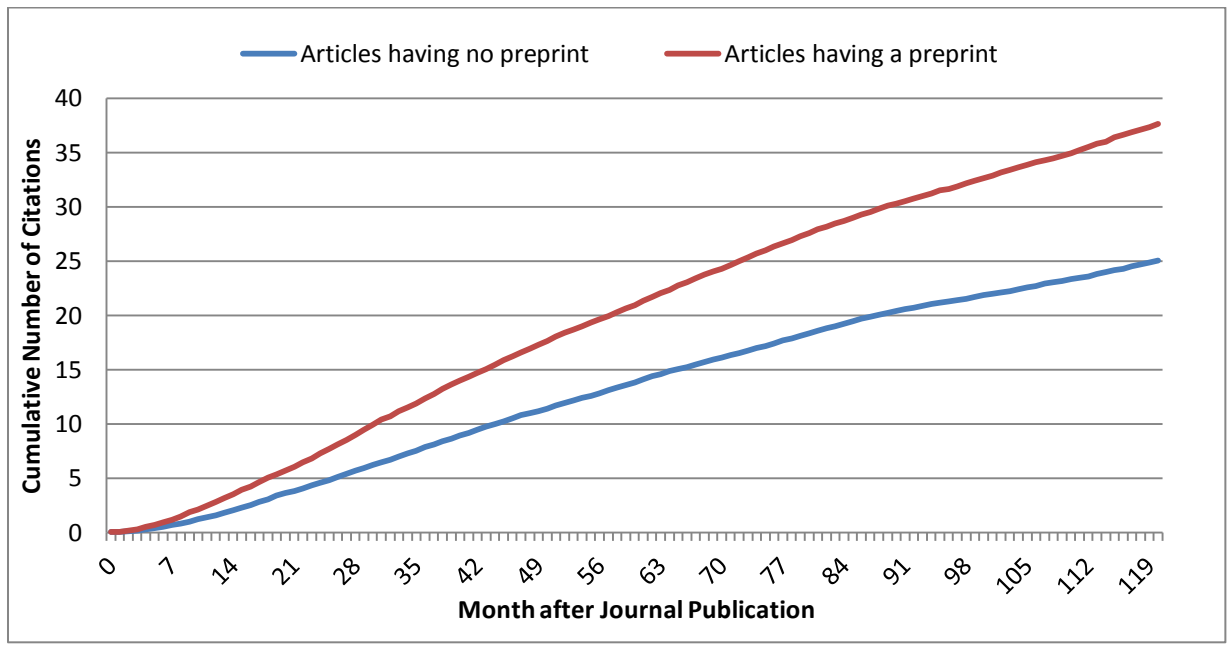

**Figure 37.** Cumulative number of citations after publication in the Astronomical Journal for articles with and without a previous preprint in arXiv. The publication year is 2000 and the citation window is ten years after publication. The citation data is based on Scopus.

### 5.3.3 Discussion

Astrophysics is a high-moving field that has just like HEP a preprint culture. It has been shown that the number of articles with a previous preprint deposited in arXiv grew linearly since 1996. Just like in HEP, scholarly communication in astrophysics requires speed. Although there is no commercial potential in astronomy and astrophysics, speed is essential because the field is characterized by competition and fast research when it comes to discoveries. Research groups often study the same objects, so that each group wants to publish results as quickly as possible to claim priority (Harley et al., 2010, p. 169). Waiting for the final version of results would allow other research groups to be far ahead. Thanks to arXiv, scientist can disseminate current findings and keep up-to-date with recent developments. Harley et al. (2010, p. 168) write that astronomers who do experimental or observational research never post in arXiv until their paper has been accepted for journal publication. Theorists, on the other hand, deposit preprints without hesitation and are grateful for any comments or pointers prior to publication. Since arXiv is available to a wide readership and the primary access point for many researchers, it is likely that somebody will read a preprint and communicate his opinion.

The distributions of the publication delay for the two presented journals suggest that authors put their paper on arXiv as they submit it to the journal. Refereeing or modifying the original manuscript takes time. As has been shown, journal publishers managed to cut down the delay over time and indeed, compared to journals in other research fields, the publication delay is short in astrophysics. On average, the arXiv community can access the results three months earlier. This gain in time is used immediately because astrophysicists do not wait until the peer-reviewed version. This was proven by the number of preprints cited before the actual article came out. It has been also shown that articles having a previous preprint in arXiv, receive on average three more citations, two years after journal publication. Schwarz & Kennicutt (2004) found that papers in The Astrophysical Journal, posted prior to publication in arXiv, were cited more than twice as often as those without a preprint in arXiv. One attempt to explain this effect is the longer visibility of preprints. Papers deposited in arXiv prior to formal publication have a longer citation lifetime. Schwarz & Kennicutt conclude that papers that are not posted in arXiv are simply ignored (ibid., p. 6).

Astrophysicists wish to be informed about new results as early as possible. ArXiv helps to get ideas into the public quickly and readers do not need to wait for peer review. Since astrophysics is a small field, scientists can judge the quality of a paper by the author's name. The final publication of an article in a journal becomes a formality. The few existing journals in astronomy and astrophysics are owned by societies and published by commercial publishers. Bertout et al. (2012, p. 4) write that the publication landscape in astronomy and astrophysics is less crowded than in other fields, because more than 90% of the original research is published by four international journals. These are the above presented Astronomy & Astrophysics and the Astronomical Journal, and in addition The Astrophysical Journal, and Monthly Notices of the Royal Astronomical Society. According to Harley et al. (2010, p. 150) the acceptance rate of papers in astrophysics is high compared to other fields. The acceptance rate is 95%, independent of the journal. It suggests that preprints deposited in arXiv are very close to the formal publication, especially because the short publication delay reveals that there is obviously no need for long revisions.

However, preprints and journal articles co-exist productively in astrophysics (Henneken et al., 2006). The early publication on arXiv gives the peer-reviewed article primacy, which might be the decisive factor for higher citation rates after journal publication.

## 5.4 Quantitative Biology

Since September 2003, arXiv maintains the section Quantitative Biology. It represents 0.8% (6,456) of all preprints, deposited in arXiv until the end of 2012.[59] Quantitative Biology has borrowed techniques from computer science, applied mathematics, biochemistry, and artificial intelligence. In the following analysis it is central to find out to which extent the biology community adopted the preprint server for the publication of their results. Therefore, two prominent journals have been chosen.

### 5.4.1 Journal of Theoretical Biology

The Journal of Theoretical Biology was founded in 1961 and covers areas such as evolutionary biology, immunology, and genetics.[60] It gives insight into biological processes and puts emphasis on mathematical and computational aspects of biology. It is published by Elsevier and its Impact Factor for the year 2011 is 2.208. Since October 2011, the publication cycle is 24 volumes a year, always on the 7[th] and 21[st] of a month. Before that, 24 issues per year were published (six volumes, with four issues each). Elsevier's Open Access policy was mentioned in the context of Nuclear Physics B. After browsing several issues, very few Open Access articles were findable. Nevertheless, there is at least one Open Access article, which also appeared as an arXiv preprint in advance, and is thus freely accessible in two ways.[61]

In the analyzed time period from 1996 to 2012, 5,617 articles have been published in the Journal of Theoretical Biology, of which 296 have a previous preprint in arXiv. This results in a share of 5.3%. One preprint can be even found in INPIRE HEP, although the title does not suggest so.[62] Besides the 296 preprints, 25 papers were deposited as postprints in arXiv. In Figure 38 we can see the publication output between 1996 and 2012 for the Journal of Theoretical Biology. It is visible that from 1996 to 2005 the number of articles published per year has been fluctuating between 214 and 300. In 2006, the number of articles published increased suddenly to 429. Since then, the number of articles published per year has been between 418 and 473. The number of preprints deposited in arXiv has been slightly growing over the years.

---

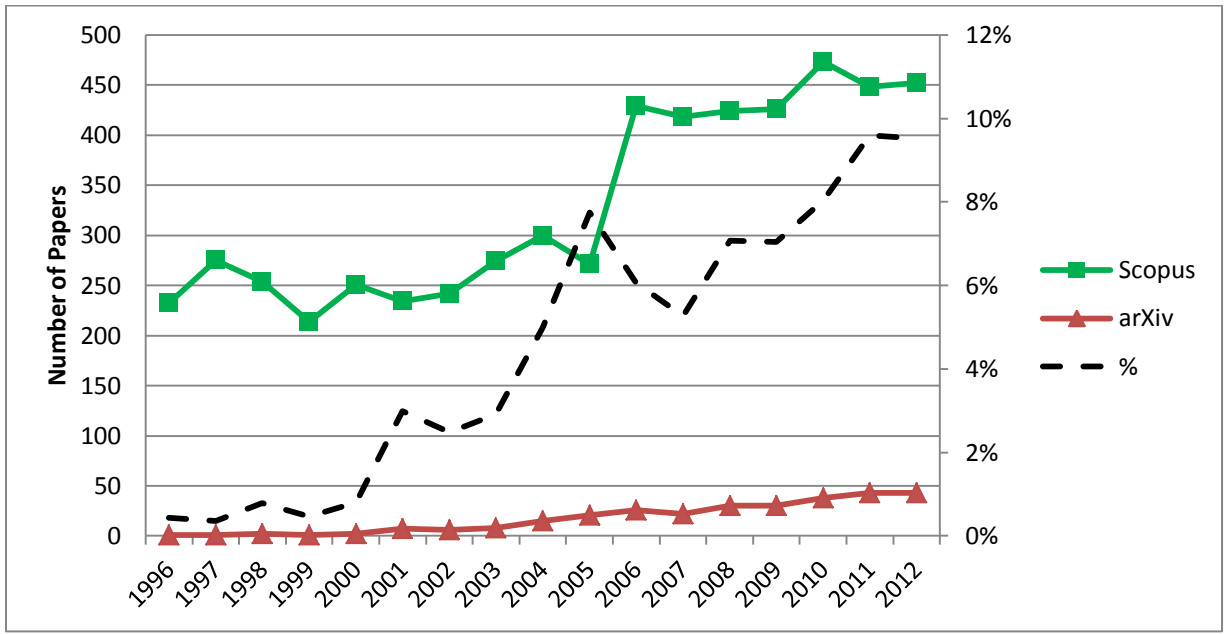

**Figure 38.** Growth of publication numbers between 1996 and 2012 for the Journal of Theoretical Biology.

In absolute numbers the amount of preprints seems low. However, the percentage graph reveals that the relative number of articles with a foregoing preprint has been growing steadily and reached nearly 10% in 2011. The median value of the publication delay of all 296 preprints is 240 days. The publication delay ranges between 1 day and 1,075 days. The publication delay of all preprints is displayed in the following figure.

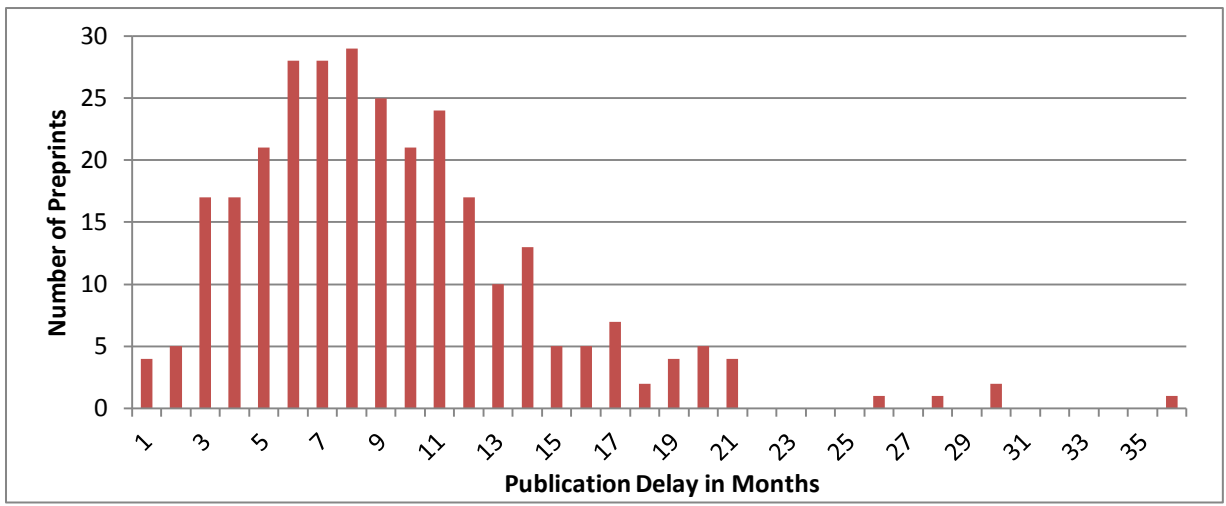

**Figure 39.** Distribution of preprints over months prior to publication. Those preprints are considered whose articles were published between 1996 and 2012 in the Journal of Theoretical Biology.

From Figure 39 we can gather that the mode of the publication delay is 8 months, reflecting the median value of 240 days of all preprints. One in ten papers (9.8%) is placed in arXiv 211 to 240 days prior to publication. Most of the preprints are scattered evenly between the third and twelfth month prior to publication. The histogram also shows that the majority of preprints are uploaded in arXiv within one year prior to publication (79.7%). For the sake of completeness it has to be mentioned that one preprint could have been counted twice to determine the median publication delay. Since there is a preprint in arXiv that is similar to two distinct journal articles published months later in the Journal of Theoretical Biology, the publication delay between the single preprint and each of the two articles could have been counted.[63] Instead, only the article with the identical title was matched.

It is of further interest to see if the primary category in arXiv overlaps with the journal's focus. Since the Journal of Theoretical Biology covers evolutionary biology, we can see in Table 2 that most of the preprints are devoted to this topic. The journal's mathematical and computational aspect of biology becomes clear in the primary category Quantitative Methods.

**Table 2.** Distribution of all preprints over the first five primary categories in arXiv for the Journal of Theoretical Biology.

| Primary category | Number of preprints |
|---|---|
| q-bio.PE (Populations and Evolutions) | 109 |
| q-bio.MN (Molecular Networks) | 30 |
| q-bio.m (Quantitative Methods) | 24 |
| q-bio.CB (Cell Behavior) | 23 |
| physics.bio-ph (Biological Physics) | 21 |

The search for the publication name in arXiv's Journal-ref. or Comments field leads to approximately 140 hits. In regard to overall 321 e-prints for this journal it shows that biologists are eager to provide a link to the peer-reviewed analogue. The next part focuses on the impact of articles published in The Journal of Theoretical Biology. Therefore, Table 3 provides an overview of the impact of articles with a preprint in arXiv, and those without. The first column shows the publication year, the second column the number of articles published in the respective year without a preprint, and the third column their average citation number within a citation window of two years after journal publication. The fourth column indicates the number of articles having a preprint, and the fifth the average citation number for those. The table starts with the year 2004 because prior to this publication year only few articles exist with a preceding preprint in arXiv.

---

**Table 3.** Average citation rates for the Journal of Theoretical Biology. The citation window is two years after the date of publication and the citation data is based on Scopus.

| Publication year | Number of articles with no preprint | Average citation number | Number of articles with preprint | Average citation number |
|---|---|---|---|---|
| 2004 | 285 | 3.51 | 15 | 4.47 |
| 2005 | 250 | 3.49 | 21 | 7.71 |
| 2006 | 403 | 3.96 | 26 | 5.12 |
| 2007 | 396 | 3.96 | 22 | 6.14 |
| 2008 | 394 | 3.98 | 30 | 3.93 |
| 2009 | 396 | 3.80 | 30 | 3.90 |
| 2010 | 435 | 3.63 | 38 | 3.63 |
| Total | 2,559 | 3.79 | 182 | 4.78 |

It is noticeable that the average citation number of articles without a preprint starts in each year with a three before the decimal point. In the column for articles with a previous preprint, the average citation numbers vary much more, on the one hand because the amount of articles is smaller, and on the other hand, it is possible that highly cited articles skew the average citation rate more than in the larger set of articles without preprints. However, Welch's t-test shows that there is an advantage for articles with a previous preprint. On average they get 0.3 times more cited within two years after publication than articles without a preprint (P=0.040). The t-test for the citation delay shows that articles with a preceding preprint are on average one month earlier cited than articles without a preprint in arXiv (P=0.024). This is well-expressed in the following figure.

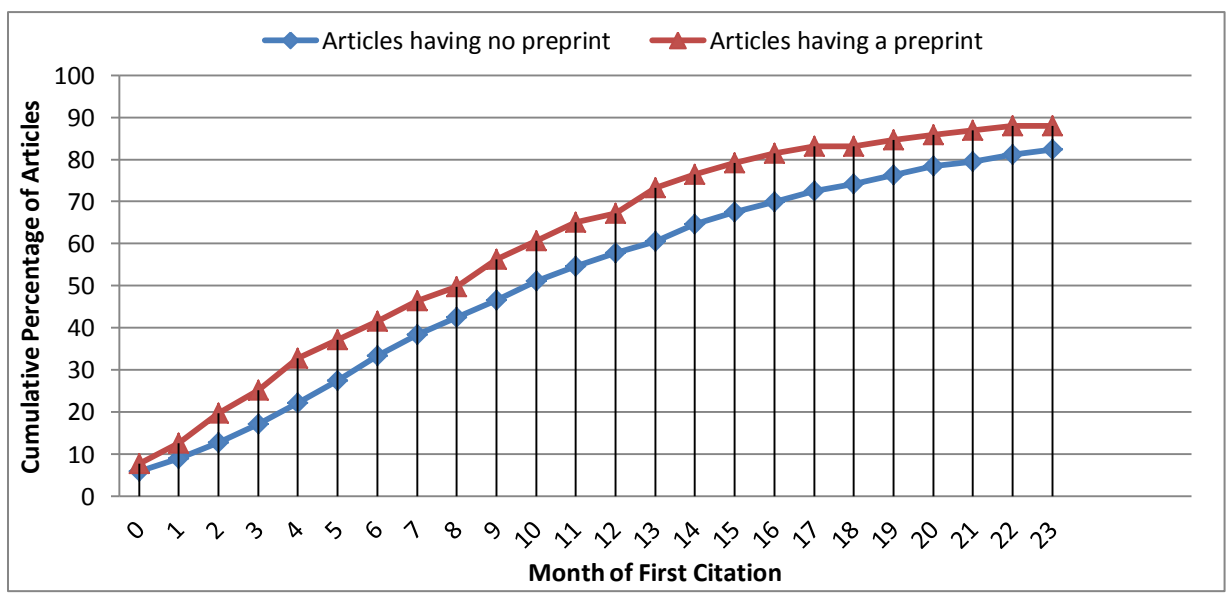

**Figure 40.** Cumulative distribution function of first citation for articles published in the Journal of Theoretical Biology between 2004 and 2010. The citation data is based on Scopus. Articles having a preprint in arXiv are on average one month earlier cited than those without a preprint.

### 5.4.2 Physical Biology

Physical Biology is an interdisciplinary online journal bringing biology and physical sciences together. The research articles focus on "quantitative characterization and understanding of biological systems at different levels of complexity."[64] It was founded in 2004 and is hosted by the Institute of Physics (IOP). Its Impact Factor for the year 2011 is 2.595. The journal is published in one volume a year consisting of four issues. IOPscience offers Open Access at a fee of $2,700.[65]

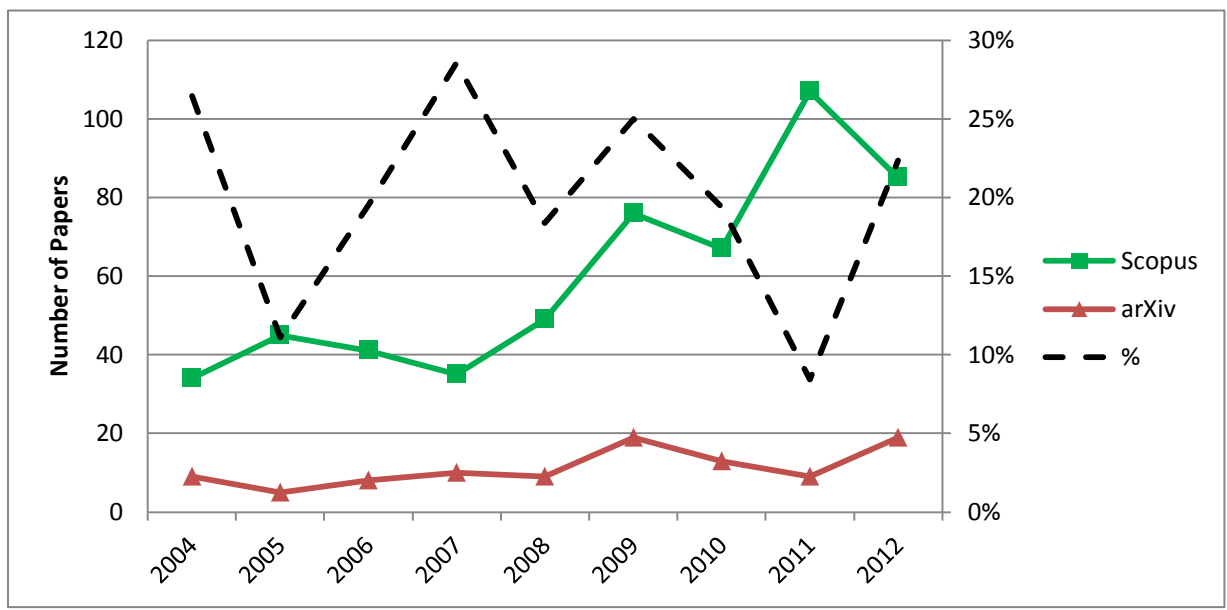

**Figure 41.** Growth of publication numbers between 2004 and 2012 for Physical Biology.

In Figure 41 the publication output for Physical Biology is illustrated. It is evident that the number of journal articles published per year has more than doubled from 34 in 2004 to 76 in 2009, and tripled to 107 in 2011. The total number of articles published between 2004 and 2012 is 539. For 101 of these articles a preprint can be found in arXiv. This accounts for 18.7%, a much higher quotient than for the Journal of Theoretical Biology. In addition, 13 postprints were posted in arXiv, which means that more than one in five papers is available in arXiv. Different from the Journal of Theoretical Biology, there is no obvious trend for preprint publication because the number of articles with a preceding preprint has been oscillating in the past five years. The next figure illustrates the publication delay. The histogram indicates that the mode value of the publication delay is one month.

---

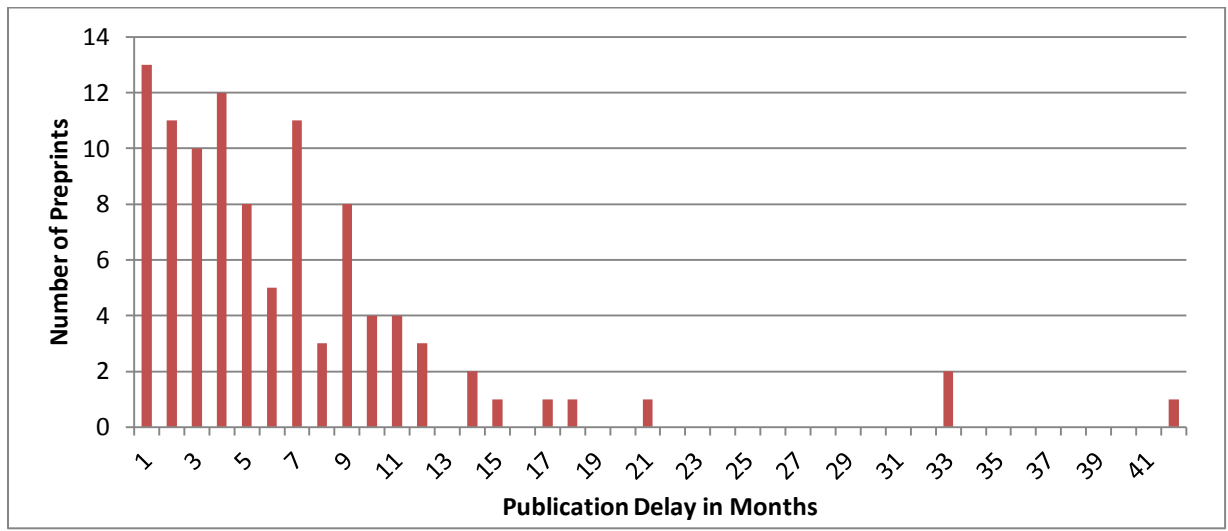

**Figure 42.** Distribution of preprints over months prior to journal publication. Those preprints are considered whose articles were published between 2004 and 2012 in Physical Biology.

A small majority of 13 preprints are published within 1 and 30 days prior to journal publication. Furthermore, it is visible that most of the preprints are published close to the date of formal publication. This can be explained by the fact that authors may place their paper in arXiv after it has been proofread and accepted for publication. It is furthermore evident that the majority of preprints are published within one year prior to the journal article (91.1%). The median publication delay of all preprints is 147 days, with a minimum of 2 days and a maximum of 1,258 days. It is accordingly much shorter than for the Journal of Theoretical Biology. Again it is of interest to see the distribution of preprints over the primary categories in arXiv.

**Table 4.** Distribution of all preprints over the first five primary categories in arXiv for Physical Biology.

| Primary category | Number of preprints |
|---|---|
| physics.bio-ph (Biological Physics) | 24 |
| q-bio.BM (Biomolecules) | 16 |
| q-bio.MN (Molecular Networks) | 14 |
| q-bio.SC (Subcellular Processes) | 12 |
| cond-mat.soft (Soft Condensed Matter) | 11 |

Apparently, the primary category of the preprints reflects the journal's scope. Most of the preprints are categorized in Biological Physics, a subcategory of Physics. It is followed by Biomolecules, Molecular Networks, and Subcellular Processes. Whereas, overall 114 e-prints exist for this journal, the search in arXiv reveals approximately 66 hits. It shows again that many biologists provide a reference to the final journal publication, which upgrades their paper and verifies it.

**Table 5.** Average citation rates for Physical Biology. The citation window is two years after the date of publication and the citation data is based on Scopus.

| Publication year | Number of articles having no preprint | Average citation number | Number of articles having a preprint | Average citation number |
|---|---|---|---|---|
| 2004 | 25 | 1.80 | 9 | 1.78 |
| 2005 | 40 | 4.45 | 5 | 3.60 |
| 2006 | 33 | 4.82 | 8 | 7.00 |
| 2007 | 25 | 4.04 | 10 | 4.40 |
| 2008 | 40 | 4.85 | 9 | 5.44 |
| 2009 | 57 | 4.95 | 19 | 6.84 |
| 2010 | 54 | 3.43 | 13 | 3.31 |
| Total | 274 | 4.18 | 73 | 4.88 |

Table 5 is structured in the same way as Table 2 for the Journal of Theoretical Biology. What we can infer from Table 5, is that in the first year of publication the citation numbers for articles having a preprint, and those that do not have a preprint are the lowest. Apparently, because it is the journal's first year after formation and it had to take root first. It is furthermore striking that the average citation rate for journal articles without a preprint in arXiv is roughly four and is not subject to strong variations. On the other hand, the average citation number for articles with a previous preprint fluctuates between 3.6 and 7 citations per article. It suggests that there are either outstanding articles among them or that they are better cited because they had a longer time to accumulate citations. Especially if e-prints have a note in arXiv on the formal publication, they are likely to be cited properly, prior to their publication in Physical Biology. However, Welch's t-test has shown that there is no significant citation advantage for articles with a previous preprint in arXiv. Additionally, there is no significant advantage to get earlier cited.

### 5.4.3   Discussion

The two presented journals suggest that Quantitative Biology is a well accepted field in arXiv. The number of preprints for both journals has been steadily growing over the years and in all likelihood the number will continue to grow. It is possible, though, that the preprint number for Quantitative Biology is comparatively high because the papers are submitted by physical scientists who are more familiar with arXiv than biologists. Especially the primary category showed that a high share of preprints for Physical Biology as well as for Journal of Theoretical Biology is classified in physics.

A statistical geneticist, Graham Coop from the University of California, told Nature that "The speed of discussion, comment, and pre-publication review allowed is needed in biology more than most fields" (Grant, 2012). Coop's view implies that the time between the submission of a manuscript and its formal journal publication is too long in biology. Most likely it is longer than the publication delays presented here because some papers are placed in arXiv after the journal's acceptance. It was shown that the median publication delay of all preprints in the Journal of Theoretical Biology is eight months, whereas it is five months for Physical Biology. The shorter delay for Physical Biology might be due to the reason that the journal was founded only in 2004; one year after Quantitative Biology was included in arXiv. The citation analysis has shown that there is no significant advantage for articles with a previous preprint to receive more citations within the first two years after journal publication.

It has to be taken into account that there are several reasons that forbid the extended use of arXiv in biology. It is a dynamic field and has commercial potential due to patentable research. Since the competition is higher than in other research fields, biologists are cautious to publish their findings prior to formal publications. This may explain the histogram of the publication delay for Physical Biology, where most of the preprints are deposited close to the date of journal publication. Also, the high number of preprints dealing with evolution in the Journal of Theoretical Biology reveals that they not only meet the journal's scope, but that patenting does not play an important role in this subfield.

Even if preprints and articles are one and the same paper in regard to content, the perception in a field such as biology may be different. Peer-reviewed journals remain the principal means of communication in biology because they guarantee correctness and quality. However, rapid review is important because it increments the growth of discoveries, which are open to the public. The common practice for journals in Biology nowadays, is to publish the article online once it is accepted. Harley et al. (2010, p. 229) write that the communication and publication means have not changed in biology but accelerated over time, particularly in subfield such as genomics and biochemistry, whereas in neurobiology speed does not play a crucial role.

Besides arXiv, there are several other preprint servers for biology. Among those are CogPrints, Nature Precedings, and E-Print Network-Biotechnology. CogPrints is a collection of self-archived papers related to psychology, biology, neuroscience, linguistics, and philosophy.

## 5.5 Library and Information Science

Since January 1993, arXiv includes the field Computer Science, which makes up 5.2% (41,771) of all submissions in arXiv, until the year 2012.[66] Computer Science comprises subfields such as Artificial Intelligence, Cryptography and Security, Databases, Digital Libraries, Hardware Architecture, Information Retrieval, Multimedia, Operating Systems, Robotics, and Social and Information Networks. For the following bibliometric analysis Library and Information Science (LIS) was chosen as a representative subject relevant to the field Computer Science.

Several journals have been examined to see if they have articles with a preceding preprint in arXiv. For the journals Library Trends, and Library Hi Tech no single preprint was found in arXiv, whereas the Journal of Documentation, and the Journal of Information Science had one preprint each, and the International Journal on Digital Libraries three preprints. These disillusioning results allow us only to analyze the following three journals in more detail: Journal of the American Society for Information Science and Technology (JASIST), Scientometrics, and Journal of Informetrics.

### 5.5.1 JASIST

The Journal of the American Society for Information Science (JASIS) was originally founded in 1950. In 2001, the journal was named JASIST. Since then, it publishes "reports of research and development in a wide range of subjects and applications in information science and technology."[67] It is published by John Wiley & Sons on behalf of the American Society for Information Science and Technology. The Impact Factor is 2.081 (2011) and the journal appears monthly, i.e., 12 issues in 1 volume a year.

The time period of interest is 2005-2012. In these eight years, 1,534 articles have been published; of which 66 have a previous preprint in arXiv. It results in a fraction of 4.3% of all articles published between 2005 and 2012. Besides these preprints, 26 postprints exist in arXiv. Table 6 presents the publication year, the total number of articles, the number of articles with a foregoing preprint, and the percentage. As we can gather from Table 6, the number of articles with a preceding preprint in arXiv has grown from 1 in 2005 to 11 in 2012.

---

As a consequence, the percentage of articles having a preprint in arXiv has been rising as well, showing a peak in 2010 where 11.1% of all articles published have a previous preprint.

**Table 6.** Overview of publication growth between 2005 and 2012 for JASIST.

| Publication year | Number of articles | Articles with preprints | % |
|---|---|---|---|
| 2005 | 139 | 1 | 0.7% |
| 2006 | 182 | 3 | 1.6% |
| 2007 | 206 | 3 | 1.5% |
| 2008 | 194 | 7 | 3.6% |
| 2009 | 212 | 10 | 4.7% |
| 2010 | 189 | 21 | 11.1% |
| 2011 | 214 | 10 | 4.7% |
| 2012 | 198 | 11 | 5.6% |
| Total | 1,534 | 66 | 4.3% |

The growth of preprints in recent years might be due to a policy change. In 2008, JASIST has implemented "green" Open Access (Kraft, 2008, p. 1539). Thus, authors who publish in JASIST have the right to self-archive preprints as long as they provide a link to the final article of the online journal. Indeed, many authors provide information on the formal publication. In arXiv's Journal-ref. and Comments search for the period 2005 to 2012, 43 e-prints were found that refer to JASIST. What was striking in the analysis of JASIST is that three preprints were also found in INSPIRE HEP. Their titles suggest that they are not related to HEP but are of general interest to the HEP community.[68]

The median value of the publication delay of all 66 preprints is 204 days, with a minimum of 9 days, and a maximum of 870 days. Furthermore, Welch's t-test has been applied to see whether there is a citation advantage for articles with a foregoing preprint or not. Only articles published between 2007 and 2011 were analyzed with a citation window of one year. As a result, articles having a preprint in arXiv are on average 1.5 times more cited than articles without a preprint, one year after journal publication (P=0.033). The same articles have been analyzed to see if there is an advantage in regard to citation delay. The t-test proves that articles with a preprint in arXiv are on average two months earlier cited than articles without a preprint (P=0.014).

---

[68] Information Resources in High- Energy Physics: Surveying the Present Landscape and Charting the Future Course. http://inspirehep.net/record/783701
Positional Effects on Citation and Readership in arXiv. http://inspirehep.net/record/826991
Last but not Least: Additional Positional Effects on Citation and Readership in arXiv. http://inspirehep.net/record/873683

### 5.5.2 Scientometrics

The journal Scientometrics was established in 1978 and is published by Springer. It comprises quantitative aspects of the science of science, and the communication in science.[69] The 2011 Impact Factor is 1.966. Currently, the journal appears in four volumes a year, each volume containing three issues.

**Table 7.** Overview of publication growth between 2005 and 2012 for Scientometrics.

| Publication year | Number of articles | Articles with preprint | % |
|---|---|---|---|
| 2005 | 118 | 4 | 3.4% |
| 2006 | 144 | 2 | 1.4% |
| 2007 | 129 | 4 | 3.1% |
| 2008 | 129 | 2 | 1.6% |
| 2009 | 193 | 6 | 3.1% |
| 2010 | 232 | 9 | 3.9% |
| 2011 | 224 | 10 | 4.5% |
| 2012 | 436 | 18 | 4.1% |
| Total | 1,605 | 55 | 3.4% |

As we can see in Table 7, the number of articles has more than tripled over time; from 118 in 2005 to 436 in 2012. Overall, Scientometrics has 1,605 articles published in the reference period, of which 55 have a foregoing preprint. In addition, 14 postprints were placed in arXiv. The number of articles with a preceding preprint has been growing continuously, and reached a maximum of 18 in 2012. The search for Scientometrics in arXiv's Journal-ref. as well as the Comments field provides 51 hits for e-prints with the restriction to 2005-2012. It shows that authors publishing in Scientometrics are diligent when it comes to providing a reference to the article version. Just as for JASIST, four preprints are included in INSPIRE HEP, which were published in Scientometrics at a later date.[70] It is worth mentioning that although the preprints are covered by INSPIRE HEP, there is no hyperlink from arXiv to INSPIRE for citation view.

The median value of the publication delay of all 55 preprints is 211 days, thus almost as high as for JASIST. The delay ranges between 1 and 902 days prior to journal publication. It is of further interest to see if there is any citation advantage for articles with a previous preprint. Therefore, articles published between 2007 and 2011 were considered with a citation window

---

of one year. Welch's t-test has revealed that articles having a previous preprint version in arXiv are on average 0.5 times more cited than those without a preprint (P=0.030). In addition, Welch's t-test proves that articles with a foregoing preprint, receive their first citation on average one month earlier (P=0.014).

### 5.5.3   Journal of Informetrics

The Journal of Informetrics was only established in 2007 and is published by Elsevier. It presents papers on fundamental quantitative aspects of information science.[71] The journal focuses on articles that describe methods (mathematical, probabilistic, and statistical), theories, techniques, and important data. The 2011 Impact Factor is 4.229, and the journal is published annually in one volume consisting of four issues.

**Table 8.** Overview of publication growth between 2007 and 2012 for the Journal of Informetrics.

| Publication year | Number of articles | Articles with preprint | % |
|---|---|---|---|
| 2007 | 33 | 5 | 15.2% |
| 2008 | 34 | 4 | 11.8% |
| 2009 | 36 | 4 | 11.1% |
| 2010 | 69 | 13 | 18.8% |
| 2011 | 67 | 11 | 16.4% |
| 2012 | 78 | 12 | 15.4% |
| Total | 317 | 49 | 15.5% |

Table 8 provides an overview of the publication output for the Journal of Informetrics since its foundation in 2007. On the whole, 317 articles have been published within six years of existence, of which 49 have a previous preprint in arXiv. These preprints account for 15.5% of all articles published. Table 8 indicates that the number of articles with a foregoing preprint has been growing over the years, just as the number of journal articles. The share of articles with a previous preprint alternates between 11.1% in 2009 and 18.8% in 2010. Although the absolute number of preprints is similar to JASIST and Scientometrics, the Journal of Informetrics reveals a relatively high percentage share. Without much doubt, this is due to the lower number of articles published per year, although this number has rapidly doubled from 2007 to 2011. In addition, 5 postprints exist in arXiv, and overall 25 e-prints provide a reference to the journal article.

---

[71] Elsevier. Journal of Informetrics. http://www.journals.elsevier.com/journal-of-informetrics/

The median value of the publication delay of all preprints is 168 days. The shortest delay is 37 days, whereas the longest delay is 820 days. The high share of preprints allows a comparison of citation numbers. The following table lists the average citation numbers for articles published since 2007.

**Table 9.** Average citation rates for the Journal of Informetrics. The citation window was set to one year after the date of publication and the citation data is based on Scopus.

| Publication year | Number of articles with no preprint | Average citation number | Number of articles with preprint | Average citation number |
|---|---|---|---|---|
| 2007 | 28 | 1.50 | 5 | 0.40 |
| 2008 | 30 | 1.27 | 4 | 2.25 |
| 2009 | 32 | 2.25 | 4 | 2.75 |
| 2010 | 56 | 1.57 | 13 | 7.00 |
| 2011 | 56 | 1.86 | 11 | 4.27 |
| Total | 202 | 1.70 | 37 | 4.32 |

From Table 9 we can infer that articles having a previous preprint are on average more cited than articles that do not have a preprint in arXiv. Welch's t-test has been again applied to see if there is a significant citation advantage. Since the journal was only established in 2007, articles published between 2007 and 2011 were analyzed and the citation window was set to one year. From the test it follows that articles having a preprint in arXiv receive on average one more citation than articles without a preprint, one year after journal publication (P=0.026). According to Welch's t-test, articles having a preprint in arXiv are on average six weeks earlier cited than articles without a preprint deposited in arXiv (P=0.029).

### 5.5.4 Discussion

To sum up, the numbers of preprints for the three presented journals are not as high as in other disciplines, but are on the rise and more than likely to grow in future. To link the publication numbers to the aspect of acceleration of scholarly communication, the publication delay of each journal was presented. The median value of the publication delay ranged between five and seven months. It has been shown that this delay is of value and might explain why articles with a previous preprint get on average one to two months earlier cited. In addition, articles with a previous preprint version in arXiv are on average 0.5 to 1.5 times more cited than articles without a preprint. This citation advantage is measurable within one year after journal publication. However, there are many reasons that substantiate these effects.

Kurtz et al. (2005) explained the advantageous effect of Open Access on citations, which can be applied to preprints as well. Firstly, they claim that the arXiv increases access to articles, especially for those who are not able to access subscription-based journals. Secondly, papers appear earlier in arXiv than on the publisher's website. During this longer time period preprints can be read and cited. Lastly, they claim that authors either deposit only their best papers in arXiv (self-selection postulate), or that only the better authors use arXiv (author postulate). These two postulates are labeled as the quality postulate. The author postulate can be supported for the results presented above if we have a closer look on the authors. Therefore, Table 10 provides an overview of the top-5 authors publishing in JASIST, Scientometrics, and the Journal of Informetrics. The data is based on overall 170 articles that were published in these three journals between 2005 and 2012, and have a preceding preprint in arXiv.

**Table 10.** Top-5 authors publishing in JASIST, Scientometrics, and Journal of Informetrics between 2005 and 2012. The table lists the author's name and the according number of preprints published in arXiv.

| JASIST | | Scientometrics | | Journal of Informetrics | |
|---|---|---|---|---|---|
| Leydesdorff, L. | 14 | Leydesdorff, L. | 8 | Bornmann, L. | 6 |
| Van Raan, A.F.J. | 4 | Waltman, L. | 4 | Leydesdorff, L. | 6 |
| Waltman, L. | 4 | Vanclay, J.K. | 3 | Waltman, L. | 4 |
| Schreiber, M. | 3 | Van Raan, A.F.J. | 3 | Rodriguez, M.A. | 3 |
| Davis, P.M. | 3 | Mryglod, O. | 2 | Vanclay, J.K. | 3 |

It is striking that the preprint numbers are affected by few authors, the so-called early adopters of arXiv. Obviously, three authors, namely Leydesdorff, Waltman, and Van Raan, account for 47 preprints published at a later date in these three journals. Every sixth preprint is attributed to Leydesdorff. Moreover, arXiv suggests that Leydesdorff is indeed an early adopter of Open Access. The search in arXiv's author field results in 195 hits for Leydesdorff.

# 6. Conclusion

The goal of the work was to examine the potential of arXiv to boost scholarly communication. Most potential to accelerate scholarly communication bears the publication delay. It was shown that it can take months or years to get a paper published in a journal because the traditional publication process is characterized by delays that vary from discipline to discipline. Whereas in HEP and astrophysics it takes two to six months for a paper to get published, one to three years can elapse in mathematics until the final article appears. Biology and LIS revealed a median publication delay of five to eight months. This paper demonstrated furthermore that the publication delay between the submission in arXiv and the publication in a Scopus-indexed peer-reviewed journal has not significantly decreased over the years. An obvious reason for this outcome may be that in the past, authors submitted their papers randomly to arXiv, whereas in recent years they do it as early as possible. Consequently, the time gap between submission and publication becomes larger. However, the lag time depends also on the subject field, the submission rate, the journal publisher, the peer-review process, the quality of the submitted paper, and the need for fast publication. Most of the time, papers are just waiting to be read. This situation could be changed with an open peer-review process.

Moreover, if the journal publication delay were not as long as it often is, arXiv would not have been so successful. Scholars use arXiv because they know that they can bypass the longsome publication process and get their results published as early as possible. Besides, in fields with high competition, such as astrophysics, preprints can establish priorities and keep other research groups away. ArXiv is not only an effective means to distribute results as early as possible; scholars can also access most up-to-date research papers. We can conclude that since the advent of arXiv, the research advancement in fields such as astrophysics and HEP is driven by the publication of preprints because this communication means is capable of increasing the speed of delivery.

Throughout the work arXiv has been presented as a flawless example of a preprint server, although it also has its drawbacks. ArXiv may put pressure on researchers and force them to keep up with colleagues. This can lead to redundant papers. Another negative effect of arXiv may be that it diminishes the incentive to publish articles Open Access, because most of the literature is freely available in arXiv. Larivière et al. (2012) analyzed papers that were published in arXiv between 1991 and March 2012. They observed that slightly less than 50% of arXiv submissions were also findable in WoS.

When research literature is available for free to this extent - who is willing to pay for journal subscriptions? Even if the share of papers available for free is high, arXiv will never displace journals. Journals are fundamental in guaranteeing quality with the peer review they perform. This has been proven in this paper by the high amount of preprints that is submitted to journals. The highest share is in HEP, where more than 90% of journal articles can be found as preprint versions in arXiv. Astrophysics and mathematics revealed a share of 50% to 60%, whereas researchers publishing in Quantitative Biology and LIS do not yet make excessive use of arXiv, and rely on refereed journals.

To Ginsparg (2011) it seemed unthinkable that free access to non-refereed papers in arXiv could coexist with refereed subscription-based publications. Nowadays, researchers make use of both sources, and see their benefits. Besides, peer review starts with arXiv because everyone can participate and share comments. As mentioned above, there is no capacity to integrate true peer review into arXiv. Moderators can only remove erroneous papers, which is the reason why one day elapses between the upload of a paper and its announcement. Nevertheless, a commenting system in arXiv could enrich the informal peer review in future.

According to O'Connell (2000, p. 6) peer review is important "both for the reader who wants to be sure that a paper has been judged worthy, and the author who needs to readily demonstrate a certain level of professional ability." Therefore, as soon as the preprint is accepted by a journal, authors should provide a note on the article in the Journal-ref. field in arXiv. It not only fulfills the journal's demand, but also upgrades the preprint, showing that it is of quality. As has been presented, the majority of authors do not update the information on the final journal article. In mathematics, only one in four authors or even fewer provide a note on the journal article. In Quantitative Biology the share was much higher, which proves that biologists rely on peer-reviewed articles. However, the highest share of authors providing a note on the final publication was in LIS, probably because authors are aware of Open Access, and all aspects of the publication process. The missing information on the formal article mainly derives from the long time that elapses between the submission and the publication, so that authors simply forget about it. Publishers are aware of long delays in the publication cycle and want to decrease any time gaps by slogans such as Elsevier's "Open to accelerate science"[72], or SpringerLink's "Accelerating the world of research".[73] Online publication of journals has reduced the time between the reception and the final publication immensely. Thanks to early view and e-first systems there is no longer need to wait for printed issues. Articles are available online, instantly after they have been proofread, and they are citable on the basis of DOIs.

---

[72] SciVerse. Accelerate Science. http://www.acceleratescience.com/
[73] Springer. SpringerLink. http://www.springerdemos.com/

The citation analysis in this work was approached in two different ways. For high-energy physics and astrophysics the analysis was based on INSPIRE HEP that tracks citations to preprints in arXiv. It has been shown that 69% to 84% of preprints in HEP receive their first citation prior to publication. In astrophysics, a quarter of all preprints received their first citation prior to publication. Furthermore, it was shown that preprints in HEP and astrophysics can accumulate many citations before they are published as a journal article. We can conclude that the arXiv accelerates citations because it makes papers early and freely available. The effect of early visibility can be supported by the fact that colleagues start reading preprints, and cite them as soon as they are accessible. It was also shown that papers published in astrophysical journals with a citation window of two years received on average three more citations by the end of two years, if they had a previous preprint in arXiv.

In mathematics, biology, and LIS the citation analysis was based on data retrieved from Scopus. Welch's t-test showed that articles that were both submitted to arXiv and at a later date published in a journal, had in the majority of cases a significant citation advantage in regard to time and total number. As mentioned above, Kurtz et al. (2005) distinguished three effects to describe a genuine advantage: the open access effect, the early view effect, and the self-selection effect. The latter effect has been illustrated in LIS, where early adopters influence the number of preprints immensely. However, if the author's performance is outstanding and not affected by the way of publishing, then the advantage of preprint publishing fades away. Briefly worded, top authors such as Leydesdorff, produce more papers than less prominent authors, and those papers are likely to receive more citations. On the other hand, the argument that arXiv users publish only high-quality papers can be neglected because most authors do it routinely, and are not aware of the impact their paper might have in future.

What almost all presented journals have in common is the growth in publication numbers and the adoption rate of preprints. Only the journal Nuclear Physics B showed a remarkable decrease in journal articles published per year. The bibliometric results presented depend very much on the discipline. High-energy physicists and astrophysicists have been at the forefront of using preprints as a rapid communication medium, due to the timeliness of their research and relatively closed groups in which they communicate research results. For researchers in disciplines such as HEP and astrophysics to deposit a preprint in arXiv is a normal step in the circulation of research outcomes. In addition, the long existing preprint culture in physics allows a co-existence with journals. It has been shown that APS is supportive of arXiv and allows submissions explicitly, just as IOPscience and Elsevier. Indeed, publishers do not have to fear arXiv as long as they are aware of the value-adding role of peer review. The arXiv has even affected the SCOAP[3] initiative, which wants to achieve Open Access for journals published in the HEP field.

The results presented also indicate that mathematicians are making a transition to publishing preprints. They are more likely to publish their research openly accessible, compared to researchers in other fields. The longevity of mathematical journal articles and the long publication delays encourage many mathematicians to publish in arXiv far ahead of the formal publication. Outside of the physical sciences and mathematics, preprints are slowly taking root. Nevertheless, the success of preprint publishing in HEP serves as a model for researchers in other disciplines. Many scholars look nowadays for new papers primarily in arXiv instead of published journals.

To sum up, the results presented provide evidence that arXiv is capable of accelerating scholarly communication. Since the pressure to publish content quickly is rising in almost every field, speed in publication is crucial. ArXiv has revolutionized the speed in scholarly communication because it makes results available prior to journal publication, and scholars do not hesitate to cite un-refereed preprints. Accelerating the access to results is indeed the primary function of a preprint server.

In conclusion, the bibliometric analysis focused on 13 journals scattered over five distinct research fields. Because of the small sample size, one has to be careful with generalizing. Future research should investigate the number, and the speed in citations and impact of preprints in arXiv that are eventually never published as journal articles. Another interesting research question is in how far a single preprint can be the origin of multiple articles, especially if several authors publish a preprint (e.g., as a research group), and individuals base their articles on this preprint. Furthermore, it would be illuminating to find out to which extent authors publish their better papers in arXiv. Likewise, it would be interesting to investigate to which extent authors are overrepresented in certain disciplines.

Also, it is worth analyzing the effect of early view on arXiv. It might be that early view diminishes the speed advantage of arXiv. Finally, it may be important to find out in how far the results in WoS are different from those in Scopus. Scopus revealed many weaknesses in the completeness of data. It may have been a better option to retrieve metadata straight from the publisher's website. Since this work documented the growing reliance on preprints in arXiv, future work should focus on other preprint servers such as CERN's document server for high-energy physics, or CogPrints in Biology.

# 7.    List of References

**Online Sources**

All online sources were last accessed 5[th] March 2013.

**Honesty Declaration**

I hereby declare that the submitted Master Thesis is my own and that all passages that are not mine have been fully and appropriately acknowledged. The parts that were taken literally from existing publications are identified as such. The data collection was conducted with the help of Daniel Lunow, a student of Computer Science and Mathematics at Humboldt University Berlin. It was him who set up the MySQL database and enabled me to query it.

_______________________________________

Signature                     Date